\documentclass[superscriptaddress,floats,prd,nofootinbib,eqsecnum]{revtex4-2}
\usepackage{amssymb,amsmath,amssymb,amsfonts,amsthm,stmaryrd,mathrsfs,physics,bbm,mathtools,bm,hyperref}
\usepackage{graphicx}
\usepackage[T1]{fontenc}

\newcommand{\makeSymbol}[1]{\mathord{\vcenter{\hbox{#1}}}}
\newcommand{\sgn}{\mathrm{sgn}}

\newcommand{\sut}{{\mathrm{SU(2)}}}
\newcommand{\cylg}{\mathbf{Cyl}_G}
\newcommand{\grade}{{\mathfrak d}}
\newcommand{\hop}{\mathbf{H} } 
\newcommand{\p}{\mathbf{P}}

\newcommand{\g}{\mathbf{g} } 

\newcommand{\hf}{\mathscr H_F}

\newcommand{\leftscript}[2]{\prescript{#2\csname prescriptcorrection\detokenize{#1}\endcsname {\mkern-8mu\relax}}{}{#1}}

\newcommand{\am}{\mathfrak a}

\def\bee{\begin{eqnarray}}
\def\eee{\end{eqnarray}}

  \newcommand{\Ff}{\mathfrak{F}}

  \newcommand{\Fj}{\mathfrak{J}}

  \newcommand{\Fq}{\mathfrak{Q}}

\newcommand{\ga}{\gamma}

\renewcommand{\o}{\omega}

\begin{document}

\title{Fermions in Loop Quantum Gravity and Resolution of Doubling Problem}

\author{Cong Zhang}
\email{cong.zhang(AT)fuw.edu.pl}
\affiliation{Institut f\"ur Quantengravitation, Friedrich-Alexander Universit\"at Erlangen-N\"urnberg, Staudtstr. 7/B2, 91058 Erlangen, Germany}

\author{Hongguang Liu}
\email{hongguang.liu(AT)gravity.fau.de}
\affiliation{Institut f\"ur Quantengravitation, Friedrich-Alexander Universit\"at Erlangen-N\"urnberg, Staudtstr. 7/B2, 91058 Erlangen, Germany}

\author{Muxin Han}
\email{hanm(AT)fau.edu}
\affiliation{Department of Physics, Florida Atlantic University, 777 Glades Road, Boca Raton, FL 33431-0991, USA}
\affiliation{Institut f\"ur Quantengravitation, Friedrich-Alexander Universit\"at Erlangen-N\"urnberg, Staudtstr. 7/B2, 91058 Erlangen, Germany}

\begin{abstract}
The fermion propagator is derived in detail from the model of fermion coupled to loop quantum gravity. As an ingredient of the propagator, the vacuum state is defined as the ground state of some effective fermion Hamiltonian under the background geometry given by a coherent state resembling the classical Minkowski spacetime. Moreover, as a critical feature of loop quantum gravity, the superposition over graphs is employed to define the vacuum state. It turns out that the graph superposition leads to the propagator being the average of the propagators of the lattice field theory over various graphs so that all fermion doubler modes are suppressed in the propagator. This resolves the doubling problem in loop quantum gravity. Our result suggests that the superposition nature of quantum geometry should, on the one hand, resolve the tension between fermion and the fundamental discreteness and, on the other hand,  relate to the continuum limit of quantum gravity.
\end{abstract}

\maketitle


\section{introduction} 

The physical world comprises gravity and matter. Even though the quantum theory of matters has been well-developed in standard model, the gravity therein is still classical. However, it has been suggested that a consistent theory of gravity coupled with quantum matters should contain gravity quantized \cite{Page:1981aj}. On the other hand, Quantum gravity (QG) also needs matter couplings because they provide probes for empirically testing QG effects \cite{Deser:1974cy,tHooft:1974toh,Goroff:1985th}. Particularly matter couplings with gravity provide a toolbox for early studies of the QG's ultraviolet (UV) behavior, and play essential roles in cosmology, black holes, asymptotic safety, QG phenomenology, etc \cite{Ashtekar:2021dab,Hawking:1975vcx,Dona:2013qba,Perez:2017krv}.

Loop Quantum Gravity (LQG), as a promising candidate of background-independent and non-perturbative QG theory, sets the stage for matter coupling with QG \cite{rovelli1988knot,ashtekar2004back,rovelli2005quantum,thiemann2007modern,han2007fundamental}. In LQG, quantum states are given by spin networks, i.e., graphs that carry quantum spins. The graphs with the quantum spins thereon give rise to quantum geometries: the quantum numbers carried by a vertex provide the quantum volume of a chunk dual to the vertex, and the quantum numbers on the edges of the vertex give the surface area of the chunk. Hence, the graphs in LQG are interpreted as spacetime quanta and are where matters can be. Indeed, matter couplings in LQG have been extensively explored (see e.g. \cite{ashtekar1989new,Thiemann:1997rt,Sahlmann:2002qj,Oriti:2006jk,Kamiski2006,Bojowald:2007nu,Bianchi:2010bn,Domagala:2010bm,Zhang:2011vg,Bodendorfer:2011ny,Kisielowski:2018oiv,Mansuroglu:2020acg,Lewandowski:2021bkt}). Notably, a framework of standard model coupled to LQG has been developed with the exciting feature of ultraviolet regularity \cite{Thiemann:1997rt}. In these models, the quantization procedure leads the matter fields to be located at the vertices or edges so that they are coupled to the quantum geometry given by the quantum spins. The picture constructed by the model shares several common features with the Lattice Field Theory (LFT), especially when the graphs are chosen to be cubic. However, in LQG, the graphs are dynamics, so it makes sense to consider superposition over graphs. It will be shown in our work that this difference could make LQG essentially differ from LFT.  

Relating the theory of LQG coupled to matter fields to the QFT in flat/curved spacetime is an exciting and challenging topic. There have been many attempts towards this issue \cite{varadarajan2000fock,Thiemann:2020cuq,assanioussi2022loop}. Our work addresses this issue by considering the propagator of LQG coupled to the chiral fermions \cite{Thiemann:1997rt,Mansuroglu:2020acg,Lewandowski:2021bkt}. In a propagator, an essential ingredient is the vacuum state. In the standard QFT, the vacuum state is the ground state of the Hamiltonian. To analog this definition in LQG, we meet the first obstacle that the Hamiltonian of QFT is background-dependent, where the background is classical. Hence, we need to propose a quantum spacetime state that resembles the classical background in QFT. A natural choice for such states is Thiemann's coherent states in LQG, which turn out to have desirable semiclassical features \cite{thiemann2001gauge,thiemann2001gaugeII,thiemann2001gaugeIII}. Thiemann's coherent states are given based on graphs like a generic state in LQG. Then an issue arises on which graphs should be selected for the coherent states. The preferred is the cubic graphs because the Hamiltonian operator has a physically satisfactory expectation value in the coherent states on cubic graphs. Given a cubic graph $\gamma$, applying the coherent state thereon peaked at the Minkowski geometry to sandwich the Hamiltonian operator, one can introduce an effective fermion Hamiltonian $\hat H^{\rm eff}_{F,\gamma}$ comprising only the matter field operators \cite{Sahlmann:2002qj}. Indeed, since the Hamiltonian contains operators concerning matter field and background, sandwiching it in the coherent state could change those background terms into their classical expectation values plus some $\hbar$-order corrections. Since the ground state $|\omega,\gamma\rangle$ of $\hat H^{\rm eff}_{F,\gamma}$ can be interpreted as the LQG analogy of QFT vacuum, the LQG vacuum on $\gamma$ will be defined as the tensor product of the coherent state and $|\omega,\gamma\rangle$. Moreover, as an essential difference between LQG and LFT, the superposition of various cubic graphs should be employed in defining the vacuum. The superposition feature of the vacuum is expected to resolve the fermion doubling.    

The fermion doubling  is a problem suffered by chiral fermions in LFT. In LFT of the chiral fermion, each fermion results in $2^m$ fermion species on $m$-dimensional lattice \cite{montvay_munster_1994,Nielsen:1981hk}. Since the fundamental discreteness of LQG makes the fermion coupling resemble the LFT to some extent, it is suspicious of the doubling problem in LQG \cite{Barnett:2015ara}. The doubling problem is a crucial issue because it reveals the tension between fermion and the fundamental discreteness of quantum spacetime, given that the fundamental discreteness is believed to be a key feature of QG \cite{Hooft_2016,1955PhRv...97..511W,Hawking:1978pog}. The confusion on fermion doubling has been long-standing in the LQG community since the first paper on LQG-fermion in 1997. The similar issue should exist in all QG approaches with discrete spacetimes. This problem is expected to be resolved by the superposition of quantum geometries \cite{Gambini:2015nra,Bianchi:2010bn,Han:2011as}. Actually, it will be shown by our work that the doubling problem can really be resolved by the the graph-superposition vacuum state.

In the Hamiltonian formalism of general relativity, the Hamiltonian is a generator of a gauge transformation, so the dynamics depict a picture of frozen evolution. A system with this feature is called the Hamiltonian-constraint system. To get the physical evolution of this system, one introduces the notion of relational evolution, where some other fields need to be employed as a physical reference frame. In the relational evolution formulation, the Dirac observables are those functions depending on the physical reference frame fields. Thus, these Dirac observables are explained as the relational evolution of some physically interesting fields with respect to the reference frame fields. The current work will introduce the Gaussian dust fields as reference frame fields so that the physical Hamiltonian governing the relational evolution equals the Hamiltonian constraint with N=1 as the lapse function. The Hamiltonian may usually be quantized in two ways: graph-changing and graph-preserving. Generally speaking,  a continuum field theory approach leads directly to the graph-changing action where the graph acted by the Hamiltonian will be changed by adding loops or vertices. Several proposals for graph-changing quantum Hamiltonian operators were considered in the literature. The graph-preserving approach is natural from the lattice discretization point of view. It makes the action of operators reducible to subspaces corresponding to the graphs. For every graph, the analytic properties of operators are much easier to study. Our work adopts the graph-preserving version but with a  small amount of difference which guarantees that the resulting operator has a well-defined adjoint. 

The Hamiltonian in the current work is complicated and contains not only the fermion operators but also the operators referring to the background metric. Indeed, the complicacy of the Hamiltonian is a primary obstacle to blocking LQG's progress. Due to the complicacy of the Hamiltonian, we apply the path integral formulation to deal with the evolution operator in the propagator. The path integral formula contains the integral of $e^{i/\hbar S}$ with $S$ denoting the effective action. As shown in the body part of the paper, $S$ is a quadratic form of the fermion field. Thus the integral concerning the fermion field takes the form of a standard fermionic Gaussian integral and is calculatable explicitly. Thus, we should pay more attention to the integral of
the holonomies and fluxes, i.e., the variable referring to the background metric. This integral can be evaluated using the stationary phase approximation. To this end, one needs to vary the action to get the equation of motion $\delta S=0$. Fortunately, this equation of motion has been well-studied in \cite{han2020effective} for the vacuum LQG. Regardless of the slight difference between the Hamiltonian in \cite{han2020effective} and the current work, the results in \cite{han2020effective} can be applied to get the solution of $\delta S=0$ adapting the boundary condition given the vacuum state. According to the stationary phase approximation algorithm, the solution endows the graphs of the vacuum state with a semiclassical dynamical geometry.

This paper is arranged as follows. In Sec. \ref{sec:one}, the classical and the quantum theories of gravity coupling fermion and Gaussian dust is reviewed, including the quantum kinematics, the coherent states, the gauge transformation and the Hamiltonian operator. In Sec. \ref{sec:pathintegral}, the path integral formulation on a single graph is introduced. Sec. \ref{sec:vacuumonagraph} defines the Minkowski vacuum. Then, in Sec. \ref{sec:propagator} we show the calculation of the propagator in detail. Finally, Sec. \ref{sec:doubling} shows how the graph superposition resolves the doubling problem. The work is summarized in Sec. \ref{sec:conclusion}.

\section{Preliminary}\label{sec:one}
\subsection{Gravity coupling fermion and  Gaussian dust}
Let $\mathcal M$ be the 4-manifold homeomorphism to $\mathbb R\times\Sigma$ with $\Sigma$ being the spatial 3-manifold. Further, we shall  assume that $\Sigma$ homeomorphism to  $\mathbb T^3$ with coordinate $x^a$ adapted to this topology. Namely, we choose the periodic boundary condition in our calculation. On $\mathcal M$, considering the model of gravity coupling Weyl fermion and the Gaussian dust, the  Lagrangian takes the form
\begin{equation}\label{eq:fullLagrangian}
\mathcal L=\mathcal L_G+\mathcal L_{F}+\mathcal L_D
\end{equation}
where $\mathcal L_{G}$ is the Einstein-Hilbert Lagrangian, $\mathcal L_{F}$ describes the Wely fermion field in gravity, and $\mathcal L_D$ is the Lagrangian of the Gaussian dust. Here the Einstein-Hilbert Lagrangian $\mathcal L_G$ is written with the tetrad field $e_\mu^I$,  which endows  the $\mathcal M$ with the metric $g_{\mu\nu}=\eta_{IJ}e^I_\mu e^J_\nu$. More precisely, $
\mathcal L_G=\frac{1}{2\kappa}R[e]$,
where $\kappa=8\pi G$, and $R[e]$ denotes the curvature of $g_{\mu\nu}=\eta_{IJ}e^I_\mu e^J_\nu$.  The fermion Lagrangian in the present work reads
\begin{equation}\label{eq:LF}
\mathcal L_F=\frac{i}{2} e \left(\psi^\dagger\sigma^I e_I^\mu\nabla_\mu\psi-c.c.\right)
\end{equation}
where $e$ is the determinant of $e_\mu^I$, $\sigma^I$  denotes the covariant Pauli matrices $(\mathbbm{1}_2,\sigma^i)$, and $\nabla_\mu\Psi$ is
\begin{equation}
\nabla_\mu\psi=\partial_\mu\psi+\frac{1}{4}\Gamma^I_{\mu J}\bar\sigma^I\sigma^J\psi.
\end{equation}
with $\bar\sigma^I=(1,-\sigma^i)$ and $\Gamma^I_{\mu J}$ being the spin connection satisfying $\dd e^I+\Gamma^I{}_J\wedge e^j=0$.  Indeed, the Lagrangian \eqref{eq:LF} implies that the left-handed Weyl fermion $\psi$ is chosen for our study.  The Gaussian dust comprises four scalar fields, denoted by $T$ and $S^a$ with $a=1,2,3$. Its Lagrangian $\mathcal L_D$ is
\begin{equation}
\mathcal L_D=-\sqrt{|\det(g)|}\left\{\frac{1}{2}\rho[g^{\mu\nu}(\partial_\mu T)(\partial_\nu T)+1]+(\partial_\mu T)(W_j\partial_\nu S^j) \right\},
\end{equation}
where the additional four scalar fields $\rho$ and $W_j$ ($j=1,2,3$) will play the role of Lagrangian multipliers and give rise to the primary constraints in the Hamiltonian theory.

Doing the Hamiltonian analysis of the Einstein-Hilbert Lagrangian, one gets a state of gravity described by the canonical pair $(K_a^i,E_i^a)$, where $E_i^a$ is the densitized triad filed on the  initial slice $\Sigma$  and $K_a^i$ is the extrinsic curvature of $\Sigma$. To introduce the Ashtekar new  variables for the loop quantization, the canonical transformation $A_a^i=\Gamma_a^i+\beta K_a^i$ should be done, where $\beta$ denotes the Barbero-Immirzi parameter, and $\Gamma_a^i$ is the SU(2) spin connection compatible with $E^a_i$. The constraint analysis of the model gives rise to the complete set of constraints containing those of the first and second classes \cite{Giesel:2012rb}. Solving the second-class constraints by introducing the Dirac brackets, one gets a total constraint system with the first-class constraints, including the Gauss constraint, the diffeomorphism constraint and the Hamiltonian constraint. In the phase space $\Gamma$ of the resulting first-class system, besides the new variables $(A_a^i, E^b_j)$, there contain the canonical pairs  $(T,P)$, $(S^j,P_j)$ and $(\xi,\xi^\dagger)$, where $T$ and $S^j$ for $j=1,2,3$ are scalar fields on $\Sigma$, their conjugate momentums $P$ and $P_j$ are density field with weight one and the half-density field $\xi$ relates to the fermion field by
\begin{equation}\label{eq:xipsi}
\xi=\sqrt[4]{|\det(E)|}\psi\Big|_{\Sigma}.
\end{equation}
The non-vanishing (anti-)Poisson brackets between these variables are 
\begin{equation}\label{eq:classicalPoisson}
\begin{aligned}
\{A_a^i(x),E^b_j(y)\}=&\kappa\beta\delta_a^b\delta^i_j\delta(x,y),\\
\{\xi_A(x),\xi_B^\dagger(y)\}_+=&-i\delta_{AB}\delta(x,y),\\
\{T(x),P(y)\}=&\delta(x,y),\\
\{S_i(x),P^j(y)\}=&\delta_i^j\delta(x,y). 
\end{aligned}
\end{equation}

The Gauss constraint $G_m$ is independent of the dust variables and reads
\begin{equation}
G_m=\frac{1}{\kappa\beta}D_a E^a_m+\frac{1}{2} \xi^\dagger\sigma_m\xi.
\end{equation}
It generates the SU(2) gauge transformation of the gravity and fermion fields. 
The diffeomorphism and Hamiltonian constraints are
\begin{equation}\label{eq:constraints}
\begin{aligned}
C_a^{\rm tot}&=H_a+P\partial_aT +P_j\partial_aS^j,\\
C^{\rm tot}&=H+\frac{P-q^{ab}\partial_aT C_b}{\sqrt{1+q^{ab}\partial_aT\partial_bT}},
\end{aligned}
\end{equation} 
where $H_a$ ($a=1,2,3$) and $H$ denote the diffeomorphism and  Hamiltonian constraints of gravity coupling fermion field, and $q_{ab}$ is the spatial metric on the spatial manifold $\Sigma$. According to the results in \cite{Lewandowski:2021bkt}, the constraints $H_a$ and $H$  are
\begin{equation}
\begin{aligned}
H_a=&\frac{1}{\kappa\beta} E^b_i F^i_{ab}+ \frac{i}{2}\Big(\xi^\dagger D_a\xi-(D_a\xi)^\dagger\xi\Big)+\beta K_a^m G_m,\\
H=&H_G+ \frac{1}{2\sqrt{q}}\Big[i(\xi^\dagger E_i^a\sigma^i D_a\xi-(D_a\xi)^\dagger E_i^a\sigma^i\xi)-\beta E^a_i K_a^i \xi^\dagger\xi-\frac{1}{\beta}(1+\beta^2)D_a E^a_i \xi^\dagger\sigma^i\xi-\beta E^a_iD_a\Big( \xi^\dagger\sigma^i\xi \Big)\Big],\\
\end{aligned}
\end{equation}
with the scalar constraint of pure gravity $H_G$:
\begin{equation}\label{eq:HG}
H_G=\frac{1}{2\kappa\sqrt{q}} E^a_i E^b_j\left(F_{ab}^m\epsilon_m{}^{ij}-2(1+\beta^2)K_{[a}^iK_{b]}^j\right).
\end{equation}
With the constraints, the action is expressed as
\begin{equation}
S=\int\dd^4 x\left(\frac{1}{\kappa\beta}E^a_i\partial_t A_a^i+\frac{i}{2}\left(\xi^\dagger\partial_t\xi-(\partial_t\xi^\dagger)\xi\right)+P_\mu\partial_t S^\mu-\lambda ^m G_m-N^a C^{\rm tot}_a -NC^{\rm tot}\right),
\end{equation}
with $P_\mu$ and $S^\mu$ as the abbreviation of $(P,P_j)$ and $(T, S_j)$ respectively, where $\lambda^m$, $N^a$ and $N$ are all Lagrangian multipliers. 

As is well-known, a totally constrained system, like the model in the present work, leads to a scenario in which the dynamical evolution is frozen. This problematic issue  can be fixed by introducing the relational evolution. To elaborate this concept, let us forget about the Gauss constraint for a moment and define $\Gamma_C$ as the constraint surface given by  $C_a^{\rm tot}=0=C^{\rm tot}$ in the phase space. Then, the physical phase space $\bar\Gamma$ is $\Gamma_C$ modulo the gauge orbits of the constraints. Since the constraints $C_a^{\rm tot}$ and $C^{\rm tot}$  generate the spacetime diffeomorphism transformations, we can introduce a gauge fixing conditions $0=T-t$ and $0=\delta_j^a S^j-x^a$ for each parameter $t$, where $x^a$ is a fiducial  coordinate  fixed once and for all  on  $\Sigma$. Let $\Gamma_t$ denote the gauge fixed surface for $t$. Then, $\sigma_t:\bar\Gamma\to\Gamma_t$ gives an  embedding of $\bar\Gamma$  into $\Gamma_C$. Given a function $F$ on $\Gamma_C$ that depends only on the gravity and fermion fields; we construct a family of Dirac observables by 
\begin{equation}\label{eq:relationalevalution}
t\mapsto \sigma_t^*F.
\end{equation}
This equation is naturally interpreted as the relational evolution of the gravity and the fermion field with respect to the Gaussian dust \cite{Giesel:2012rb,dapor2013relational}. The physical Hamiltonian governing this evolution is
\begin{equation}
\hop=\int\dd^3 x \sqrt{1+q^{ab}\partial_aT\partial_bT}\, H+q^{ab}\partial_aTH_b=\int\dd^3 x H.
\end{equation}
It  is worth noting  that the derivation of  $\hop$ needs  the fact that $F$ depends only on the gravity and  the fermion field and  that $\partial_a T=0$.

\subsection{quantum kinematics on cubic lattices}
The current work aims at restoring the results of the quantum fermion field on the Minkowski background from fermion coupling quantum gravity. To this end, cubic lattices will be chosen for our calculation because cubic lattices can reflect the continuum properties of the Hamiltonian better than the others \cite{Dapor:2017gdk,Liegener:2020dcg,Zhang:2021qul,Zhang:2020mld}. Due to this, we do not lose any essentials by concerning ourselves with only the cubic graphs. Thus, unless  otherwise stated, all graphs mentioned below are cubic. 

For a graph $\gamma\subset\Sigma$, the collections of edges and vertices  are denoted by $E(\gamma)$ comprising $|E(\gamma)|$ elements and $V(\gamma)$ containing $|V(\gamma)|$ element, respectively. Given $\gamma$, a function $\psi$ on $\sut^{|E(\gamma)|}$ gives rise to a cylindrical function $\psi_\gamma$ of the connection $A_a^i$, that is
\begin{equation}
\psi_\gamma(A)=\psi(\{h_e(A)\}_{e\in E(\gamma)}),
\end{equation}
where $h_e(A)$ is the holonomy of $A$ along $e$, i.e., 
\begin{equation}\label{eq:holonomy}
h_e(A)=\mathcal P\exp\int_e A=1+\sum_{n=1}^\infty\int_0^1\dd t_n\int_0^{t_n}\dd t_{n-1}\cdots\int_0^{t_2}\dd t_1A(t_1)\cdots A(t_n).
\end{equation} 
The space of the cylindrical functions on $\gamma$ is denoted by $\cylg^{(\gamma)}$.  Two cylindrical functions $\psi_{\gamma}^{(1)},\psi_{\gamma}^{(2)}\in \cylg^{(\gamma)}$ have the inner product
\begin{equation}\label{eq:innerproductg}
\langle \psi_{\gamma}^{(1)}|\psi_{\gamma}^{(2)}\rangle=\int_{\sut^{|E(\gamma)|}}\dd\mu_H(\vec h) \psi^{(1)}(\vec h)^*\,\psi^{(2)}(\vec h)
\end{equation}
where $\dd\mu_H$ is the Haar measure on $\sut^{|E(\gamma)|}$. 
The Cauchy completion of  the inner product space $\cylg^{(\gamma)}$ gives rise to the Hilbert space ${\mathcal H}_\gamma^G$. ${\mathcal H}_\gamma^G$ by definition is isometric to $L^2(\sut^{|E(\gamma)|},\dd\mu_H)$.

The  fermion Hilbert space  $\mathcal H_\gamma^F$ associated with $\gamma$ is the tensor product of Hilbert spaces $\mathcal H_v^F$ located at vertices $v\in V(\gamma)$, i.e.,
$$\mathcal H_\gamma^F=\bigotimes_{v\in V(\gamma)}\mathcal H_v^F,$$ 
where each  $\mathcal H_v^F$ is the Hilbert space of a fermionic oscillator of two degrees of freedom, and spanned by the orthonomal basis $\{|0,0\rangle_v,|0,1\rangle_v,|1,0\rangle_v,|1,1\rangle_v\}$; that is, the inner product of $\mathcal H_v^F$ reads
\begin{equation}
{}_v\langle i_1,i_2|j_1,j_2\rangle_v=\delta_{i_1,j_1}\delta_{j_1,j_2}. 
\end{equation}
The Hilbert space $\mathcal H_v^F$ is graded where each $|i_1,i_2\rangle_v$ carries  the degree $\grade(i_1,i_2)=(-1)^{i_1+i_2}$.
For  the graded objects, the rule to deal with them, roughly speaking, is that an extra sign factor is added whenever the order of a product of two objects are exchanged.  This rule is also applied when the tensor product of $\mathcal H_v^F$ and $\mathcal H_{v'}^F$ for different $v,v'\in V(\gamma)$ is considered (refer to \cite{Lewandowski:2021bkt} for more details on the tensor product of graded vector spaces).  

With the Hilbert spaces ${\mathcal H}_\gamma^G$ and $\mathcal H_\gamma^F$, the Hilbert space ${\mathcal H}_\gamma$ of the entire system on $\gamma$ is
\begin{equation}\label{eq:entireHilg}
{\mathcal H}_\gamma={\mathcal H}_\gamma^G\otimes\mathcal H_\gamma^F.
\end{equation}
A vector $\Psi\in{\mathcal H}_\gamma$ 
takes the general form
\begin{equation}
\Psi=\bigotimes_{v\in V(\gamma)}\left(\sum_{i,j}\psi_{v;ij}|i,j\rangle_v\right),\text{ with } \psi_{v;ij}\in\mathcal H_\gamma^G.
\end{equation}
The inner product between $\Psi^{(a)}:=\bigotimes_{v\in V(\gamma)}\left(\sum_{i,j}\psi^{a}_{v;ij}|i,j\rangle_v\right)$ (for $a=1,2$) is
\begin{equation}
\langle\Psi^{(1)}|\Psi^{(2)}\rangle=\prod_{v\in V(\gamma)}\left(\sum_{i,j}\int\dd\mu_h \psi^{(1)}_{v;ij}(\{h_e\}_{e\in E(\gamma)})^*\, \psi^{(2)}_{v;ij}(\{h_e\}_{e\in E(\gamma)}) \right).
\end{equation}

As shown in \cite{ashtekar2004back},  ${\mathcal H}_{\gamma}$ admits the spin network decomposition 
\begin{equation}
{\mathcal H}_{\gamma}=\bigoplus_{\vec j,\vec l}\left(\mathcal H_{\gamma}^{G,(\vec j,\vec l)}\otimes \mathcal H_\gamma^F\right)
\end{equation}
where $\mathcal H_{\gamma}^{G,(\vec j,\vec l)}$ is the spin network subspace of ${\mathcal H}_{\gamma}^G$ defined by the assignments $\vec j=\{j_1,j_2,\cdots,j_{|E(\gamma)|}\}$ and $\vec l=\{l_1,\cdots,l_{|V(\gamma)|}\}$ which, respectively, assign to each edge and vertex of $\gamma$ an irreducible representation of $\sut$. Taking advantage of this decomposition, we define the subspace  $\widetilde{\mathcal H}_{\gamma}$ of $\mathcal H_{\gamma}$ as 
\begin{equation}\label{eq:decompositionG}
\widetilde{\mathcal H}_{\gamma}=\bigoplus_{\vec j',\vec l}\left(\mathcal H_{\gamma}^{G,(\vec j',\vec l')}\otimes\mathcal H_\gamma^F\right).
\end{equation}  
where $\vec j'$ denotes such assignments that the spin associated to every  edge is nonvanishing. Indeed, by introducing $\widetilde{\mathcal H}_{\gamma}$, we remove the states in ${\mathcal H}_\gamma$ which can be represented on a smaller graph. The introduction of $\widetilde{\mathcal H}_{\gamma}$ follows the standard procedure in LQG, where one removes the  spin network states carrying vanishing spins for the direct sum decomposition of the total Hilbert space. A difference here is that we concern ourselves with only the cubic lattices so that there is no spurious vertex in our graphs. As a consequence, our case allows the assignments $\vec l$ of trivial irreducible representations to vertices of $\gamma$. 
Taking advantage  of the Hilbert spaces $\widetilde{\mathcal H}_{\gamma}$ on all (cubic) graphs $\gamma$, we  get the total Hilbert space of the entire model as
\begin{equation}\label{eq:directsum}
\mathcal H=\bigoplus_{\gamma}\widetilde{\mathcal H}_{\gamma}. 
\end{equation}

The flux operators $\hat p_k^{v,e}$ on ${\mathcal H}_\gamma^G$ is defined by
\begin{equation}\label{eq:Js}
(\hat p^{v,e}_k\psi)(h_{e'}(A),\cdots,h_e(A),\cdots,h_{e''}(A))=\left\{
\begin{aligned}
&it\left.\frac{\dd}{\dd\epsilon}\right|_{\epsilon=0}\psi(h_{e'}(A),\cdots,e^{-\epsilon\tau^k}h_e(A),\cdots,h_{e''}(A)),\ v=s_e,\\
&it\left.\frac{\dd}{\dd\epsilon}\right|_{\epsilon=0}\psi(h_{e'}(A),\cdots,h_e(A)e^{\epsilon\tau^k},\cdots,h_{e''}(A)),\ v=t_e,
\end{aligned}
\right.
\end{equation}
where $\tau^k=(-i/2)(\text{Pauli matrix})^k$, and $s_e$ and $t_e$ denote the source and target points of $e$ respectively, and the dimensionless parameter $t$ is $t=\kappa\hbar/a^2$ with $a$ being some unit of length.
The multiplication operators $D^\iota_{ab}(h_e)$ for all $\iota\in\frac12\mathbb Z_{>0}$ and $-\iota\leq a,b\leq \iota$ acts on $\mathcal H_\gamma^G$ as
\begin{equation*}
(D^\iota_{ab}(h_e)\psi)(A)=D^\iota_{ab}(h_e(A))\psi(A)
\end{equation*}
where $D^\iota_{ab}(h_e(A))$ are the entries of the Wigner-D matrix $D^\iota(h_e(A))$ of $h_e(A)\in $SU(2). The commutators between theses operators are
\begin{equation}\label{eq:commutators}
\begin{aligned}
[D^\iota(h_{e}),D^\iota(h_{e'})]&=0=[\hat p_i^{s_e,e},p_j^{t_{e'},e'}]\\
[\hat p_k^{s_e,e},\hat p_j^{s_{e'},e'}]&=it\delta_{ee'}\epsilon_{kj}{}^l\hat p_l^{s_e,e},\\
[\hat p_k^{t_e,e},\hat p_j^{t_{e'},e'}]&=it\delta_{ee'}\epsilon_{kjl}\hat p_l^{t_e,e},\\
[D^\iota(h_{e'}),\hat p_j^{s_e,e}]&=it\delta_{ee'}D'{}^\iota(\tau^j)D^\iota(h_e),\\
[D^\iota(h_{e'}),\hat p_j^{t_e,e}]&=-it\delta_{ee'}D^\iota(h_e)D'{}^\iota(\tau^j).
\end{aligned}
\end{equation}
where $D'{}^\iota(\tau^j)$ is the corresponding representation matrix of $\tau^j$.

On the  Hilbert  space $\mathcal H_\gamma^F$, there are the operators $\hat\zeta_{v,A}$ and $\hat\zeta_{v,A}^\dagger$  for $A=\pm$ and $v\in V(\gamma)$. For each $v\in V(\gamma)$, the space $\mathcal H_v^F$ is where the operators $\hat\zeta_{v,A}$ and $\hat\zeta_{v,A}^\dagger$ act.  The operators $\hat\zeta_{v,A}$ and $\hat\zeta_{v,A}^\dagger$ are also graded with the degrees $
\grade(\hat\zeta_{v,A})=\grade(\hat\zeta_{v,A}^\dagger)=1$. 
Their action reads 
\begin{equation}
\begin{array}{lll}
\hat\zeta_{v,+}^\dagger|0,i_2\rangle_v=|1,i_2\rangle,& \hat\zeta_{v,+}^\dagger|1,i_2\rangle_v=0,& \forall i_2=0,1,\\
\hat\zeta_{v,+}|0,i_2\rangle_v=0,& \hat\zeta_{v,+}|1,i_2\rangle_v=|0,i_2\rangle_v,& \forall i_2=0,1,\\
\hat\zeta_{v,-}^\dagger|i_1,0\rangle_v=(-1)^{i_1}|i_1,1\rangle_v,& \hat\zeta_{v,-}^\dagger|i_1,1\rangle_v=0,&\forall i_1=0,1,\\
\hat\zeta_{v,-}|i_1,0\rangle_v=0,&\hat\zeta_{v,-}|i_1,1\rangle_v=(-1)^{i_1}|i_1,0\rangle_v,& \forall i_1=0,1,
\end{array}
\end{equation}
where the extra sign factors are caused by the operators and the states being graded objects. Therefore, the rule for dealing with the graded objects is also applied to the graded operators.  For instance, the rule leads to
\begin{equation}
\hat\zeta_{v,A}^\dagger |j_1,j_2\rangle_{v'}\otimes|i_1,i_2\rangle_{v}\otimes |k_1,k_2\rangle_{v''}=(-1)^{\grade(\hat\zeta_{v,A}^\dagger)\grade(j_1,j_2)}|j_1,j_2\rangle_{v'}\otimes(\hat\zeta_{v,A}^\dagger|i_1,i_2\rangle_{v})\otimes (|k_1,k_2\rangle_{v''}
\end{equation} 
where we exchange the order between $|j_1,j_2\rangle_{v'}$ and $\hat\zeta_{v,A}^\dagger$ on the right-hand side.

With $\hat\zeta_{v,A}$ and $\hat\zeta_{v,A}^\dagger$, we introduce the operator  
\begin{equation}\label{eq:thetazeta}
\hat  \theta_A(v)=\sqrt{\hbar}\,\hat\zeta_{v,A}.
\end{equation}
Clearly, they satisfy the anti-commutation relation
\begin{equation}\label{eq:quantumbracket}
[\hat\theta_A(v),\hat\theta_B^\dagger(v')]_+=\hbar\delta_{AB}\delta_{v,v'}
\end{equation}
where $\delta_{v,v'}$ is the Kronecker delta. In comparison with  the classical Poisson brackets \eqref{eq:classicalPoisson}, equation \eqref{eq:quantumbracket} changes the Dirac delta $\delta(x,y)$ to Kronecker delta $\delta_{v,v'}$. Indeed, this change results from the canonical transformation \cite{Thiemann:1997rt},
\begin{equation}
\theta(x)=\int_{\mathbb T^3}\dd^3 y\sqrt{\delta(x,y)}\xi(y).
\end{equation}

\subsection{coherent states}\label{sec:coherents}
\subsubsection{LQG coherent state}\label{sec:coherentsG}


Given $g\in$SL($2,\mathbb C$), it labels an SU(2) heat kernel coherent state $\psi^t_g$ defined by 
\begin{equation}
\psi^t_{g}(h)=\frac{t^{3/4} e^{-\frac{ p^2}{2t}}}{\sqrt{2}\pi^{\frac14} e^{ t/8}}\sqrt{\frac{\sinh(p)}{p} } \sum_j d_je^{-\frac{t}{2}j(j+1)}\chi_j(g h^{-1}),
\end{equation}
where $\chi_j(gh^{-1})$ is the trace of $gh^{-1}$ in the $j$-representation and $p=\sqrt{\vec p\cdot\vec p}$ with $g=e^{-i\vec p\cdot\vec\tau}e^{\vec\theta\cdot\vec\tau}$.
The inner product between two coherent states $\psi_{g_1}^t$ and $\psi_{g_2}^t$ is
\begin{equation}\label{eq:ginnerproduct}
\langle\psi_{g_1}^t|\psi_{g_2}^t\rangle= \frac{\eta \sqrt{\sinh(p_1)\sinh(p_2)}}{\sqrt{p_1p_2}\sinh(\eta)}e^{-\frac{p_1^2+p_2^2-2\eta^2}{2t}}(1+O(t^{\infty}))
\end{equation}
where $\eta$ is defined by  $\tr(g_1^\dagger g_2)=2\cosh(\eta)$ and requiring  the imaginary part of $\eta$, i.e. $\Im(\eta)$, is in $[0,\pi]$ \cite{thiemann2001gaugeIII}. These coherent states form an overcomplete basis of $L^2(\sut,\dd\mu_H)$ with the measure 
\begin{equation}\label{eq:measure}
\dd\nu_t(e^{i\vec p\cdot\vec\tau}e^{\vec\theta\cdot\vec\tau})=\frac{2}{\pi t^3}\dd\mu_H(e^{\vec\theta\cdot\vec\tau})\dd ^3\vec p.
\end{equation}
where $\dd\mu_H$  denotes the SU(2) Haar measure, namely, 
\begin{equation}\label{eq:completerelation}
\int \dd\nu_t(g)|\psi^t_{g}\rangle\langle\psi^t_{g}|=\mathbbm{1}_{L^2(\sut,\dd\mu_H)},
\end{equation}

Putting on each edge an SU(2) coherent state, one constructs a type of LQG coherent state by their tensor product \cite{thiemann2001gauge}. Precisely, given $\g=\{g_e\}_{e\in E(\gamma)}$, the LQG coherent state $|\Psi_\g^t,\gamma\rangle\in{\mathcal H}_\gamma^G$ is 
\begin{equation}\label{eq:coherentstateg}
|\Psi_\g^t,\gamma\rangle=\bigotimes_{e\in E(\gamma)}\psi_{g_e}^t.
\end{equation} 
According to \eqref{eq:completerelation}, the coherent states defined  by \eqref{eq:coherentstateg} form an overcomplete basis of ${\mathcal H}_\gamma^G$, i.e., 
\begin{equation}\label{eq:completerelationG2}
\int\mathcal D[\g,\gamma] |\Psi_\g^t,\gamma\rangle\langle \Psi_\g^t,\gamma|=\mathbbm{1}_{{\mathcal H}_\gamma^G},
\end{equation}
with the measure $\mathcal D[\g,\gamma]=\prod_{e\in E(\gamma)}\dd\nu_t(g_e)$. Let us decompose $g_e\in $SL($2,\mathbb C$) as  $g_e=e^{ip_j(e)\tau^j}u(e)$ with $u\in \sut$.
Then, the expectation values of the holonomy and flux operators in the coherent states are
\begin{equation}\label{eq:expectationvalue}
\begin{aligned}
\langle \Psi_\g^t,\gamma| \hat p_j^{s_e,e} |\Psi^t_{\g},\gamma\rangle=&p_j(e)+O(t),\\ 
\langle \Psi_\g^t,\gamma|\hat p^{t_e,e} |\Psi^t_{\g},\gamma\rangle=&-p^{\mathfrak t}_j(e)+O(t),\\
\langle \Psi_\g^t,\gamma|D^{\frac{1}{2}}_{ab}(h_e) |\Psi^t_{\g},\gamma\rangle=&D^{\frac12}_{ab}(u(e))+O(t),
\end{aligned}
\end{equation}
where $p^{\mathfrak t}_j(e)$, given by $ p^{\mathfrak t}_j(e)\tau^j=u(e)^{-1}p_k(e)\tau^k u(e)$,  is the $j$th component of $u(e)^{-1}p_k(e)\tau^k u(e)$.

Given a graph $\gamma$, we need to work in $\widetilde{\mathcal H}_\gamma^G$, while the coherent states \eqref{eq:coherentstateg} are not elements in $\widetilde{\mathcal H}_\gamma^G$. To construct the coherent states  in $\widetilde{\mathcal H}_\gamma^G$, let us employ  the projection $\p_\gamma: \mathcal H_\gamma^G\to \widetilde{\mathcal H}_\gamma^G$ and define $|\widetilde{\Psi_\g^t},\gamma\rangle\in \widetilde{\mathcal H}_\gamma^G$ as
\begin{equation}\label{eq:coherentstateg1}
|\widetilde{\Psi_\g^t},\gamma\rangle:=\p_\gamma |\Psi_\g^t,\gamma\rangle=\bigotimes_{e\in E(\gamma)}\widetilde{\psi_{g_e}^{t}}.
\end{equation} 
where $\widetilde{\psi_{g_e}^{t}}$ is given by
\begin{equation}\label{eq:psiprime}
\widetilde{\psi_g^{t}}=\frac{t^{3/4} e^{-\frac{ p^2}{2t}}}{\sqrt{2}\pi^{\frac14} e^{ t/8}}\sqrt{\frac{\sinh(p)}{p} } \sum_{j\neq 0} d_je^{-\frac{t}{2}j(j+1)}\chi_j(g h^{-1}).
\end{equation} 
By  definition, $\widetilde{\psi_g^{t}}$ is just $\psi_g^{t}$ with removing the component of $j=0$. For $p^2\gg t$, the norm of this component is much smaller than 1, i.e., the norm of $\psi_g^t$.  As a consequence,  $\widetilde{\psi_g^{t}}$ and, thus,  $|\widetilde{\Psi_\g^t},\gamma\rangle$ inherit most properties of $\psi_g^{t}$ and $|\Psi_\g^t,\gamma\rangle$. 
In particular, $|\widetilde{\Psi_\g^t},\gamma\rangle$ are normalized  up to some $O(t^\infty)$ term, i.e.,
\begin{equation}\label{eq:compare2inner}
\langle\widetilde{\Psi_{\g}^t},\gamma|\widetilde{\Psi_{\g}^t},\gamma\rangle=1+O(t^{\infty}).
\end{equation}
Moreover, the expectation values of monomials of holonomies, fluxes and volume operators in the states $\widetilde{\psi_{g}^{t}}$ and $|\widetilde{\Psi_{\g}^t},\gamma\rangle$ coincide with the classical values. That  is to say, the states $\widetilde{\psi_{g}^{t}}$ and $|\widetilde{\Psi_{\g}^t},\gamma\rangle$ have the desirable semiclassical limit. This is  implied by the
\begin{equation}\label{eq:differenceexpect}
\langle\Psi_\g^t,\gamma|\hat  M|\Psi_\g^t,\gamma\rangle=\langle \widetilde{\Psi_\g^t},\gamma|\hat  M| \widetilde{\Psi_\g^t},\gamma\rangle+O(t^\infty). 
\end{equation}
where $\hat M$  a general monomial of fluxes, holonomies and volume operator. The derivations for \eqref{eq:compare2inner} and \eqref{eq:differenceexpect} are quite  technical and put in Appendix \ref{app:semiclassicalpsit},  in order to keep the flow of our argument.. 

\subsubsection{the fermion coherent state and the coherent state of the entire system}

Give  a vertex $v\in V(\gamma)$, the fermion coherent state $|\phi_{\nu_o}\rangle_v$ at $v$ is
\begin{equation}
|\phi_{\nu_o}\rangle_v=\exp(-\frac{1}{2}\nu_o^\dagger\nu_o+\sum_{A=\pm}\hat\zeta^\dagger_{v,A}\nu_{o,A} )|0,0\rangle_v.
\end{equation}
where $\nu_o=(\nu_{o,+},\nu_{o,-})^T$ is a Grassmann-valued 2-vector, and $\nu_o^\dagger=(\nu_{o,+}^*,\nu_{o,-}^*)$ with $\nu_{o,A}^*$ being the complex conjugate of $\nu_{o,A}$. It is easy to verify that $|\phi_{\nu_o}\rangle$ is an eigenstate  of $\hat\zeta_v$, i.e., 
\begin{equation}\label{eq:coherenteigen}
\begin{aligned}
\hat\zeta_{v,A}|\psi_{\nu_o}\rangle_v=\nu_{o,A}|\psi_{\nu_o}\rangle_v.
\end{aligned}
\end{equation}
In addition, the inner product between  $|\phi_{\nu_o}\rangle_v$ and $|\phi_{\nu_o'}\rangle_v$  reads
\begin{equation}\label{eq:finnerproduct}
{}_v\langle\phi_{\nu_o}|\phi_{\nu'_o}\rangle_v=e^{-\frac{1}{2}\nu_o^\dagger\nu_o-\frac{1}{2}\nu'_o{}^\dagger\nu'_o+\nu^\dagger_o\nu'_o}.
\end{equation}
As usual coherent states,  $|\phi_{\nu_o}\rangle_v$ forms a overcomplete basis of $\mathcal H_v^F$,
\begin{equation}\label{eq:completerelationF}
\begin{aligned}
\int \dd\mu_H(\nu_o) |\phi_{\nu_o}\rangle_v\,{}_v\langle\phi_{\nu_o}|=\mathbbm{1}_{\mathcal H_v^F},
\end{aligned}
\end{equation}
where the measure $\dd\mu_H(\nu(v))$ is
\begin{equation}\label{eq:fermionmeasure}
\dd\mu_H(\nu_o)=\prod_{A=\pm}\dd\nu_{o,A}^*\dd\nu_{o,A}. 
\end{equation}
Let $\nu:v\mapsto \nu(v)=(\nu_+(v),\nu_-(v))^T$ be a 2-component Grassmann-vector-valued field on $V(\gamma)$. Taking advantage of the fermion coherent state at each single vertex, we achieve the total fermion coherent state
\begin{equation}
|\Phi_\nu,\gamma\rangle=\bigotimes_{v\in V(\gamma)}|\phi_{\nu(v)}\rangle_v.
\end{equation}
According to \eqref{eq:completerelationF}, the overcomplete condition is
\begin{equation}\label{eq:completerelationF2}
\int\mathcal D[\nu,\gamma]|\Phi_\nu,\gamma\rangle\langle\Phi_\nu,\gamma|=\mathbbm{1}_{\mathcal H_\gamma^F}
\end{equation}
with
\begin{equation}
\mathcal D[\nu,\gamma]=\prod_{v\in V(\gamma)}\prod_{A=\pm}\dd\nu^*_A (v)\dd\nu_A(v).
\end{equation}

A coherent state of the entire system is labelled by a pair $Z=(\g,\nu)$. The coherent state $|Z,\gamma\rangle\in \mathcal H_\gamma$ is just the tensor product 
\begin{equation}
|Z,\gamma\rangle=|\Psi_\g,\gamma\rangle \otimes |\Phi_\nu,\gamma\rangle.
\end{equation}
According to \eqref{eq:completerelationG2} and \eqref{eq:completerelationF2},  the coherent state $|Z,\gamma\rangle$ admits the overcomplete condition
\begin{equation}\label{eq:resolutionofidentityTotal}
\mathbbm{1}_{\mathcal H_\gamma}=\int \mathcal D[Z,\gamma] |Z,\gamma\rangle\langle Z,\gamma|
\end{equation}
where $\mathcal D[Z,\gamma]$ is
\begin{equation}
\mathcal D[Z,\gamma]=\mathcal D[\nu,\gamma]\mathcal D[\g,\gamma].
\end{equation}
Moreover, the inner product between $|Z,\gamma\rangle$ and $|Z',\gamma\rangle$, by \eqref{eq:ginnerproduct} and  \eqref{eq:finnerproduct}, is
\begin{equation}\label{eq:innerproductall}
\begin{aligned}
\langle Z,\gamma|Z',\gamma\rangle=&\prod_{e\in E(\gamma)}\frac{\eta_e \sqrt{\sinh(p_e)\sinh(p'_e)}}{\sqrt{p_ep'_e}\sinh(\eta_e)}\times \\
&\exp(-\sum_{e\in E(\gamma)}\frac{p_e^2+(p_e')^2-2\eta_e^2}{2t}-\sum_{v\in V(\gamma)}\frac{\nu^\dagger(v)\nu(v)+\nu'{}^\dagger(v)\nu'(v)-2\nu^\dagger(v)\nu'(v)}{2})
\end{aligned}
\end{equation}

\subsection{Gauge transformation and Hamiltonian operator}\label{Gauge transformation and the Hamiltonian constraint}
 The Gauss constraint in terms of $\hat\zeta_v$ reads
\begin{equation}
\hat{G}_{v,m}=\hbar\left(\sum_{e \text{ at } v}\frac{\hat p^{v,e}_m}{t}+\hat\zeta_{v}^\dagger\frac{\sigma_m}{2}\hat\zeta_{v}\right).
\end{equation}
The gauge transformation generating the Gauss constraint reads \cite{Lewandowski:2021bkt}, 
\begin{equation}\label{eq:gaugetransformation}
\begin{aligned}
u\cdot\Psi=\bigotimes_{v\in V(\gamma)} \left( \begin{pmatrix}
|1,0\rangle_v,|0,1\rangle_v
\end{pmatrix}D^{\frac{1}{2}}(u_v)
\begin{pmatrix}
 \psi_{v;10}^{(u)}\\
 \psi_{v;01}^{(u)}
\end{pmatrix}+\psi_{v;00}^{(u)}|0,0\rangle+ \psi_{v;11}^{(u)}|1,1\rangle\right)
\end{aligned}
\end{equation}
where $u:v\mapsto u_v\in \sut$ is an SU(2) valued function on $V(\gamma)$ and $\psi_{v;ij}^{(u)}\in\mathcal H_\gamma^G$ is
\begin{equation}
\psi_{v;ij}^{(u)}(\{h_e\}_{e\in E(\gamma)})=\psi_{v;ij}(\{u(s_e)^{-1}h_eu(t_e)\}).
\end{equation}
Applying the gauge transformation \eqref{eq:gaugetransformation} to the coherent state $|Z,\gamma\rangle$, one has 
\begin{equation}\label{eq:gaugatransformation}
\begin{aligned}
u |Z,\gamma\rangle=\left(\bigotimes_{e\in E(\gamma)}\psi_{u(s_e)g_e u(t_e)^{-1}}\right)\otimes \left(\bigotimes_{v\in V(\gamma)}|\phi_{u_v\cdot\nu_v}\rangle_v\right)
\end{aligned}
\end{equation}

 Given a graph $\gamma$, the Hamiltonian $\hat \hop_\gamma$ associated with it is 
\begin{equation}\label{eq:hgamma}
\hat \hop_\gamma=\p_\gamma(\hat H_\gamma^G+\hat H_\gamma^F)\p_\gamma,
\end{equation}
where $\hat H_\gamma^G$ is the graph preserving Hamiltonian constraint operator of vacuum LQG, and $\hat H_\gamma^F$ is the graph preserving fermionic  Hamiltonian containing the interaction term of gravity and fermion. One can refer to \cite{ashtekar2004back,thiemann2007modern} for more details on $\hat H_\gamma^G$ whose expression is given by
\begin{equation}
\begin{aligned}
\hat{H}^G_\gamma=\frac{1}{2}\sum_{v\in V(\gamma)}\left( \hat{H}^E_\gamma(v)+( \hat{H}^E_\gamma(v))^\dagger -(1+\beta^2)(\hat{H}^L_\gamma(v)+(\hat{H}^L_\gamma(v))^\dagger)\right)
\end{aligned}
\end{equation}
with 
\begin{equation}\label{eq:heg}
\begin{aligned}
 \hat{H}^E_\gamma(v)&=\frac{-1}{4i\kappa\beta\ell_P^2}\sum_{e(I),e(J),e(K) \text{ at } v}\epsilon^{IJK}\tr(h_{\alpha_{IJ}}h_{e(K)}^{-1}[h_{e(K)},\hat V_v]).\\
\hat{H}^L_\gamma(v)&=-\frac{4\kappa}{i \beta ^7 \ell_P^{10}}\sum_{e(I),e(J),e(K) \text{ at } v}\varepsilon^{IJK}\tr( [h_{e(I)},[\hat V_\gamma,\hat H^E_\gamma]]h_{e(I)}^{-1} [h_{e(J)},[\hat V_\gamma,\hat H^E_\gamma]]h_{e(J)}^{-1}[h_{e(K)},\hat V_v]h_{e(K)}^{-1})
\end{aligned}
\end{equation}
where $\alpha_{IJ}$ is the minimal loop in $\gamma$ containing edges $e(I)$ and $e(J)$, $\hat V_v$ is the volume operator at the vertex $v$ so that the total volume is $\hat V_\gamma=\sum_{v\in V(\gamma)}\hat V_v$ and $\hat H^E_\gamma=\sum_{v\in V(\gamma)}\hat H^E_\gamma(v)$. For $\hat H_\gamma^F$, we employ the one introduced  in \cite{Lewandowski:2021bkt} but need to adapt it to make it graph-preserving. One has
\begin{equation}
\hat H_\gamma^F=\sum_{v\in V(\gamma)} \hat H_F(v)
\end{equation}
where $\hat H_F(v)$ is
\begin{equation}\label{eq:HF}
\hat H_F(v)=\frac12\widehat{\sqrt{V_v^{-1}}}\left(i\left(\hat H_F^{(1)}(v)-\hat H_F^{(1)}(v)^\dagger\right)-\beta\hat H_F^{(2)}(v)-\frac{1+\beta^2}{\beta}\hat H_F^{(3)}(v)-\beta(\hat H_F^{(1)}(v)+\hat H_F^{(1)}(v)^\dagger)\right)\widehat{\sqrt{V_v^{-1}}}
\end{equation}
with the explicit expression of $\hat H_F^{(i)}(v)$ for $i=1,2,3$ as 
\begin{equation}
\begin{aligned}
\hat H_F^{(1)}(v)=&\frac{\kappa\hbar\beta}{2t}\sum_{e\text{ at }v}\hat\theta^\dagger(v)\sigma^i\left(h_e\hat\theta(t_e)-\hat\theta(v) \right) \hat p_i^{v,e},\\
\hat H_F^{(2)}(v)=&\frac{1}{i\hbar \beta^2}\left[\hat H^E_\gamma(v),\hat V_v\right]\hat\theta^\dagger(v)\hat\theta(v),\\
\hat H_F^{(3)}(v)=&\frac{\kappa\hbar\beta}{2t} \left(\sum_{e \text{ at } v} \hat p_i^{v,e}\right)\hat\theta^\dagger(v)\sigma^i\hat\theta(v).
\end{aligned}
\end{equation}
It is worth noting that  the extra factors $1/2$ in $\hat H_F^{(1)}(v)$ and $\hat H_F^{(3)}(v)$ compared to the results in \cite{Lewandowski:2021bkt} come from the fact that due to each edge containing two vertices,  
$\hat H_F^{(1)}(v)$ and $\hat H_F^{(3)}(v)$ are counted twice in the summation over $v\in V(\gamma)$ in $\hat H_\gamma^F$.
Moreover, Eq. \eqref{eq:HF}  uses the inverse volume operator defined by \cite{yang2016new}
\begin{equation}\label{eq:inversev}
\begin{aligned}
&\hat V_v^{-\frac12}=-\frac{16}{3}\left(\frac{1}{i\ell_P^2 \beta}\right)^3\sum_{e(I),e(J),e(K) \text{ at } v} \epsilon^{IJK}\tr\left([\hat{h}_{e(I)},\hat V_v^{1/2}]\hat{h}_{e(I)}^{-1}[\hat{h}_{e(J)},\hat{V}_v^{1/2}]\hat{h}_{e(J)}^{-1}[\hat{h}_{e(K)},\hat{V}_v^{1/2}]h_{e(K)}^{-1}\right).
\end{aligned}
\end{equation}
Indeed, $\hat V_v^{-\frac12}$ is not always positive semiclassically, where the sign relates to the orientation $\mathrm{sgn}(\det(e))$. However, the sign is canceled in the semiclassical limit of $\hat H_F(v)$, since it always contains a pair of $\hat V_v^{-\frac12}$.

The Hamiltonian $\hat \hop_\gamma$ given by \eqref{eq:hgamma} is the same as the regular graph-preserving Hamiltonian up to the projection $\p_\gamma$ which is introduced based on the following considerations. At first, one needs to consider that a general operator $\hat A$ cannot annihilate edges of graphs if we require a densely defined adjoint operator $\hat A^\dagger$ in the regular LQG kinematic Hilbert space, which contains not only cubic but also all other graphs. To explain this point, let  us
assume an operator $\hat A$ which changes each  graph $\gamma$ to a  new graph $\hat A[\gamma]$ by erasing an edge $e_\gamma$ of $\gamma$, i.e., $\hat A[\gamma]=\gamma-\{e_\gamma\}$. Fix a graph $\mathring{\gamma}$ and  consider the action of the adjoint $\hat A^\dagger$ on $A[\mathring{\gamma}]$. By definition, the result of $A[\mathring \gamma]$ acted by $\hat A^\dagger$, denoted by $\hat A^\dagger[A[\mathring \gamma]]$, is a linear combination of such graphs  $\gamma$ that $A[\gamma]=A[\mathring\gamma]$. Consequently, $\hat A^\dagger[A[\mathring \gamma]]$ cannot be defined because there are uncountable infinitely many such  graphs $\gamma$. As an example, let us consider an operator $\hat A$ which always erases a segment of a loop, like
 \begin{equation}\label{eq:examplea}
\hat A:\makeSymbol{\raisebox{0.0\height}{\includegraphics[width=0.1\textwidth]{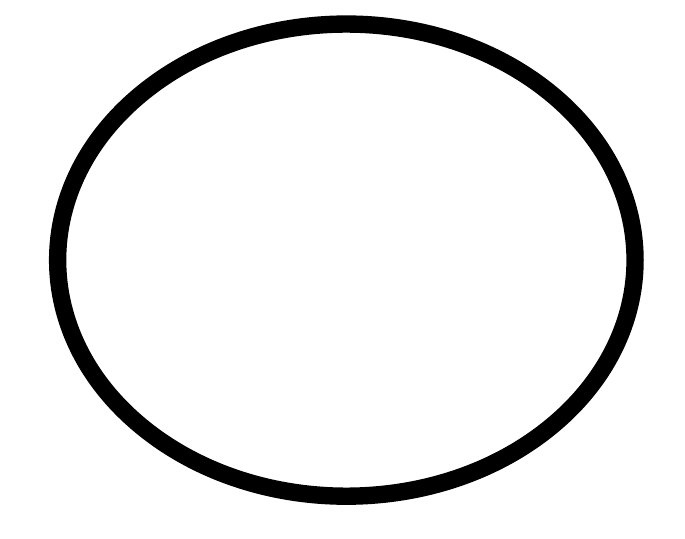}}}\quad\xrightarrow{\quad}\quad   \makeSymbol{\raisebox{0.0\height}{\includegraphics[width=0.09\textwidth]{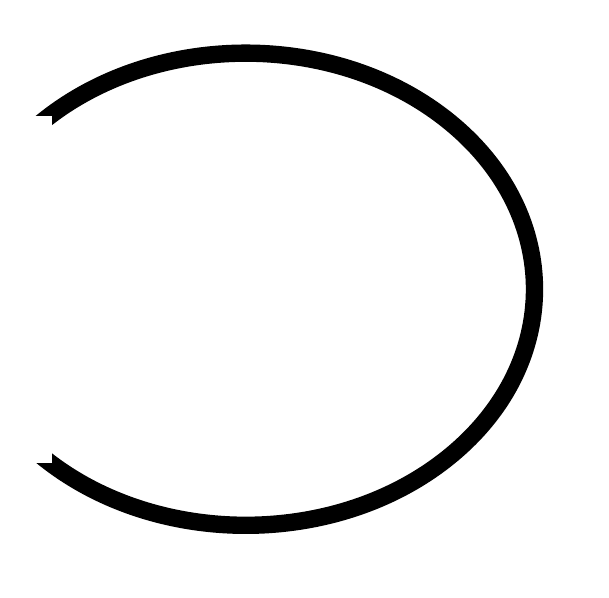}}}.
\end{equation}
Then, one has
\begin{equation}
\hat A^\dagger:\makeSymbol{\raisebox{0.0\height}{\includegraphics[width=0.09\textwidth]{loop2}}}\quad\xrightarrow{\quad}\quad   \makeSymbol{\raisebox{0.0\height}{\includegraphics[width=0.1\textwidth]{loop}}}+\makeSymbol{\raisebox{0.0\height}{\includegraphics[width=0.13\textwidth]{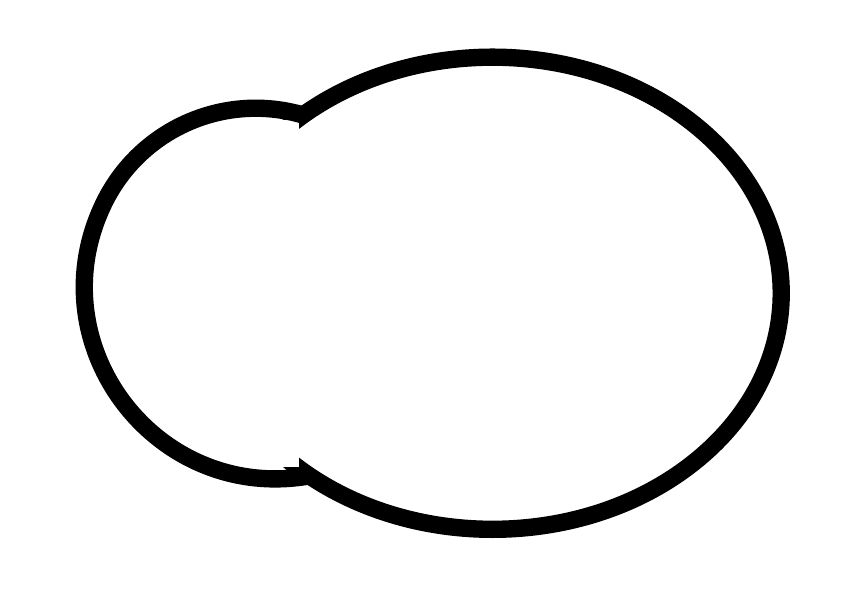}}}+\makeSymbol{\raisebox{0.0\height}{\includegraphics[width=0.15\textwidth]{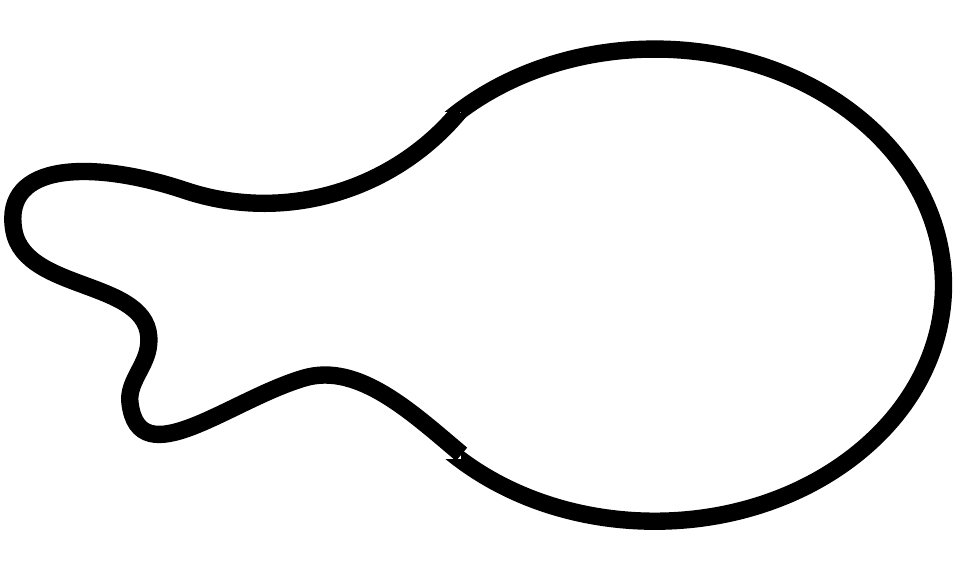}}}+\cdots.
\end{equation}
Indeed, in $\hat H_\gamma^E(v)$, the loop holonomy $h_{\alpha_{IJ}}$ (see \eqref{eq:heg}) to regularize the curvature plays the same role as $\hat A$ in the above example for the loop which has a segment $e$ carrying a  small spin $j_e=1/2$. Second, the operators $\hat H_\gamma^G$ and $\hat H_\gamma^F$ can change graphs just by the holonomy  operators  therein, because these holonomies shift spins and, thus, can annihilate edges carrying spin $1/2$. Thus, the projection in $\hat\hop_\gamma$ makes  $\hat\hop_\gamma$ different from $\hat H_\gamma^G+\hat H_\gamma^F$ by their actions on spin network states with small spins, while these states are purely quantum. Thus, the difference between $\hat\hop_\gamma$ and $\hat H_\gamma^G+\hat H_\gamma^F$ can only be presented at the quantum geometry level.

Taking advantage of $\hat{\hop}_\gamma$, the Hamiltonian operator associated with a single graph, we define the total physical Hamiltonian which could act on the total Hilbert space $\mathcal H$ as
\begin{equation}\label{eq:definehop}
\hat \hop=\sum_\gamma\hat\hop_\gamma.
\end{equation}
Due to projection operator, $\hat\hop_\gamma$ kills the Hilbert spaces $\widetilde{\mathcal H}_{\gamma'}$ for all $\gamma'\neq \gamma$. Thus, the action of $\hat\hop$ on $\widetilde{\mathcal H}_\gamma$ is the same as that of $\hat\hop_\gamma$. This fact will bring us great convenience for our future calculations.

\section{path integral formulation on a single graph}\label{sec:pathintegral}
For getting the transition amplitude between arbitrary states in  $\widetilde{\mathcal H}_\gamma$ for a given $\gamma$, it is sufficient to calculate 
\begin{equation}
A_{Z',Z}^\gamma=\langle Z',\gamma|\exp[-\frac{i}{\hbar}T\hat \hop_\gamma]|Z,\gamma\rangle,
\end{equation} 
due to $\widetilde{\mathcal H}_\gamma$ being a subspace of $\mathcal H_\gamma$ in which $|Z,\gamma\rangle$ forms an overcomplete basis. Note that despite $|Z,\gamma\rangle$ being not in $\widetilde{\mathcal H}_\gamma$, $\hat \hop_\gamma$ still has a well-defined action on  it.
Dividing the time interval $[0,T]$ into $N$ slices with length $\delta \tau $ such that $T=N\delta \tau $, inserting the resolution of identity \eqref{eq:resolutionofidentityTotal}, and employing the approximation $\exp(-\frac{i}{\hbar}\delta\tau\hat \hop_\gamma)\cong1-\frac{i}{\hbar}\delta\tau\hat \hop_\gamma$, we finally get 

\begin{equation}
\begin{aligned}
A_{Z',Z}^{\gamma,N}=&\int\left(\prod_{i=1}^{N-1}\mathcal{D}[ Z_i,\gamma] \right) \prod_{i=0}^{N-1}\langle  Z_{i+1},\gamma| Z_i,\gamma\rangle\exp\left[-\frac{i }{\hbar}\sum_{i=0}^{N-1} \delta \tau  \frac{\langle Z_{i+1},\gamma|\hat \hop_\gamma|Z_i,\gamma\rangle}{\langle Z_{i+1},\gamma| Z_i,\gamma\rangle}\right]\\
\end{aligned}
\end{equation}
with $Z_0=Z$ and $Z_N=Z'$. By employing \eqref{eq:innerproductall}, we get
\begin{equation}\label{eq:innermeasure}
\begin{aligned}
\langle Z_{i+1},\gamma| Z_i,\gamma\rangle=&\xi(\g_{i+1},\g_i) \exp\left(-\sum_{e\in E(\gamma)}\frac{p_{i+1}(e)^2+p_{i}(e)^2-2\eta_{i+1,i}(e)}{2t}\right)\\
& \exp\left(-\sum_{v\in V(\gamma)}\left(\frac{1}{2}\nu_{i+1}^\dagger(v)\nu_{i+1}(v)+\frac{1}{2}\nu_{i}^\dagger(v)\nu_{i}(v)-\nu_{i+1}^\dagger(v)\nu_{i}(v)\right)\right)
\end{aligned}
\end{equation}
where $p_i(e)$ parametrize  $\g_{i}(e)$ as above, $\eta_{i+1,i}(e)$ is given correspondingly by $\tr(\g_{i+1}(e)^\dagger \g_i(e))=2\cosh(\eta_{i+1,i}(e))$, and $\xi(\g_{i+1},\g_i)$ is
\begin{equation}
\begin{aligned}
\xi(\g_{i+1},\g_i)=&\prod_{e\in E(\gamma)} \frac{\eta_{i+1,i}(e) \sqrt{\sinh(p_{i+1}(e))\sinh(p_i(e))}}{\sqrt{p_{i+1}(e)p_{i}(e)}\sinh(\eta_{i+1,i}(e))}.
\end{aligned}
\end{equation}
Substituting \eqref{eq:innermeasure} into the expression of $A_{Z',Z}^{\gamma,N}$, we have
\begin{equation}\label{eq:Adis}
A_{Z',Z}^{\gamma,N}=\int_{\mathcal R(Z',Z)} \prod_{i=1}^{N-1}\mathcal{D}[Z_i,\gamma]\Xi(\g,\gamma)e^{S_N[\g,\nu]/t},
\end{equation}
where  $\mathcal R_\gamma(Z',Z)$ is the set of all polygon paths satisfying $Z[0]=Z$ and $Z[T]=Z'$, $\Xi[\g]$ is given by $\Xi[\g]=\prod_{i=0}^{N-1}\xi(\g_{i+1},\g_i)$, and  $S_N$ is given by
\begin{equation}\label{eq:SNdefine}
\begin{aligned}
S_N[\g,\nu]=&\sum_{i=0}^{N-1}K_\gamma(Z_{i+1},Z_i)-\frac{i  \kappa}{a^2} \delta \tau  \sum_{i=0}^{N-1}\frac{\langle Z_{i+1},\gamma|\hat \hop_\gamma|Z_i,\gamma\rangle}{\langle Z_{i+1},\gamma|Z_i,\gamma\rangle},\\
\end{aligned}
\end{equation}
with
\begin{equation}\label{eq:kineticterm}
\begin{aligned}
K_\gamma(Z_{i+1},Z_i)=&\sum_{e\in E(\gamma)}\left(\eta_{i+1,i}(e)^2-\frac{1}{2}\left(p_{i+1}(e)^2+p_{i}(e)^2\right)\right)\\
&+t\sum_{v\in V(\gamma)}\left(\nu_{i+1}^\dagger(v)\nu_{i}(v)-\frac{1}{2}\left(\nu_{i+1}^\dagger(v)\nu_{i+1}(v)+\nu_{i}^\dagger(v)\nu_{i}(v)\right)\right).
\end{aligned}
\end{equation}
Taking advantage of $A_{Z',Z}^{\gamma,N}$, we have
\begin{equation}\label{eq:limitA}
A_{Z',Z}^{\gamma}=\lim_{N\to\infty}A_{Z',Z}^{\gamma,N}.
\end{equation}
With $A_{Z',Z}^{\gamma}$, It can be obtained for the transition amplitude between any initial state $|\Psi_{\rm in},\gamma\rangle$ and final state $|\Psi_{\rm out},\gamma\rangle$. Here, $|\Psi_{\rm in},\gamma\rangle$ and $|\Psi_{\rm out},\gamma\rangle$ can be in either $\mathcal H_\gamma$ or $\widetilde{\mathcal H}_\gamma$.  More precisely, the transition amplitude  from $|[\Psi_{\rm in}],\gamma\rangle$ to 
$|[\Psi_{\rm out}],\gamma\rangle$ is
\begin{equation}\label{eq:transitionamplitudegeneral}
\begin{aligned}
&\langle[\Psi_{\rm out}],\gamma|\exp[\frac{i}{\hbar}  T\hat\hop_\gamma]|[\Psi_{\rm in}],\gamma\rangle\\
=&\mathcal N_\gamma^{(\rm in)}\mathcal N_\gamma^{(\rm out)}\int\dd\mu_H(u)\mathcal D[Z',\gamma]\mathcal D[Z,\gamma] \langle\Psi_{\rm out},\gamma|Z',\gamma\rangle A_{Z',Z}^\gamma\langle Z,\gamma|u|\Psi_{\rm in},\gamma\rangle,
\end{aligned}
\end{equation}
where $|[\Psi_{\rm in}],\gamma\rangle$ and  
$|[\Psi_{\rm out}],\gamma\rangle$ denote the gauge invariant correspondence of $|\Psi_{\rm in},\gamma\rangle$ and  
$|\Psi_{\rm out},\gamma\rangle$, i.e.,
\begin{equation}
|[\Psi_s],\gamma\rangle=\mathcal N_\gamma^{(s)}\int\dd \mu_H(u)\, u|\Psi_s,\gamma\rangle,\ \text{for } s={\rm in,out},
\end{equation}
with $\mathcal N_\gamma^{(s)}$ being the normalization factor
\begin{equation}\label{eq:Ns}
\mathcal N_\gamma^{(s)}=\left|\int \dd \mu_H(u) \langle\Psi_s,\gamma|u|\Psi_s,\gamma\rangle\right|^{-\frac{1}{2}}.
\end{equation}

The current work will concern ourselves with the $n$-point correlation functions which are objects taking the form
\begin{equation}\label{eq:B}
\begin{aligned}
B=&\langle Z',\gamma| \prod_{j=1}^n \exp[\frac{i}{\hbar}\hat \hop_\gamma T_j] \hat O_{j}\exp[-\frac{i}{\hbar}\hat \hop_\gamma T_j]|Z,\gamma\rangle\\
=&\langle Z',\gamma| \exp[\frac{i}{\hbar}\hat\hop_\gamma T_n ]\prod_{j=1}^{n} \hat O_{j}\exp[-\frac{i}{\hbar}\hat \hop_\gamma(T_{j}-T_{j-1})]|Z,\gamma\rangle
\end{aligned}
\end{equation}
where the sequence of moments $T_k$ for $0\leq k\leq n$ satisfy $T_n\geq T_{n-1}\geq\cdots\geq T_0= 0$ and $\hat O_k$ is an operator  polynomial of holonomies, volume operators and fermion field operators. The flow of evolution contained in $B$ is forward between $T_0$ and $T_n$ but backward between $T_n$ and $T_{n+1}$. Since a backward evolution is equivalent a forward one with a minus Hamiltonian, we can rewrite $B$ as
\begin{equation}\label{eq:B2}
\begin{aligned}
B=\langle Z',\gamma| \exp[-\frac{i}{\hbar}\hat\hop_\gamma^{(T)} (T_{n+1}-T_n) ]\prod_{j=1}^{n} \hat O_{j}\exp[-\frac{i}{\hbar}\hat \hop_\gamma^{(T)}(T_{j}-T_{j-1})]|Z,\gamma\rangle
\end{aligned}
\end{equation}
with $T_{n+1}=2 T_n\geq T_n$, where the time-dependent Hamiltonian $\hat\hop_\gamma^{(T)}$ is
\begin{equation}
\hat\hop_\gamma^{(T)}=\left\{
\begin{aligned}
\hat\hop_\gamma,&\quad T<T_n\\
-\hat\hop_\gamma,&\quad T>T_n.
\end{aligned}
\right.
\end{equation}
In \eqref{eq:B2}, the evolution flow becomes  forward  all the time, while the price is the time-dependent Hamiltonian $\hat\hop_\gamma^{(T)}$. 
Now $B$ can be calculated by applying the standard path integral formulation. We only need to proceed with steps analogous to calculating $A^\gamma_{Z,Z'}$; we divide the  interval $[T_{k-1},T_k]$ for all $1 \leq k\leq n+1$ into  $N_k$ slices such that $T_k-T_{k-1}=N_k\delta\tau$ and then insert resolution of identity. Since the Hamiltonian in the present case becomes time dependent, the final path integral formula will involve an action slightly different from the previous one  in $A^\gamma_{Z,Z'}$. With straightforward derivation, one obtains the final result taking the form. 
\begin{equation}\label{eq:Bint}
\begin{aligned}
B=\lim_{N_i\to\infty}\int_{\mathcal R(Z',Z)}\prod_{i=1}^{N+n-1} \mathcal D[Z_i,\gamma]
\left(\prod_{k=1}^{n} \frac{\langle Z_{\mathfrak{f}(k)+1 },\gamma|\hat O_k|Z_{\mathfrak{f}(k)},\gamma\rangle}{\langle Z_{\mathfrak{f}(k)+1},\gamma|Z_{\mathfrak{f}(k)},\gamma\rangle}\right) \Xi[\g]e^{\frac{1}{t} \widetilde S_{\vec N}[\g,\nu]}
\end{aligned}
\end{equation}
where $N=\sum_{j=1}^{n+1}N_j$, an extra resolution of identity is inserted at each moment $T_k$  to sandwich $\hat O_k$, $\mathfrak{f}(k)=k-1+\sum_{i=1}^{k}N_i$ is employed
to count the resolution of identity before $\hat O_k$ and the action $\widetilde S_{\vec N}[\g,\nu]$ for $\vec N=(N_1,N_2,\cdots,N_{n+1})$ is
\begin{equation}\label{eq:action1}
\begin{aligned}
\widetilde S_{\vec N}[\g,\nu]=\sum_{j=0}^{N+n-1} K_\gamma(Z_{j+1},Z_j)-\frac{i  \kappa}{a^2}  \sum_{j=0}^{N+n-1}\delta \tau_j \frac{\langle Z_{j+1},\gamma|\hat \hop_\gamma|Z_j,\gamma\rangle}{\langle Z_{j+1},\gamma|Z_j,\gamma\rangle}
\end{aligned}
\end{equation}
with $Z_0=Z$, $Z_{N+n}=Z'$ and $\delta\tau_j$ given by
\begin{equation}\label{eq:deltatauj}
\delta\tau_j=\left\{
\begin{array}{cl}
    0 & j=\mathfrak f(k),\ \forall 1\leq k\leq n, \\
    \delta\tau & j\leq   \mathfrak f(n)-1\text{ and } j\neq \mathfrak f(k),\  \forall 1\leq k\leq n,\\
   -\delta\tau &   \mathfrak f(n)+1\leq j\leq N+n-1.
\end{array}
\right.
\end{equation}
In \eqref{eq:deltatauj}, $\delta\tau_j$ takes $-\delta\tau$  due to the Hamiltonian taking $-\hat\hop_\gamma$ after $T_n$; and $\delta\tau_{\mathfrak f(k)}$ takes $0$ because the matrix element of $\hat O_k$ contributes only the kinetic term to the action.  

Finally, with \eqref{eq:Bint}, one can formulate $\exp[-\frac{i}{\hbar}\hat\hop_\gamma^{(T)} (T_{n+1}-T_n) ]\prod_{j=1}^{n} \hat O_{j}\exp[-\frac{i}{\hbar}\hat \hop_\gamma^{(T)}(T_{j}-T_{j-1})]$ sandwiched by any initial and final states, analogous to \eqref{eq:transitionamplitudegeneral}.

\section{Minkowski  vacuum on cubic lattices}\label{sec:vacuumonagraph}
We seek some vacuum state in which the quantum gravity state resembles the classical  Minkowski geometry. A natural candidate is a state comprising the coherent states (see Sec. \ref{sec:coherents}) peaked at the Minkowski geometry. To employ these coherent states, an issue that the coherent states are graph-dependent arises so that one seems have to choose some preferred graphs.  Then physical results might depend on the choice of graphs.   To minimize the dependence on the choice of graphs, we propose considering the superposition of graphs. Furthermore, since the concerning background  is semiclassical, it is reasonable to prioritize those graphs that reflect the semiclassical properties better than others. Taking into account of the topology of $\Sigma$ and the desirable form of the expectation value of the Hamiltonian operator on cubic lattice \cite{Zhang:2021qul}, we would consider superposition states in which cubic lattices carry significant weight. As a typical representative of such states, the one comprising only cubic graphs is thus employed.

As discussed above, given a cubic graph $\gamma$,  the vacuum state $|\Omega,\gamma\rangle\in \widetilde{\mathcal H}_{\gamma}$ on $\gamma$ should take the form 
\begin{equation}\label{eq:omega}
|\Omega,\gamma\rangle=\p_\gamma|\Psi_{\g_M^\gamma}^t,\gamma\rangle\otimes |\omega,\gamma\rangle=|\widetilde{\Psi_{\g_M^\gamma}^t},\gamma\rangle\otimes |\omega,\gamma\rangle
\end{equation}
where $|\Psi_{\g_M^\gamma}^t,\gamma\rangle\in \mathcal H_\gamma^G$ is the coherent state peaking at the Minkowski geometry on $\gamma$ and $ |\omega,\gamma\rangle$ is some state in $\mathcal H_\gamma^F$. To define $\g_M^\gamma$ precisely, we need to introduce a Euclidean metric $\delta_{ij}$ on $\Sigma$ such that the  perimeter of $\Sigma$ along each axis is $\ell_o$. Then, for $\gamma$ being the cubic graph with edges parallel to the axes and the fiducial lattice spacing $\mu$ in the fiducial metric, $\g_M^\gamma$ is  given by
\begin{equation}\label{eq:euclideanbackground}
\g_M^\gamma(e)=e^{-i\mathring{p}_{\gamma}\tau_{e}}
\end{equation}
where $\tau_e=\tau_j$ for those edges $e$ along the $j$-axis direction and $\mathring{p}_{\gamma}=P\mu^2/(a^2\beta)$ with $P\delta_{ij}$ being the dynamical interior geometry on $\gamma$ endowed by $|\widetilde{\Psi_{\g_M^\gamma}^t},\gamma\rangle$. In \eqref{eq:omega}, the state is defined in $\widetilde{\mathcal H}_\gamma$, setting the stage for superposition over graphs. Moreover, as discussed  at the end of Sec. \ref{sec:coherentsG}, the projection $\p_\gamma$ will not destroy the semiclassical features of $|\Psi_{\g_M^\gamma}^t,\gamma\rangle$. That is to say,  $|\widetilde{\Psi_{\g_M^\gamma}^t},\gamma\rangle$ can play the role of  $|\Psi_{\g_M^\gamma}^t,\gamma\rangle$ to endow the graph $\gamma$ with a semiclassical Minkowski geometry. For $|\omega,\gamma\rangle$, its role indicate that $|\Omega,\gamma\rangle$ is a vacuum. Thus $|\omega,\gamma\rangle$ is defined as the ground state of the effective fermion Hamiltonian  $\hat H_{F,\gamma}^{\rm eff}$ in the  semiclassical Minkowski background, where $\hat H_{F,\gamma}^{\rm eff}$ is given by
\begin{equation}\label{eq:effectivematterH}
\langle \Phi|\hat H_{F,\gamma}^{\rm eff}|\Phi'\rangle=-\langle\widetilde{ \Psi_{\g_M^\gamma}^t}\otimes \Phi,\gamma|\p_\gamma\hat H_\gamma^F\p_\gamma| \widetilde{\Psi_{\g_M^\gamma}^t}\otimes \Phi',\gamma\rangle,\ \forall \Phi,\Phi'\in \mathcal H_\gamma^F. 
\end{equation} 
This definition together with \eqref{eq:differenceexpect} leads $\hat H_{F,\gamma}^{\rm eff}$ to be
\begin{equation}\label{eq:HeffMaction}
\begin{aligned}
\hat H_{F,\gamma}^{\rm eff}=\frac{i\hbar }{2}\frac{1}{a\sqrt{\mathring{p}_{\gamma}\beta}}\sum_{v\in V(\gamma)}\sum_{e \in E_v(\gamma)} \left(\hat\zeta^\dagger_v\sigma^e\hat\zeta_{v+\delta_e}-\hat\zeta^\dagger_{v+\delta_e}\sigma^e\hat\zeta_v \right)
\end{aligned}
\end{equation}
where $E_v(\gamma)=\{e_x^+(v),e_y^+(v),e_z^+(v)\}$ is the set of the three  edges starting from $v$ and directing to the positive directions, and $\delta_e$ and $\sigma^e$ for the edge $e$ along the $k$th direction are defined by $(\delta_e)_i:=\delta_{k,i}$ and $\sigma^e:=\sigma^k$. 

To diagonalize $\hat H_{F,\gamma}^{\rm eff}$, we introduce the operators 
\begin{equation}
\hat{\xi}_{\vec k}=\Theta(\vec k)\widehat{\widetilde \zeta}_{\vec k}
\end{equation}
where $\widehat{\widetilde \zeta}_{\vec k}$ is the Fourier transformation of $\hat\zeta_v$, i.e., $$\widehat{\widetilde \zeta}_{\vec k}=\sqrt{\frac{\mu^3}{\ell_o^3}}\sum_{v\in V(\gamma)}\hat\zeta_v e^{-i\frac{2\pi}{\ell_o} \vec k\cdot\vec x_v}$$ with $\vec x_v\in (\mu\mathbb{Z})^3$ denoting the coordinate of $v$, 
and $\Theta(\vec k)$ is 
\begin{equation}
\Theta(\vec k)=
\begin{pmatrix}
\frac{\sin(\frac{2\pi\mu}{\ell_o}k^3)- \mathfrak{s}(\vec k)}{\sin(\frac{2\pi\mu}{\ell_o}k^1)-i\sin(\frac{2\pi\mu}{\ell_o}k^2)}\sqrt{\frac{\mathfrak{s}(\vec k)+\sin(\frac{2\pi\mu}{\ell_o}k^3)}{2\mathfrak{s}(\vec k)}},\sqrt{\frac{\mathfrak{s}(\vec k)+\sin(\frac{2\pi\mu}{\ell_o}k^3)}{2\mathfrak{s}(\vec k)}}\\
\frac{\sin(\frac{2\pi\mu}{\ell_o}k^3)+ \mathfrak{s}(\vec k)}{\sin(\frac{2\pi\mu}{\ell_o}k^1)-i\sin(\frac{2\pi\mu}{\ell_o}k^2)}\sqrt{\frac{\mathfrak{s}(\vec k)-\sin(\frac{2\pi\mu}{\ell_o}k^3)}{2\mathfrak{s}(\vec k)}},\sqrt{\frac{\mathfrak{s}(\vec k)-\sin(\frac{2\pi\mu}{\ell_o}k^3)}{2\mathfrak{s}(\vec k)}}
\end{pmatrix}
=:
\begin{pmatrix}
\Theta(\vec k)_{+,+}&\Theta(\vec k)_{+,-}\\
\Theta(\vec k)_{-,+}&\Theta(\vec k)_{-,-}
\end{pmatrix}
,
\end{equation}
with 
\begin{equation}
\mathfrak{s}(\vec k)=\sqrt{\sum_{m=1}^3\sin^2(\frac{2\pi\mu}{\ell_o}k^m)}.
\end{equation}
Here, the total number $\ell_o/\mu=N$ of vertices along each direction is assumed to be even\footnote{As shown below, the lattice we consider  will be the refinement of an initial cubic lattice $\ga_0$. Given the initial cubic lattice $\ga_0$ with $N_0^3$ vertices, after the lattice refinements of $n$-steps, the refined cubic lattice $\ga_n$ has $N^3_n$ vertices, where $N_n=2^nN_0$ is always even for $n\geq1$. Therefore we are mostly interested in the case with an even number of vertices in each direction on the lattice. Moreover, the assumption also guarantees that the vacuum state is Bosonic.}. The  range of $k^a$ for $a=1,2,3$ is chosen as $k^a\in K[\gamma]=[-\frac{\ell_o}{2\mu},\frac{\ell_o}{2\mu}-1]\cap\mathbb{Z}$, i.e. $\vec k\in K[\gamma]^3$. Moreover, the matrix $\Theta(\vec k)$ is actually introduced to diagonalize the matrix $\sum_{j=1}^3\sin(\frac{2\pi\mu}{\ell_o}k^j)\sigma^j$, i.e., 
\begin{equation}\label{eq:diaksigma0}
\sum_{j=1}^3\sin(\frac{2\pi\mu}{\ell_o}k^j)\sigma^j=\Theta(\vec k)^\dagger
\begin{pmatrix}
-\mathfrak{s}(\vec k)&0\\
0&\mathfrak{s}(\vec k)
\end{pmatrix}
\Theta(\vec k). 
\end{equation}
A straightforward calculation gives the properties of the operators $\hat\xi_{\vec k,A}$
\begin{equation}\label{eq:operatorxik}
\begin{aligned}
[\hat\xi_{\vec k,A},\hat\xi_{\vec k',B}]_+&=0=[\hat\zeta_{\vec k,A}^\dagger,\hat\xi_{\vec k',B}^\dagger]_+,\\
[\hat\xi_{\vec k,A}^\dagger,\hat\xi_{\vec k',B}]_+&=\delta_{AB}\delta_{\vec k,\vec k'},\\
\hat\xi_{\vec k,A}|O,\gamma\rangle&=0,
\end{aligned}
\end{equation}
where $|O,\gamma\rangle$ denotes the unphysical vacuum state, i.e., 
\begin{equation}
|O,\gamma\rangle=\bigotimes_{v\in V(\gamma)}|0,0\rangle_v.
\end{equation}  
 The effective Hamiltonian operator $\hat H_{F,\gamma}^{\rm eff}$ in terms of $\hat\xi_{\vec k}$ is
\begin{equation}
\begin{aligned}
\hat H_{F,\gamma}^{\rm eff}=\frac{\hbar}{a\sqrt{\mathring{p}_{\gamma}\beta}}\sum_{\vec k\in K[\gamma]^3}\left(\mathfrak{s}(\vec k)\hat\xi^\dagger_{\vec k,+}\hat\xi_{\vec k,+}-\mathfrak{s}(\vec k) 
\hat\xi^\dagger_{\vec k,-}\hat\xi_{\vec k,-}\right).
\end{aligned}
\end{equation}
Due to the algebra \eqref{eq:operatorxik}, the operator $\hat\xi^\dagger_{\vec k,A}\hat\xi_{\vec k,A}$ for each  $A=\pm$ has the eigenvalues $0$ and $1$. Thus the ground  state $|\omega,\gamma\rangle$ of $\hat H_{F,\gamma}^{\rm eff}$ satisfies
\begin{equation}\label{eq:groundstate1}
\hat \xi_{\vec k,-}^\dagger|\omega,\gamma\rangle=0=\hat \xi_{\vec k,+}|\omega,\gamma\rangle,
\end{equation}
and takes eigenvalue
\begin{equation}
\omega(\gamma)=-\frac{\hbar}{a\sqrt{\mathring{p}_{\gamma}\beta}}\sum_{\vec k\in K[\gamma]^3}\mathfrak{s}(\vec k). 
\end{equation}
According to \eqref{eq:groundstate1}, $|\omega,\gamma\rangle$ can be expressed explicitly as
\begin{equation}\label{eq:vacuumomega1}
|\omega,\gamma\rangle=\sgn(\gamma)\prod_{\vec k\in K[\gamma]^3}\hat \xi_{\vec k,-}^\dagger |O,\gamma\rangle,
\end{equation}
with $\sgn(\gamma)$ being either $1$ or $-1$ such that
\begin{equation}\label{eq:innerproductvacuumFermion}
\langle \omega,\gamma|\Phi_\nu,\gamma\rangle=\exp(-\sum_{v\in V(\gamma)}\frac{1}{2}\nu^\dagger(v)\nu(v))\prod_{\vec k\in K[\gamma]^3}\sum_{B=\pm}\Theta(\vec k)_{-,B} \widetilde\nu_{B}(\vec k).
\end{equation}
where $\tilde\nu$ is the Fourier transformation of $\nu$, 
\begin{equation}
\widetilde\nu(\vec k)=\sqrt{\frac{\mu^3}{\ell_o^3}}\sum_{v\in V(\gamma)}\nu(v) e^{-i\frac{2\pi}{\ell_o}\vec k\cdot\vec x_v}.
\end{equation}
Indeed, one can verify easily that
\begin{equation}
\begin{aligned}
\langle \omega,\gamma|\Phi_\nu,\gamma\rangle
=&\text{(a sign factor)}\times \sgn(\gamma)\exp(-\sum_{v\in V(\gamma)}\frac{1}{2}\nu^\dagger(v)\nu(v))\prod_{\vec k\in K[\gamma]^3}\sum_{B=\pm}\Theta(\vec k)_{-,B} \widetilde\nu_{B}(\vec k)
\end{aligned}
\end{equation}
which allows us to choose a convention of $\sgn(\gamma)$ such that \eqref{eq:innerproductvacuumFermion} holds. The assumption of the even number of vertices in each lattice ensures that there are an even number of modes $\vec k$ in $K[\gamma]^3$. Therefore, the  ground state given by  \eqref{eq:vacuumomega1} is Bosonic.

\begin{figure}[ht]
	\begin{center}
	\includegraphics[width=0.4\textwidth]{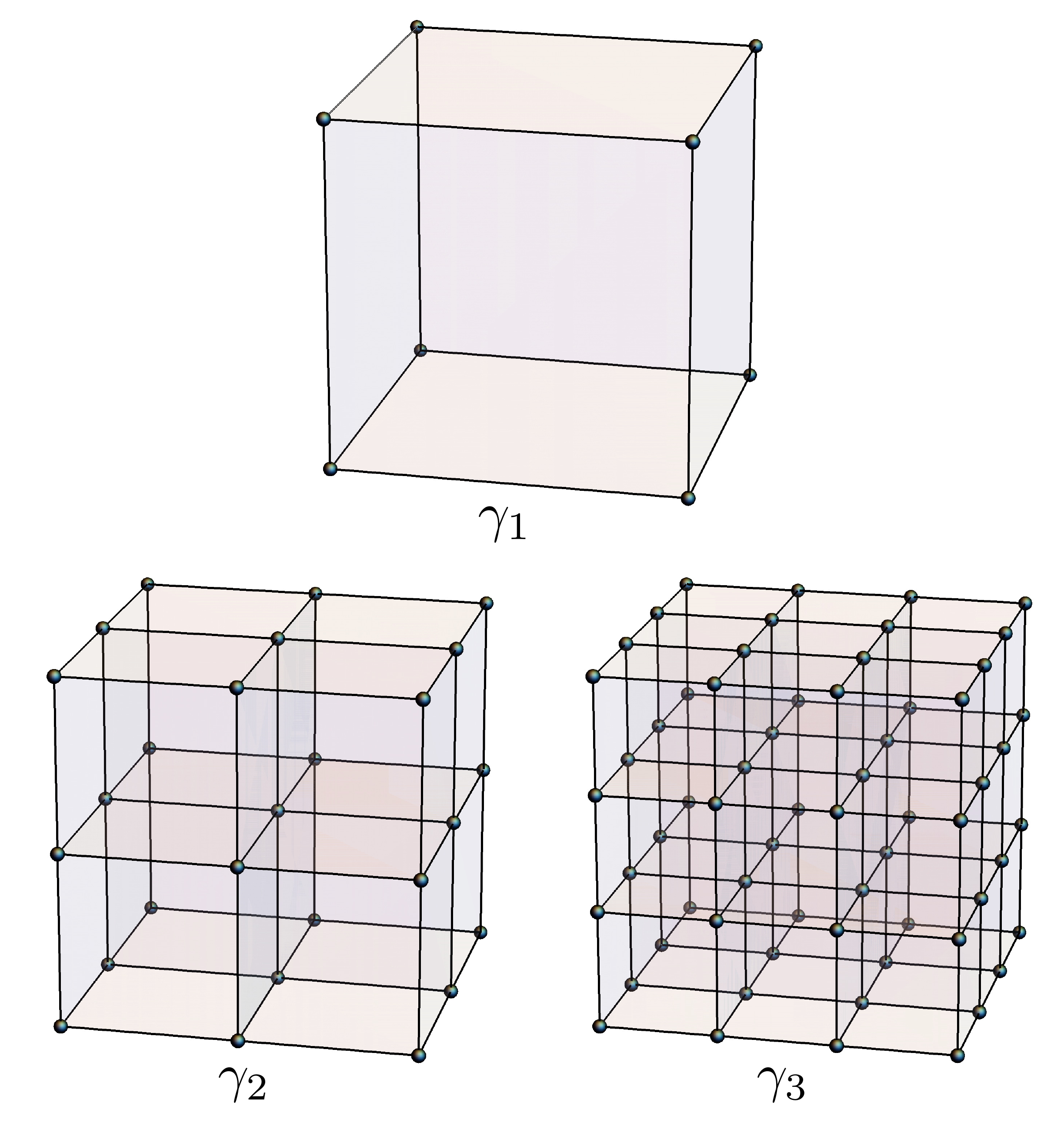}
	\caption{ Lattice refinement of a cell of $\gamma_1$ to $\gamma_2$ and $\gamma_3$. The fiducial lattice spacing of $\gamma_n$ is $\mu/n$ such that
	the number of vertices $N_n$ along each direction on $\gamma_n$ satisfies $N_n=n N_1$, with $N_1$ be that of $\gamma_1$.
	}
	\label{refinement}
	\end{center}
\end{figure}

The final vacuum state $|\Omega\rangle$ will be the superposition of  $|\Omega,\gamma\rangle$ on various cubic graphs. To define $|\Omega\rangle$ precisely, let us fix an initial lattice $\gamma_1=\gamma$ with fiducial lattice spacing $\mu_1\equiv \mu$ and define $\gamma_n$ as a lattice refinement of $\gamma_1$ such that its the fiducial lattice spacing is $\mu_1/n$ (see Fig. \ref{refinement}). Let $L\gg 1$ be the total number of the refinements. Then, the final vacuum state is defined as 
\begin{equation}
\begin{aligned}
|\Omega\rangle=\sum_{n=1}^nw_n|\Omega,\gamma_n\rangle
\end{aligned}
\end{equation}
where $w_n$ is the satisfying $\sum_{n=1}^L|w_n|^2=1$. Here the vacuum state is defined for the future calculation of the  propagator . The requirement that $\gamma_n$ is a lattice refinement of $\gamma_1$ guarantees that all graphs in the superposition contribute  to the final result.  Note that $|\Omega\rangle$ is not gauge invariant. The gauge invariant vacuum $|[\Omega]\rangle$ is given by the group averaging of $|\Omega\rangle$, i.e., 
\begin{equation}\label{eq:vaccumstate}
|[\Omega]\rangle=\sum_{n=1}^Lw_n|[\Omega],\gamma_n\rangle
\end{equation}
where $|[\Omega],\gamma_n\rangle$ is the normalized gauge invariant projection of $|\Omega,\gamma\rangle$,  i.e, 
\begin{equation}
|[\Omega],\gamma_n\rangle=\mathcal N_n \int\dd\mu_H(u) u|\Omega,\gamma\rangle
\end{equation}
with $\mathcal N_n$ denoting the normalization factor.

  \section{propagator on Minkowski background}\label{sec:propagator}
  Another ingredient for propagator is the field operator. Due to \eqref{eq:xipsi} and \eqref{eq:thetazeta}, it is natural to define the fermion field operator as 
  \begin{equation}\label{eq:psi}
  \hat\psi_A(v)=\sqrt{\hbar}\hat V_v^{-\frac12}\hat\zeta_{v,A}, \forall A=\pm,
  \end{equation}
  where $\hat V_v^{-\frac12}$ is given in \eqref{eq:inversev}. As the statement below \eqref{eq:inversev}, there is a sign ambiguity  in $\hat V_v^{-\frac12}$. However, as far as the  propagator  is concerned, $\hat\psi(v)$ and, thus, $\sqrt{\hbar}\hat V_v^{-\frac12}$ alway appear in pair. Thus, the sign ambiguity does not make any essential. For convenience, we introduce the notation
\begin{equation}\label{eq:etatauv}
\begin{aligned}
\hat \psi_A(\tau,v)=&\exp[\frac{i}{\hbar}  \tau \hat\hop] \hat\psi_{A}(v) \exp[-\frac{i}{\hbar}  \tau \hat\hop],\\
\hat \psi_A^\dagger(\tau,v)=&\exp[\frac{i}{\hbar}  \tau \hat\hop] \hat\psi_{A}^\dagger(v) \exp[-\frac{i}{\hbar}  \tau \hat\hop].
\end{aligned}
\end{equation}
Applying $\hat\psi_A(\tau,v)$ and $\hat\psi^\dagger_A(\tau, v)$, the operator $\hat G_{ \vec A}(\vec \tau,\vec v)$ for the  propagator  is given by
\begin{equation}
\begin{aligned}
\hat G_{\vec A}(\vec \tau,\vec v):=\mathcal{T}\left(\hat\psi_{A_1}(\tau_1,v_1)\hat\psi_{A_{2}}^\dagger(\tau_{2},v_{2}) \right)
\end{aligned}
\end{equation}
where the vectors $\vec A$, $\vec\tau$ and $\vec v$ denote $(A_1,A_2)$, $(\tau_1,\tau_2)$ and $(v_1,v_2)$ respectively and $\mathcal T$ is the time ordering operator to put the operators at latter time to the left. It should be spelled out that we concern ourselves with only the case where $v_1$ and $v_2$ are both in $V(\gamma_1)$ so that all graphs have contribution to the final results. Note that the operator $\hat G_{\vec A}(\vec\tau,\vec v)$ is not gauge invariant. To get a gauge invariant operator, we need to choose a smeared function $\hat F^{\vec A}$ to do contraction with  $\hat G_{\vec A}(\vec\tau,\vec v)$. A concrete example of $\hat F^{\vec A}\equiv\hat F^{(A_1,A_2)}$ is 
\begin{equation}\label{eq:concreteF}
\hat F^{\vec A}=\sum_{\gamma}\p_\gamma D^{\frac{1}{2}}_{A_2A_1}(h_{v_2\to v_1})\p_\gamma 
\end{equation}
where $h_{v_2\to v_1}$ denotes a holonomy along a path in $\gamma_1$ from $v_2$ to $v_1$. Here that the path of the holonomy $h_{v_2\to v_1}$ is required to be in $\gamma_1$ ensures that  $\hat F^{\vec A}$ is a well-defined operator in all $\widetilde{\mathcal H}_{\gamma_n}$.  It is worthing noting that our calculation shown below does not depend on the explicit expression 
of $\hat F^{\vec A}$. It will be obtained that, as far as the leading order is concerned, the final result is just the classical value of $\hat F^{\vec A}$ times some kernel $G_{\vec A}(\vec\tau,\vec v)$, where the kernel $G_{\vec A}(\vec\tau,\vec v)$ restores the  propagator  function of the regular QFT.  Using $\widehat{G[F]}(\vec\tau,\vec v)$ to  denote $\hat F^{\vec A}\hat G_{\vec A}(\vec\tau,\vec v)$, we define the  propagator  as
 \begin{equation}\label{eq:twopoint1}
G[F](\vec\tau,\vec v)=\langle[\Omega]|\widehat{G[F]}(\vec \tau,\vec v)|[\Omega]\rangle.
 \end{equation}

Substituting \eqref{eq:vaccumstate} into \eqref{eq:twopoint1}, one gets
\begin{equation}\label{eq:twopointtot}
G[F](\vec\tau,\vec v)=\sum_{m,m'=1}^Lw_{m'}^*w_m\langle [\Omega],\gamma_{m'}|\widehat{G[F]}(\vec\tau,\vec v)|[\Omega],\gamma_m\rangle.
\end{equation}
In the operator $\widehat{G[F]}(\vec\tau,\vec v)$ the evolution operator $\exp[-\frac{i}{\hbar}\tau\hat\hop]$ is involved. According to \eqref{eq:definehop} and \eqref{eq:hgamma}, each Hilbert space $\widetilde{\mathcal H}_\gamma$ is preserved by  $\hat\hop$. As a consequence, the evolution operator $\exp[-\frac{i}{\hbar}\tau\hat\hop]$ preserves each single  $\widetilde{\mathcal H}_\gamma$. This fact, together with  $v_1,v_2\in V(\gamma_{1})$ and the definition of $\hat F^{\vec A}$, leads to  that $\widehat{G[F]}(\vec\tau,\vec v)|[\Omega],\gamma_m\rangle$ is in $ \widetilde{\mathcal{H}}_{\gamma_m}$. Since $\widetilde{\mathcal H}_{\gamma_m}$ is orthogonal to $\widetilde{\mathcal H}_{\gamma_{m'}}$ for $m\neq m'$, \eqref{eq:twopointtot} can be simplified as
\begin{equation}\label{eq:twopointtot2}
G[F](\vec\tau,\vec v)=\sum_{m=1}^L|w_{m}|^2\langle [\Omega],\gamma_{m'}|\widehat{G[F]}(\vec\tau,\vec v,\gamma_m)|[\Omega],\gamma_m\rangle.
\end{equation}
where $\widehat{G[F]}(\vec\tau,\vec v,\gamma):=\p_\gamma D^{\frac{1}{2}}_{A_2A_1}(h_{v_2\to v_1})\hat G_{\vec A}(\vec\tau,\vec v,\gamma)\p_\gamma $ is given by
\begin{eqnarray}
\hat G_{\vec A}(\vec\tau,\vec v,\gamma):=&\mathcal{T}\left(\hat\psi_{A_1}(\tau_1,v_1,\gamma)\hat\psi_{A_{2}}^\dagger(\tau_2,v_2,\gamma) \right)\label{eq:Ggammadefinition}\\
\hat \psi_A(\tau,v,\gamma)=&\exp[\frac{i}{\hbar}  \tau\hat\hop_\gamma] \hat \psi_A(v)\exp[-\frac{i}{\hbar}  \tau\hat\hop_\gamma],\\
\hat \psi_A(\tau,v,\gamma)^\dagger=&\exp[\frac{i}{\hbar}  \tau\hat\hop_\gamma] \hat \psi_A^\dagger(v)\exp[-\frac{i}{\hbar}  \tau\hat\hop_\gamma].
\end{eqnarray}
According to  \eqref{eq:twopointtot2}, the  propagator  $G[F](\vec\tau,\vec v)$ is completely encoded in 
\begin{equation}
\begin{aligned}
G[F](\vec\tau,\vec v,\gamma):=&\langle [\Omega],\gamma|\widehat{G[F]}(\vec\tau,\vec v,\gamma)|[\Omega],\gamma\rangle\\
=&\mathcal N_\gamma\int\dd \mu_H(u)\langle\Omega,\gamma| \widehat{G[F]}(\vec\tau,\vec v,\gamma) u|\Omega,\gamma\rangle
\end{aligned}
\end{equation}
where  $u:V(\gamma)\ni v\mapsto u(v)\in \sut$ denotes the gauge transformation field and the normalization factor is given by
\begin{equation}\label{eq:ngamma}
\mathcal N_\gamma^{-1}=\int\dd \mu_H(u)\langle\Omega,\gamma|u|\Omega,\gamma\rangle.
\end{equation}

We need to apply the path integral formulation to calculation $G[F](\vec\tau,\vec v,\gamma)$. As in regular QFT, the path integral will be calculated by stationary phase approximation which needs to solve the equation of motion obtained by vanishing the variation of the effective action.  In LQG, without considering  the fermion field, the equation of motion and its solution have been well studied in \cite{han2020effective}, despite the different boundary states used therein. Actually, without fermion, the results in \cite{han2020effective} can be applied directly  to calculate the transition amplitude between the coherent state $|\Psi_{\g_M^\gamma}^t,\gamma\rangle$. However, in the current work, our boundary state contains $\p_\gamma |\Psi_{\g_M^\gamma}^t,\gamma\rangle$ rather than $|\Psi_{\g_M^\gamma}^t,\gamma\rangle$. Thus, to relate our work with the results in \cite{han2020effective}, we introduce $|\Omega_0,\gamma\rangle$ as
\begin{equation}
|\Omega_0,\gamma\rangle=|\Psi_{\g_M^\gamma}^t,\gamma\rangle\otimes|\omega,\gamma\rangle
\end{equation}
so that our vacuum state satisfy 
\begin{equation}
|\Omega,\gamma\rangle=\p_\gamma|\Omega_0,\gamma\rangle.
\end{equation}
Then, we define  $G^0[F](\vec\tau,\vec v,\gamma)$ as 
\begin{equation}
\begin{aligned}
G^0[F](\vec\tau,\vec v,\gamma)&:=\mathcal N_\gamma\int\dd \mu_H(u)\langle\Omega_0,\gamma| \widehat{G[F]}(\vec\tau,\vec v,\gamma) u|\Omega_0,\gamma\rangle.
\end{aligned}
\end{equation}
By \eqref{eq:concreteF}, the projection contained in $\hat F^{\vec A}$ leads directly to 
\begin{equation}
G^0[F](\vec\tau,\vec v,\gamma)=G[F](\vec\tau,\vec v,\gamma).
\end{equation}
Nonetheless, we would do the following discussion to see that $G^0[F](\vec\tau,\vec v,\gamma)$ and $G[F](\vec\tau,\vec v,\gamma)$ are the same up to a $O(t^\infty)$ term for a general $\hat F^{\vec A}$ without $\p_\gamma$ contained.  We  have
\begin{equation}\label{eq:diffgg0}
\begin{aligned}
&G^0[F](\vec\tau,\vec v,\gamma)-G[F](\vec\tau,\vec v,\gamma)\\
=&\mathcal N_\gamma\int\dd \mu_H(u)\langle\Omega_0,\gamma| \widehat{G[F]}(\vec\tau,\vec v,\gamma) u(1-\p_\gamma)|\Omega_0,\gamma\rangle\\
&+\mathcal N_\gamma\int\dd \mu_H(u)\langle\Omega_0,\gamma|(1-\p_\gamma) \widehat{G[F]}(\vec\tau,\vec v,\gamma) u|\Omega_0,\gamma\rangle\\
&+\mathcal N_\gamma\int\dd \mu_H(u)\langle\Omega_0,\gamma|(1-\p_\gamma) \widehat{G[F]}(\vec\tau,\vec v,\gamma) u(1-\p_\gamma)|\Omega_0,\gamma\rangle\\
\end{aligned}
\end{equation}
Note that our fermion field operators $\hat\psi_A(v)$ and $\hat\psi^\dagger_A(v)$ are constructed on discrete graphs. This fact leads to that  $\hat\psi_A(v)$ and $\hat\psi^\dagger_A(v)$ are bounded. Indeed, they are bounded  by $1$, i.e., $\|\hat\psi_A(v)\|=\|\hat\psi^\dagger_A(v)\|<1$. As a consequence, $\hat  G_{\vec A}(\vec\tau,\vec v,\gamma)$ is bounded by $1$. We thus have, for instance, 
\begin{equation}
\begin{aligned}
&\Big|\langle\Omega_0,\gamma| \widehat{G[F]}(\vec\tau,\vec v,\gamma) u(1-\p_\gamma)|\Omega_0,\gamma\rangle\Big|^2\\
\leq &\left\|(1-\p_\gamma|\Omega_0,\gamma\rangle\right\|^2 \left\|\hat F^{\vec A} |\Omega_0,\gamma\rangle\right\|^2.
\end{aligned}
\end{equation}
The term $\left\|\hat F^{\vec A} |\Omega_0,\gamma\rangle\right\|^2$ is just the expectation value of $\hat F^{\vec A}(\hat F^{\vec A})^\dagger$ in the state $|\Omega_0,\gamma\rangle$ which is finite. For the term $\left\|(1-\p_\gamma|\Omega_0,\gamma\rangle\right\|^2$, according to \eqref{eq:compare2inner}, it is a $O(t^\infty)$ term. Thus, we get 
\begin{equation}
\Big|\langle\Omega_0,\gamma| \widehat{G[F]}(\vec\tau,\vec v,\gamma) u(1-\p_\gamma)|\Omega_0,\gamma\rangle\Big|^2=O(t^\infty).
\end{equation}
The same arguments are applied to the other terms in \eqref{eq:diffgg0}. It thus concludes that 
\begin{equation}\label{eq:GG0}
G[F](\vec\tau,\vec v,\gamma)=G^0[F](\vec\tau,\vec v,\gamma)+O(t^\infty). 
\end{equation}
Moreover, proceeding the same step analogous to \eqref{eq:diffgg0},  $\mathcal N_\gamma^{-1}$ can be  simplified as
\begin{equation}\label{eq:ngamma2}
\begin{aligned}
\mathcal N_\gamma^{-1}=&\int \dd\mu_H(u)\langle\Omega_0,\gamma|\p_\gamma u\p_\gamma|\Omega_0,\gamma\rangle\\
=&\int \dd\mu_H(u)\langle\Omega_0,\gamma| u|\Omega_0,\gamma\rangle+O(t^\infty). 
\end{aligned}
\end{equation}
We finally have
\begin{equation}\label{eq:gfapp}
\begin{aligned}
G[F](\vec\tau,\vec v,\gamma)=\langle [\Omega_0],\gamma|\widehat{G[F]}(\vec\tau,\vec v,\gamma)|[\Omega_0],\gamma\rangle+O(t^\infty)
\end{aligned}
\end{equation}
where $|[\Omega_0],\gamma\rangle$ denotes the normalized gauge invariant projection of $|\Omega_0,\gamma\rangle$. The right hand side of \eqref{eq:gfapp}, i.e.,  $\langle [\Omega_0],\gamma|\widehat{G[F]}(\vec\tau,\vec v,\gamma)|[\Omega_0],\gamma\rangle$ is calculable by employing the results in \cite{han2020effective}, as shown in the following.

\subsection{Calculation of $G^0[F](\vec\tau,\vec v,\gamma)$}
To perform the time ordering operation in \eqref{eq:Ggammadefinition}, we introduce the map $\mathfrak i:\{1,2\}\to \{1,2\} $ to reordering  $\tau_1$ and $\tau_2$ such that $\tau_{\mathfrak i(1)}> \tau_{\mathfrak i(2)}$. Then, we have
\begin{equation}
\begin{aligned}
\hat G_{\vec A}(\vec\tau,\vec v,\gamma)=&\sgn(\vec \tau)\exp[\frac{i}{\hbar}\tau_{\mathfrak i(2)} \hat\hop_\gamma ]\hat Y_{A_{\mathfrak i(2)}}^{(v_{\mathfrak i(2)})}\exp[-\frac{i}{\hbar}(\tau_{\mathfrak i(2)}-\tau_{\mathfrak i(1)})\hat\hop_\gamma]\hat Y_{A_{\mathfrak i(1)}}^{(v_{\mathfrak i(1)})}\exp[-\frac{i}{\hbar}\tau_{\mathfrak i(1)}\hat\hop_\gamma]
\end{aligned}
\end{equation}
where $\sgn(\vec\tau)$ is the sign factor generated by time ordering, i.e., $\sgn(\vec\tau)=1$ for $\tau_1>\tau_2$, and $-1$ otherwise, and $\hat Y_A^{(v)}$ denotes either $\hat\psi_{A_1}(v_1)$ or $\hat\psi_{A_2}^\dagger$, explicitly given by
\begin{equation}\label{eq:definehaty}
\hat Y_{A_i}^{(v_i)}=\left\{
\begin{aligned}
&\hat\psi_{A_1}(v_1),&&i=1,\\
&\hat\psi_{A_2}^\dagger (v_2),&& i=2.
\end{aligned}
\right.
\end{equation}
Then, $G^0[F](\vec\tau,\vec v,\gamma)$ can be calculated directly by applying \eqref{eq:Bint}. One have
\begin{equation}
G^0[F](\vec\tau,\vec v,\gamma)=\lim_{N_i\to 0} G^0_{\vec N}[F](\vec\tau,\vec v,\gamma)
\end{equation}
with
\begin{equation}\label{eq:G01}
\begin{aligned}
&G^0_{\vec N}[F](\vec\tau,\vec v,\gamma)\\
=&\mathcal N_\gamma\int\dd\mu_H(u)\int \prod_{i=0}^{N+3} \mathcal D[Z_i,\gamma]
\frac{\langle Z_{N+3},\gamma|\hat F^{\vec A}|Z_{N+2},\gamma\rangle}{\langle Z_{N+3},\gamma|Z_{N+2},\gamma\rangle}
\left(\sgn(\vec \tau)\prod_{j=1}^{2} \frac{\langle Z_{\mathfrak{f}(j)+1 },\gamma|\hat Y_{A_{\mathfrak i(j)}}^{(v_{\mathfrak i(j)})}|Z_{\mathfrak{f}(j)},\gamma\rangle}{\langle Z_{\mathfrak{f}(j)+1},\gamma|Z_{\mathfrak{f}(j)},\gamma\rangle}\right) \times\\
&\Xi[\g]e^{\frac{1}{t}\widetilde S_{\vec N}[\g,\nu]} \langle\Omega_0,\gamma|Z_{N+3},\gamma\rangle\langle Z_0,\gamma|u|\Omega_0,\gamma\rangle.
\end{aligned}
\end{equation}
Eq. \eqref{eq:G01} can be obtained with the same steps for  \eqref{eq:Bint} except for some subtleties. At first, due to the operator $\hat F^{\vec A}$, we need to insert one more resolution of identity than  in \eqref{eq:Bint}. This fact gives rise to an additional kinetic term $K_\gamma(Z_{N+3},Z_{N+2})$ in the action. In other words, the action  $\widetilde S_{\vec N}[\g,\nu]$ in \eqref{eq:G01}  is not the same as the one obtained by directly applying  \eqref{eq:Bint} for our concrete case. Instead, they differ due to the additional kinetic term $K_\gamma(Z_{N+3},Z_{N+2})$ in \eqref{eq:G01}. However, this slight difference will not cause any essential in our following calculation. Second, we use the convention that $N_k$ denotes 
the number of slices dividing the interval $[\tau_{\mathfrak i(k-1)},\tau_{\mathfrak i(k)}]$ for $1\leq k\leq 3$, where $\tau_{\mathfrak  i(0)}=0$ and $\tau_{\mathfrak i(3)}=2\tau_{\mathfrak i(2)}$ as did for \eqref{eq:Bint}. Under this convention, for the case $\tau_2<\tau_1$ so that $\tau_{\mathfrak i(1)}=\tau_2$ and $\tau_{\mathfrak i(2)}=\tau_1$, the number of slices dividing $[0,\tau_2]$ is denoted by $N_1$ and the number of slices dividing $[\tau_2,\tau_1]$ is denoted by $N_2$. 

To evaluate $G_{\vec N}^0[F]$, let us consider $\langle\Omega_0,\gamma|Z_{N+3},\gamma\rangle$ and $\langle Z_0,\gamma|u|\Omega_0,\gamma\rangle$  at first. By definition, $\langle\Omega_0,\gamma|Z_{N+3},\gamma\rangle$  can be simplified as 
\begin{equation}\label{eq:innerproductvacuum1G}
\begin{aligned}
\langle \Omega_0,\gamma|Z_{N+3},\gamma\rangle=&\left(\prod_{e\in E(\gamma)}\frac{\mathring\eta_{N+3}(e) \sqrt{\sinh(p_{N+3}(e))\sinh(\mathring{p}_\gamma)}}{\sqrt{p_{N+3}(e)\mathring{p}_\gamma}\sinh(\mathring\eta_{N+3}(e))}\right)\left(\prod_{\vec k\in K[\gamma]^3}\sum_{A=\pm}\Theta(\vec k)_{-,A}\widetilde{\nu}_{N+3,A}(\vec k)
\right)\times\\
&\exp(-\sum_{e\in E(\gamma)}\frac{p_{N+3}(e)^2+(\mathring{p}_\gamma)^2-2\mathring\eta_{N+3}(e)^2}{2t}-\frac{1}{2}\sum_{v\in V(\gamma)}\nu_{N+3}^\dagger(v)\nu_{N+3}(v))
\end{aligned}
\end{equation}
where $\mathring\eta_{N+3}(e)$ is given by $2\cosh(\mathring\eta_{N+3}(e))=\tr(\g_{N+3}(e) e^{-i\mathring{p}_\gamma\tau_e})$, and $\widetilde{\nu}_{N+3}$ is the Fourier transformation of $\nu_{N+3}$.  Similarly, we have
\begin{equation}\label{eq:innerproductvacuum2G}
\begin{aligned}
\langle Z_0,\gamma|u|\Omega_0,\gamma\rangle=&\left(\prod_{e\in E(\gamma)}\frac{\mathring\eta_0^u(e) \sqrt{\sinh(p_0(e))\sinh(\mathring{p}_\gamma)}}{\sqrt{p_0(e)\mathring{p}_\gamma}\sinh(\mathring\eta_0^u(e))}\right)\left(\prod_{\vec k\in K[\gamma]^3}\sum_{A=\pm }\Theta(\vec k)_{-,A}\widetilde{\nu}^u_{0,A}(\vec k)\right)^*\times\\
&\exp(-\sum_{e\in E(\gamma)}\frac{p_0(e)^2+(\mathring{p}_\gamma)^2-2(\mathring\eta_0^u(e))^2}{2t}-\frac{1}{2}\sum_{v\in V(\gamma)}\nu_0^\dagger(v)\nu_0(v))
\end{aligned}
\end{equation}
where $\mathring\eta_0^u(e)$ is given by $2\cosh(\mathring\eta_0^u(e))=\tr(g_0(e)^\dagger u e^{-i\mathring{p}_\gamma\tau_e})$ and $\widetilde{\nu}^u_0$ denote the Fourier transformation of $\nu^u_0$, with $\nu^u_0$ given by $(\nu^u_0)_A(v)=(u(v)^{-1})_{A}{}^B\nu_B(v)$ being  the gauge transform of $\nu$.  
Substituting \eqref{eq:innerproductvacuum1G}, \eqref{eq:innerproductvacuum2G} into \eqref{eq:G01}, we have
\begin{equation}\label{eq:G02}
\begin{aligned}
G^0_{\vec N}[F](\vec\tau,\vec v,\gamma)=&\mathcal N_\gamma\int\dd\mu_H(u)\mathcal \int \prod_{i=0}^{N+3} \mathcal D[Z_i,\gamma]
\frac{\langle Z_{N+3},\gamma|\hat F^{\vec A}|Z_{N+2},\gamma\rangle}{\langle Z_{N+3},\gamma|Z_{N+2},\gamma\rangle}
\left(\sgn(\vec \tau)\prod_{j=1}^{2} \frac{\langle Z_{\mathfrak{f}(j)+1 },\gamma|\hat Y_{A_{\mathfrak i(j)}}^{(v_{\mathfrak i(j)})}|Z_{\mathfrak{f}(j)},\gamma\rangle}{\langle Z_{\mathfrak{f}(j)+1},\gamma|Z_{\mathfrak{f}(j)},\gamma\rangle}\right)\times\\
& \Xi'[\g,u]e^{\frac{1}{t}\widetilde S_{\vec N}'[\g,\nu,u]}
\prod_{\vec k\in K[\gamma]^3}\sum_{A=\pm}\Theta(\vec k)_{-,A}\widetilde{\nu}_{N+3,A}(\vec k)\left(\prod_{\vec k\in K[\gamma]^3}\sum_{A=\pm }\Theta(\vec k)_{-,A}\widetilde{\nu}^u_{0,A}(\vec k)\right)^*
\end{aligned}
\end{equation}
with
\begin{equation}\label{eq:tildeSN}
\begin{aligned}
\widetilde S_{\vec N}'[\g,\nu,u]=&\widetilde S_{\vec N}[\g,\nu]-\sum_{e\in E(\gamma)}\frac{p_{N+3}(e)^2+(\mathring{p}_\gamma)^2-2\mathring\eta_{N+3}(e)^2}{2}-\frac{t}{2}\sum_{v\in V(\gamma)}\nu_{N+3}^\dagger(v)\nu_{N+3}(v)\\
&-\sum_{e\in E(\gamma)}\frac{p_0(e)^2+(\mathring{p}_\gamma)^2-2(\mathring\eta_0^u(e))^2}{2}-\frac{t}{2}\sum_{v\in V(\gamma)}\nu_0^\dagger(v)\nu_0(v)
\end{aligned}
\end{equation}
and
\begin{equation}
\Xi'[\g,u]=\Xi[\g]\prod_{e\in E(\gamma)}\frac{\mathring\eta_{N+3}(e) \sqrt{\sinh(p_{N+3}(e))\sinh(\mathring{p}_\gamma)}}{\sqrt{p_{N+3}(e)\mathring{p}_\gamma}\sinh(\mathring\eta_{N+3}(e))}\frac{\mathring\eta_0^u(e) \sqrt{\sinh(p_0(e))\sinh(\mathring{p}_\gamma)}}{\sqrt{p_0(e)\mathring{p}_\gamma}\sinh(\mathring\eta_0^u(e))}.
\end{equation}
To further simplify the expression \eqref{eq:G02}, we turn to the term 
\begin{equation*}
\sgn(\vec \tau)\prod_{j=1}^{2} \frac{\langle Z_{\mathfrak{f}(j)+1 },\gamma|\hat Y_{A_{\mathfrak i(j)}}^{(v_{\mathfrak i(j)})}|Z_{\mathfrak{f}(j)},\gamma\rangle}{\langle Z_{\mathfrak{f}(j)+1},\gamma|Z_{\mathfrak{f}(j)},\gamma\rangle}=\sgn(\vec\tau)  \frac{\langle Z_{\mathfrak{f}(2)+1 },\gamma|\hat Y_{A_{\mathfrak i(2)}}^{(v_{\mathfrak i(2)})}|Z_{\mathfrak{f}(2)},\gamma\rangle}{\langle Z_{\mathfrak{f}(2)+1},\gamma|Z_{\mathfrak{f}(2)},\gamma\rangle} \frac{\langle Z_{\mathfrak{f}(1)+1 },\gamma|\hat Y_{A_{\mathfrak i(1)}}^{(v_{\mathfrak i(1)})}|Z_{\mathfrak{f}(1)},\gamma\rangle}{\langle Z_{\mathfrak{f}(1)+1},\gamma|Z_{\mathfrak{f}(1)},\gamma\rangle}.
\end{equation*}
By definition of $\sgn(\vec\tau)$, it can be canceled if we can reorder the right hand sight such that the term with $\hat Y_{A_1}^{(v_1)}$ is always put on the right. To write  the result in a compact form, let us introduce a function $\mathfrak n(i)$ defined as follows:  if $\tau_1<\tau_2$, $\mathfrak{n}(i)=\mathfrak{f}(i)$ for  all $i=1,2$; otherwise, $\mathfrak{n}(1)=\mathfrak{f}(2)$ and $\mathfrak{n}(2)=\mathfrak{f}(1)$. Then, a straightforward calculation gives us  
\begin{equation}
\begin{aligned}
&\sgn(\vec \tau)\prod_{j=1}^{2} \frac{\langle Z_{\mathfrak{f}(j)+1 },\gamma|\hat Y_{A_{\mathfrak i(j)}}^{(v_{\mathfrak i(j)})}|Z_{\mathfrak{f}(j)},\gamma\rangle}{\langle Z_{\mathfrak{f}(j)+1},\gamma|Z_{\mathfrak{f}(j)},\gamma\rangle}
=\left(\prod_{j=1}^{2n}\frac{\langle Z_{\mathfrak n(j)+1},\gamma|\hat V_{v_j}^{-\frac12} |Z_{\mathfrak n(j)},\gamma\rangle}{\langle Z_{\mathfrak n(j)+1},\gamma|Z_{\mathfrak n(j)},\gamma\rangle}\right) \mathcal F_{\vec A} 
\end{aligned}
\end{equation}
where  $\mathcal F_{\vec A}=\hbar \nu_{\mathfrak n(1),A_1}(v_1) \nu_{\mathfrak n(2)+1,A_{2}}^*(v_{2})$ and we used \eqref{eq:definehaty}, \eqref{eq:psi} and \eqref{eq:coherenteigen}.
Hence $G_{\vec N}^0[F]$ can be expressed as
\begin{equation}\label{eq:G03}
\begin{aligned}
G^0_{\vec N}[F](\vec\tau,\vec v,\gamma)=& \mathcal N_\gamma \int\dd\mu_H(u)\mathcal \int \prod_{i=0}^{N+3} \mathcal D[Z_i,\gamma]
\frac{\langle Z_{N+3},\gamma|\hat F^{\vec A}|Z_{N+2},\gamma\rangle}{\langle Z_{N+3},\gamma|Z_{N+2},\gamma\rangle}
\left(\prod_{j=1}^{2}\frac{\langle Z_{\mathfrak n(j)+1},\gamma|\hat V_{v_j}^{-\frac12}|Z_{\mathfrak n(j)},\gamma\rangle}{\langle Z_{\mathfrak n(j)+1},\gamma|Z_{\mathfrak n(j)},\gamma\rangle}\right)\times\\
& \Xi'[\g,u]e^{\frac{1}{t}\widetilde S_{\vec N}'[\g,\nu,u]}
\mathcal F_{\vec A}\prod_{\vec k\in K[\gamma]^3}\sum_{A=\pm}\Theta(\vec k)_{-,A}\widetilde{\nu}_{N+3,A}(\vec k)\left(\prod_{\vec k\in K[\gamma]^3}\sum_{A=\pm }\Theta(\vec k)_{-,A}\widetilde{\nu}^u_{0,A}(\vec k)\right)^*
\end{aligned}
\end{equation}

To calculate  \eqref{eq:G03}, let us  divide  $\widetilde S_{\vec N}'[\g,\nu,u]$  into two parts as
\begin{equation}
\widetilde S_{\vec N}'[\g,\nu,u]=S_{\vec N}^G[\g,u]+S_{\vec N}^F[\g,\nu]
\end{equation}
where $S_{\vec N}^G[\g,u]$ is the effective action of the vacuum LQG and $S_{\vec N}^F[\g,\nu]$ is the effective action of fermion coupled to gravity. The explicit expression of $S_{\vec N}^G[\g,u]$ and $S_{\vec N}^F[\g,\nu]$ can be obtained by referring to \eqref{eq:tildeSN}, \eqref{eq:action1}, and \eqref{eq:kineticterm}. We have
\begin{equation}\label{eq:sng}
\begin{aligned}
S_{\vec N}^G[\g,u]=&\sum_{j=-1}^{N+3}\sum_{e\in E(\gamma)}\left(\eta_{i+1,i}(e)^2-\frac{1}{2}\left(p_{i+1}(e)^2+p_i(e)^2\right)\right)\\
&-\frac{i \kappa}{a^2} \sum_{\substack{j=0}}^{N+2}\delta \tau_j \frac{\langle \g_{j+1},\gamma|\p_\gamma \hat H_\gamma^G\p_\gamma |\g_j,\gamma\rangle}{\langle\g_{j+1},\gamma|\g_j,\gamma\rangle}
\end{aligned}
\end{equation}
with $\eta_{N+4,N+3}(e)=\mathring\eta_{2N+2n+1}(e)$ and $\eta_{0,-1}(e)=\mathring\eta_{0}^{u}(e)$ and
\begin{equation}\label{eq:Sf}
\begin{aligned}
S_{\vec N}^F[\g,\nu]=&t\left(\sum_{j=0}^{N+2}\sum_{v\in V(\gamma)}\nu_{j+1}^\dagger(v)\nu_{j}(v)-\sum_{j=0}^{N+3}\sum_{v\in V(\gamma)}\nu_{j}^\dagger(v)\nu_{j}(v)\right)\\
&-\frac{i  \kappa}{a^2} \sum_{j=0}^{N+2}\delta\tau_j \frac{\langle Z_{j+1},\gamma|\p_\gamma \hat H_\gamma^F\p_\gamma|Z_j,\gamma\rangle}{\langle Z_{j+1},\gamma|Z_j,\gamma\rangle}.
\end{aligned}
\end{equation}
Note that $S_{\vec N}^F$ is a quadratic form of the fermion field $\nu_j(v)$ due to $\hat H_\gamma^F$ is quadratic in terms of the operator $\hat \theta(v)=\sqrt{\hbar}\hat\zeta_v$ by \eqref{eq:HF}. Thus, the integral of fermion fields in \eqref{eq:G03} is the standard Gaussian integral, which can be performed in principle. Let us use $I[\g,u]$ to denote the result of the fermion integral. It depends on $\g$ and $u$ because, for instance, the holonomies and fluxes are contained in the coefficient matrix of $S_{\vec N}^F$. Using $I[\g,u]$, we can write $G_{\vec N}^0[F]$ in the following form 
\begin{equation}\label{eq:GFin}
\begin{aligned}
G_{\vec N}^0[F]=\int\dd\mu_H(u)\int\prod_{i=0}^{N+3}\mathcal D[\g_i,\gamma] W[\g,u] I[\g,u] e^{\frac{1}{t} S_{\vec N}^G[\g,u]}
\end{aligned}
\end{equation}
where $W[\g,u]$ represents some function of $\g$ and $u$, and its explicit expression does not affect the present discussion. Since we concern ourselves with the leading order of $G_{\vec N}^0[F]$  in $t$, the integral of \eqref{eq:GFin} will be calculated by  the stationary  phase approximation method. To this end, we need to expand $S_{\vec N}^G[\g,u]$ to power series of $t$, i.e., 
$S_{\vec N}^G[\g,u]=S_0+tS_1+\cdots$. Then the stationary  phase approximation method needs us to study the equation of motion $\delta S_0=0$. The equation of motion has been well-studied in \cite{han2020effective}. According to the result of \cite{han2020effective},  in the continuous limit, i.e., $\delta\tau\ll 1$, $\delta S_0=0$ gives rise to the Hamilton's equation with respect to the Hamiltonian given by the leading order, i.e., $O(t^0)$-term, of the Hamiltonian operator's expectation value in the coherent states. In our work, the Hamiltonian is $\p_\gamma\hat H_\gamma^G\p_\gamma$ whose expectation value is the same as that of $\hat H_\gamma^G$ in the leading order. According to \cite{han2020effective}, the leading order of the expectation value of $\hat H_\gamma^G$ in the coherent state is the same as its classical value  $H_\gamma^G$. The classical value $H_\gamma^G$  is $H_G$ (see \eqref{eq:HG}) regularized with the holonomies and  fluxes on $\gamma$. According to these facts and the boundary condition resulting from the gravity part $|\Psi_{\g_M^\gamma}^t,\gamma\rangle$ of the boundary states $|\Omega_0,\gamma\rangle$, the solution to the equation of motion $\delta S_0=0$ is $\g_j=\g_M^\gamma$ for all $0\leq j\leq N+3$ and $u=\mathbbm{1}$. Hence, the solution endows $\gamma$ with a Minkowski geometry at any moment $\tau_j\equiv j \delta\tau$. Therefore, as far as the leading order of $G_{\vec N}^0[F]$ in $t$ is concerned, \eqref{eq:GFin} is 
\begin{equation}\label{eq:Ggeneral}
G_{\vec N}^0[F]=\left(\int\dd\mu_H(u)\int\prod_0^{N+3} \mathcal D[\g_i,\gamma] W[\g,u]e^{\frac{1}{t}S_{\vec N}^G[\g, u]}\right)I(\g_M^\gamma,\mathbbm 1)(1+O(t)). 
\end{equation}
Returning to our specific calculation, \eqref{eq:Ggeneral} implies 
\begin{equation}\label{eq:GNABT}
\begin{aligned}
G^0_{\vec N}[F](\vec\tau,\vec v,\gamma)=&\mathcal N_\gamma  \langle\Psi_{\g_M^\gamma}^t,\gamma|\hat F^{\vec A}|\Psi_{\g_M^\gamma}^t,\gamma\rangle
\left(\prod_{j=1}^{2}\langle\Psi_{\g_M^\gamma}^t,\gamma|\hat V_{v_j}^{-\frac12}|\Psi_{\g_M^\gamma}^t,\gamma\rangle\right)\\
&A_\gamma^{G,\vec N}\mathfrak{F}^{\vec N}_{\vec A}(\vec\tau,\vec v,\gamma)(1+O(t))
\end{aligned}
\end{equation}
where we assumed reasonably that $\hat F^{\vec A}$ does not contain any fermion operators, $A_\gamma^{G,\vec N}$ reads 
 \begin{equation}
A_\gamma^{G,\vec N}=\int\dd\mu_H(u)\prod_{i=0}^{N+3}\mathcal D[\g_i,\gamma] \Xi'[\g,u]e^{ S_{\vec N}^G[\g,u]/t}
\end{equation}
and $\mathfrak{F}^{\vec N}_{\vec A}(\vec\tau,\vec v,\gamma)$ is
\begin{equation}\label{eq:GNABF}
\begin{aligned}
\mathfrak{F}^{\vec N}_{\vec A}(\vec\tau,\vec v,\gamma)=&\int \prod_{i=0}^{N+3}\mathcal D[\nu_i,\gamma] e^{ \frac{1}{t}S_{\vec N}^F[\g_M^\gamma,\nu]}
\mathcal F_{\vec A}\prod_{\vec k\in K[\gamma]^3}\sum_{A=\pm}\Theta(\vec k)_{-,A}\widetilde{\nu}_{2N+2n+1,A}(\vec k)\times\\
&\left(\prod_{\vec k\in K[\gamma]^3}\sum_{A=\pm }\Theta(\vec k)_{-,A}\widetilde{\nu}_{0,A}(\vec k)\right)^*.
\end{aligned}
\end{equation}

According to the discussion below \eqref{eq:B2}, the limit of $A_\gamma^{G,\vec N}$ as $\delta\tau\to 0$ has the physical interpretation that the initial $\int\dd\mu_H(u) u|\Psi_{\g_M^\gamma}^t,\gamma\rangle$ evolves by the Hamiltonian $\p_\gamma\hat H_\gamma^G\p_\gamma$ from $\tau=0$  to $\max(\tau_1,\tau_2)$, then continues evolving, however, by the minus Hamiltonian $-\p_\gamma\hat H_\gamma^G\p_\gamma$ to $2\max(\tau_1,\tau_2)$, and meets the final state $\langle \Psi_{\g_M^\gamma}^t,\gamma|$. Thus, one has
\begin{equation}\label{eq:AN}
\lim_{N_i\to\infty} A_\gamma^{G,\vec N}=\int\dd\mu_H(u) \langle \Psi_{\g_M^\gamma},\gamma| u |\Psi_{\g_M^\gamma},\gamma\rangle \equiv \mathcal N_\gamma'. 
\end{equation} 
According to \eqref{eq:GNABF}, $\mathfrak{F}^{\vec N}(\vec\tau,\vec v,\gamma)$ is a standard Fermionic Gaussian integral, and thus is calculable. We put the step-by-step derivation and the results in Appendix \ref{app:calculatingF}. With that result, we can introduce  $\mathfrak{F}(\vec\tau,\vec v,\gamma)$ as
\begin{equation}\label{eq:FN2}
\mathfrak{F}(\vec\tau,\vec v,\gamma):=\lim_{N_i\to\infty}\mathfrak{F}^{\vec N}(\vec\tau,\vec v,\gamma)
\end{equation}
For the factor $\mathcal N_\gamma$ in \eqref{eq:GNABT}, inserting a resolution of identity in \eqref{eq:ngamma2} gives rise to
\begin{equation}\label{eq:ngamma3}
\mathcal N_\gamma^{-1}=\int\dd\mu_H(u)\int\mathcal D[Z,\gamma]\langle\Omega_0,\gamma|Z,\gamma\rangle\langle Z,\gamma|u|\Omega_0,\gamma\rangle+O(t^\infty). 
\end{equation} 
Then, applying \eqref{eq:innerproductvacuum1G} and \eqref{eq:innerproductvacuum2G} with letting $Z_{N+3}=Z=Z_0$, one may simplify the integral in \eqref{eq:ngamma3} to a Gaussian like one. The stationary phase approximation analysis as did for the integral \eqref{eq:GFin} gives that the integrand is peaked at $u=\mathbbm{1}$. Following  the arguments analogous to \eqref{eq:Ggeneral}, one gets $$\mathcal N_\gamma^{-1}=\langle\Psi_{\g_\gamma^M},\gamma|u|\Psi_{\g_\gamma^M},\gamma\rangle(1+O(t))=\mathcal N_\gamma'(1+O(t))$$
which leads to
\begin{equation}\label{eq:NNgp}
\mathcal N_\gamma\mathcal N_\gamma'=1+O(t).
\end{equation}
Combining \eqref{eq:NNgp}, \eqref{eq:FN2}, \eqref{eq:AN}, and the result of the expectation value of $\hat V_{v_j}^{-\frac12}$, and taking  the limit $N_i\to \infty$, we get
\begin{equation}
G^0[F](\vec\tau,\vec v,\gamma)=\frac{1}{(a^2\mathring{p}_\gamma\beta)^{3/2}}\langle\hat F^{\vec A}\rangle \mathfrak{F}_{\vec A}(\vec\tau,\vec v,\gamma)(1+O(t)). 
\end{equation}
Recalling  the relation \eqref{eq:GG0} and substituting the explicit expression of $\mathfrak{F}_{\vec A}(\vec\tau,\vec v,\gamma)$ given by \eqref{eq:FN}, we get the  propagator  on $\gamma$
\begin{equation}\label{eq:ggamma}
\begin{aligned}
G(\vec\tau,\vec v,\gamma)=&\frac{\hbar}{V_{\rm tot}}\sum_{\vec k \in K[\gamma]^3}\exp(i\frac{2\pi}{\ell_o}\vec k\cdot \left(\vec x_{v_1}-\vec x_{v_{2}}  \right))\times\\
&\Bigg(\theta(\tau_1-\tau_2)\frac{\mathfrak{s}(\vec k)\sigma^0-\sum_{m=1}^3\sin(\frac{2\pi\mu}{\ell_o}k^m)\sigma^m}{2\mathfrak{s}(\vec k)} e^{-i\omega(\vec k)(\tau_1-\tau_2)}\\
&-\theta(\tau_2-\tau_1)\frac{\mathfrak{s}(\vec k)\sigma^0+\sum_{m=1}^3\sin(\frac{2\pi\mu}{\ell_o}k^m)\sigma^m}{2\mathfrak{s}(\vec k)}e^{-i\omega(\vec k)(\tau_2-\tau_1)}\Bigg)
\end{aligned}
\end{equation}
where $V_{\rm tot}=(P\ell_o)^{3/2}$ is the total volume of $\gamma$ under the geometry endowed by the coherent state $|\Psi_{\g_M^\gamma},\gamma\rangle$ and $\omega(\vec k)$, by \eqref{eq:omegak1}, is
\begin{equation}\label{eq:omega2}
\omega(\vec k)=\frac{\mathfrak{s}(\vec k)}{\sqrt{P}\mu}
\end{equation}
with $\mu$ being the lattice spacing under the fiducial metric. Note that
$G(\vec\tau,\vec v,\gamma)$ is the matrix $\{G_{\vec A}(\vec\tau,\vec v,\gamma)\}_{A_1,A_2=\pm}$ with $\{G_{\vec A}(\vec\tau,\vec v,\gamma)$ given by $G[F](\vec\tau,\vec v,\gamma)=G_{\vec A}(\vec\tau,\vec v,\gamma) \langle\hat F^{\vec A}\rangle$. Moreover, $V_{\rm tot}$ is independent of graphs and actually can be interpreted as the volume of $\Sigma$ under the dynamics metric $P\delta_{ij}$. Furthermore, $\sqrt{P}\mu$ in \eqref{eq:omega2} is the lattice spacing of $\gamma$ under the dynamical metric $P\delta_{ij}$. 
 With $G(\vec\tau,\vec v,\gamma)$, the final  propagator  after superposition over graphs is
 \begin{equation}\label{eq:finaresult}
 G(\vec\tau,\vec v)=\sum_{m=1}^L|w_m|^2G(\vec\tau,\vec v,\gamma_m). 
 \end{equation}
 
 \section{Fermion doubling and its resolution}\label{sec:doubling}
\begin{figure}[ht]
	\begin{center}
	\includegraphics[width=0.5\textwidth]{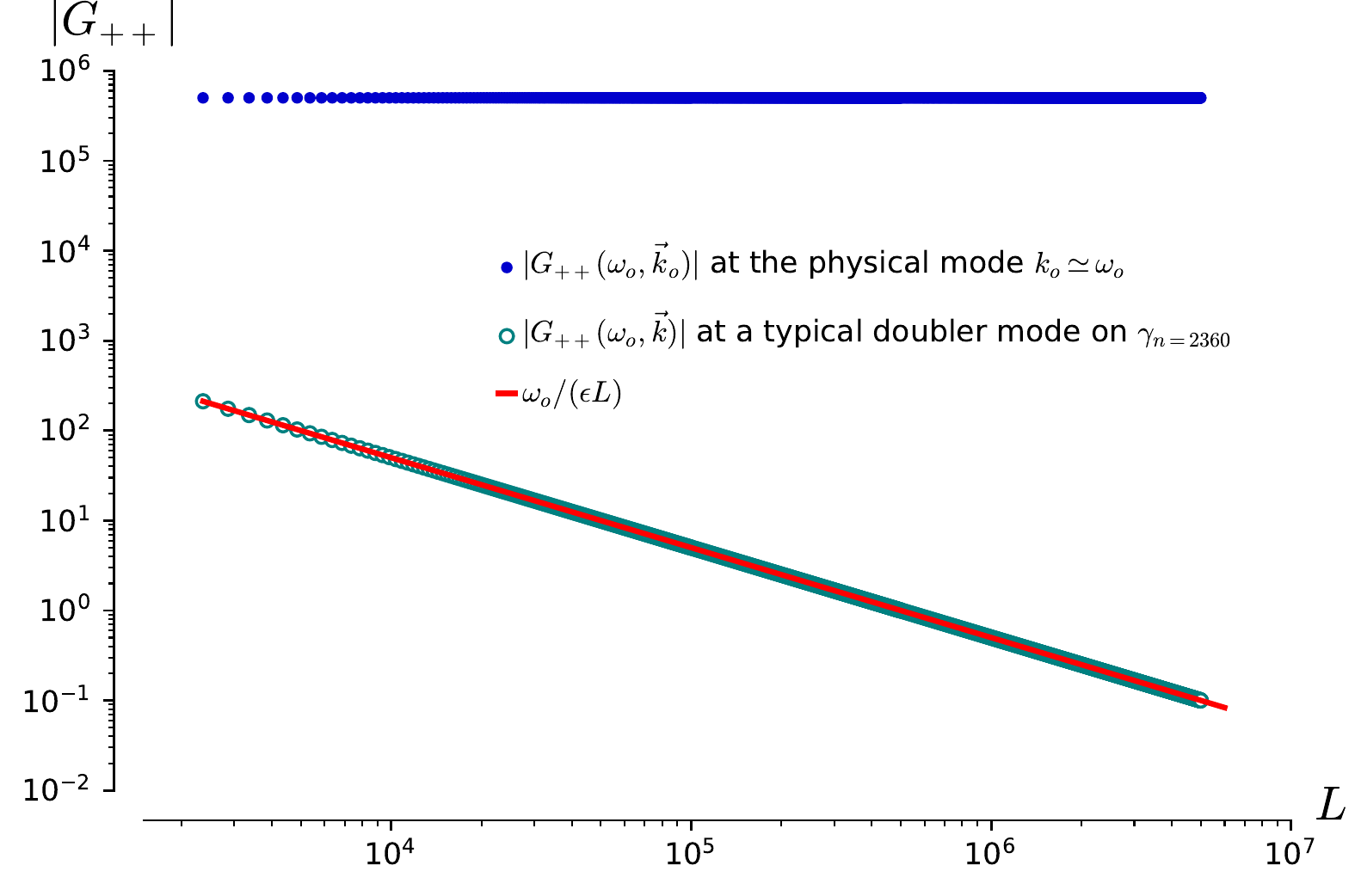}
	\caption{Log-log plots of  $|G_{++}(\o_o,\vec{k}_o)|$ at the physical mode $k_o\simeq \o_o$ (blue dots), and $|G_{++}(\o_o,\vec{k})|$ at a typical doubler mode on $\ga_{n}$ with $n=2360$ (green circles). The parameters are $\o_o=50,a_1=2^{-50}$ $\epsilon=10^{-4}$. The red line draws $\o_o/(\epsilon L)$ as a function of $L$. In this calculation, we choose $\hbar=1$. 
	}
	\label{poles}
	\end{center}
\end{figure}

 For the convenience of our further discussion, let us rescale the coordinate on $\Sigma$ from $x^i$ to $x'{}^i=\sqrt{P}x^i $ so that the dynamical metric becomes $\delta_{ij}$ in the prime coordinate. Besides, we rescale $k^j$ by $ k'{}^j =2\pi k^j/(\sqrt P\ell_o)$ so that $\vec k'$ takes values in the fundamental Brillouin zone (FBZ) of $\gamma$ which is denoted by $\mathrm{FBZ}(\gamma)$. Here, $\vec k'\in \mathrm{FBZ}(\gamma)$ means $-\pi/a\leq k'{}^j\leq \pi/a$ and $k'{}^j=-\pi/a+2n\pi/{\sqrt{P}\ell_o}$ for all $j=1,2,3$, where $n\in \mathbb Z$ and $a=\sqrt{P}\mu$ being the lattice spacing under the dynamical metric. In what follows, for convenience we will still use the unprimed letters to denote the rescaled variables.   With the rescaled variables,   $G(\vec\tau,\vec v,\gamma)$ in \eqref{eq:ggamma} can be written as 
 \begin{equation}
 \begin{aligned}
 G(\vec \tau,\vec v,\gamma)=\frac{\hbar}{V_{\rm tot}}\sum_{\vec k\in \mathrm{FBZ}(\gamma)}\int\frac{\dd \omega}{2\pi i} G_{\gamma}(\omega,\vec k) e^{-i\omega(\tau_1-\tau_2)+i\vec k\cdot(\vec x_{v_1}-\vec x_{v_2})}
 \end{aligned}
 \end{equation}
 with $G_\gamma(\omega,\vec k)$ given by
 \begin{equation}
 \begin{aligned}
 G_\gamma(\omega,\vec k)=\frac{-\omega\sigma^0+\frac{1}{a}\sum_{m=1}^3\sin(a k^m )\sigma^m}{\omega^2-\frac{1}{a^2}\sum_{m=1}^3\sin^2(a k^m )+i\epsilon}
 \end{aligned}
 \end{equation}
 As usual in QFT, $G_\gamma(\omega,\vec k)$ has the doubling problem. A physical  mode at, saying, $(\omega,\vec k)=(\omega,k,0,0)$ satisfying $\omega=\frac{1}{a}\sin(ak)>0$, implies another spurious doubler  mode at $(\omega,\frac{\pi}{a}-k,0,0)$, so the fermion species is doubled in each  direction on the lattice. The fermion doubling problem causes the trouble
of the continuum limit of fermions on lattices, and is intractably linked to chirality by the Nielsen-Ninomiya no-go theorem \cite{Nielsen:1981hk}. However, in LQG, the lattices are dynamical and define the quantum states, so that the superposition over lattices makes sense. As shown in the previous sections, once we consider the vacuum given by the superposition of graphs,  the resulting  propagator  is the average of those on the lattices.  More precisely, according to \eqref{eq:finaresult}, the propagator after the superposition of graph is 
\begin{equation}\label{eq:propagator}
G(\vec\tau,\vec v)=\frac{\hbar}{V_{\rm tot}}\sum_{\vec k\in\mathrm{FBZ}(\gamma_L)}\int\frac{\dd\omega}{2\pi i} \sum_{m=1}^L|w_m|^2 \chi_m(\vec k)G_{\gamma_m}(\omega,\vec k)e^{-i\omega(\tau_1-\tau_2)+i\vec k\cdot(\vec x_{v_1}-\vec x_{v_2})}. 
\end{equation} 
where $\chi_m(\vec k)$ is the characteristic function given by
\begin{equation}
\chi_m(\vec k)=\left\{
\begin{array}{cc}
1,\ &\vec k\in\mathrm{FBZ}(\gamma_m),\\
0,\ &\text{othereise}.
\end{array}
\right.
\end{equation}
Eq. \eqref{eq:propagator}  implies the propagator in the Fourier space is 
\begin{equation}\label{eq:average}
G(\omega,\vec k)= \sum_{m=1}^L|w_m|^2 \chi_m(\vec k)G_{\gamma_m}(\omega,\vec k). 
\end{equation}
The assumptions of Nielsen-Ninomiya theorem is clearly violated by summing over lattices. Thus, the superposition over graphs may help to resolve the doubling problem. This can be seen from the following heuristic discussion.
For each graph $\gamma_m$, the physical  mode, saying $(\omega_o,\vec k)=(\omega_o, k,0,0)$, occurs around $k=\frac{1}{a_m}\arcsin(a_m\omega_o)\approx \omega_o$. Since the physical mode is independent of the graphs, the average in \eqref{eq:average} will keep the value of $G(\omega_o,\vec k)$ around the physical  mode unchanged. However, for the doubler  modes, the one associated with $\gamma_m$ occurs at $k=\pi/{a_m}-\frac{1}{a_m}\arcsin(a_m\omega_o)\approx  \pi/a_m-\omega_o$ which depends on the graph. Therefore, the  average  in \eqref{eq:average}  will suppress $G(\omega_o,\vec k)$ around the doubler  modes.  This discussion explains the mechanism that the superposition over graphs resolves the doubling problem. The issues can also be checked by a numerical investigation. In the numerical experiment, we choose $w_m=1/\sqrt L$ for all $m$ and set the parameters as $\omega_o=50,a_1=2^{-50},L=1.2\times10^5,\epsilon=10^{-4}$. Then, the results show that $|G_{++}(\omega_o,\vec k)|<4.2$ at the doubler modes on all $\{\ga_n\}_{n=1}^L$, while $|G_{++}(\omega_o,\vec k_o)|\simeq 5\times 10^5$ (equals $\o_o/\epsilon$) at the physical mode. Moreover, as shown in Fig.\ref{poles}, $L$ goes large, $G(\omega_o,\vec{k})$ remains large and constant at the physical mode $k_o$, while it is suppressed at the doubler mode. The coincidence between $G(\omega_o,\vec{k}_n)$ for the doubler modes $\vec k_n$ and $\omega_o/(\epsilon L)$ shown in FIG. \ref{poles} indicates that $G(\omega_o,\vec{k})$ at the doubler mode is indeed only dominated by one term in the sum. In absence of the  doubling mode, $G(\o_o,\vec{k})$ is peaked at the physical mode $\vec{k}=\vec{k}_o$. In any neighborhood of $\vec{k}_o$ and with sufficiently large $L$, $G(\o_o,\vec{k})$ approximates the continuum fermion propagator arbitrarily well, due to the well-known result $\lim_{L\to\infty}\frac{1}{L}\sum_{n=1}^L f_n=\lim_{n\to\infty}f_n$ for any sequence $\{f_n\}$. 

 \section{outlook and discussion}\label{sec:conclusion}
In this paper, we investigate the propagator of the LQG coupling fermion field. Since the LQG graphs own the physical meanings, the vacuum state in LQG can thus be defined as a superposition of the states on various graphs. Applying the path integral formulation, we calculate the propagator step by step. Finally, it is shown that the graph-superposition feature results in the LQG fermion propagator being an average of the LFT fermion propagators over lattices. In the resulting fermion propagator, the doubler modes are suppressed by the average, but the physical mode remains unchanged. This result supports the fermion in LQG being free of the doubling problem. Moreover, according to our results, the superposition nature of quantum geometry should be a key to resolving the tension between fermion and the fundamental discreteness of QG. In addition, the superposition of lattices bringing the fermion propagator close to its continuous limit suggests that the LQG's continuous limit should also consider the superposition of lattices. This picture is similar to that suggested by group field theory \cite{Finocchiaro:2020fhl}.

In LQG, a state on $\gamma$ can be rewritten on a larger graph $\gamma'$ by letting the extra edges in $\gamma'$ but not in $\gamma$ take vanishing spins. Thus, the states $|\Omega,\gamma_n\rangle$ for all $1\leq n\leq L$ can be thought of as a state on $\gamma_L$. Consequently, the superposition vacuum state $|\Omega\rangle$ is a state on $\gamma_L$. The action of the Hamiltonian on $|\Omega,\gamma_n\rangle$ is defined by considering the state as a one on $\gamma_n$. Thus the Hamiltonian acts on $|\Omega,\gamma_n\rangle$ in such a way that the objects, like loops, in the Hamiltonian operator, are chosen as the minimal ones in $\gamma_n$. However, a minimal loop in $\gamma_n$ is not the minimal one in $\gamma_L$. 
That is to say, if one repeats the above calculation but regards the state as a one on $\gamma_L$, the Hamiltonian becomes nonlocal. This argument relates the doubling problem resolution proposed by the current work to the locality violation. It will be left as our future work to derive the nonlocal Hamiltonian.

Interestingly, the propagator \eqref{eq:average} suggests that the quantum geometry provides a soft UV cut-off to fermions. Let us scale $\vec{k}$ large so that $\vec{k}$ is outside $\mathrm{FBZ}(\ga_m)$ for certain $m$. Then the factor $\chi_{n}(\vec k)$ makes all terms with $n\leq m$ in \eqref{eq:average} vanish. Consequently, as we scale $\vec{k}$ larger and larger, fewer and fewer terms in the summation of the propagator survive and, thus, the propagator approaches $0$. This result is consistent with the expectation that QG should regularize the UV behavior of matter fields.

Finally, let us discuss the path integral formulation introduced in Sec. \ref{sec:pathintegral}. The path integral formula is derived following the standard procedure. Then, an issue of whether the path integral formula is well-defined arises. This is a generally existing issue in the path integral formulation and can be explored from the following two perspectives. The first is to consider the path integral as the limit of a $N$-dimensional integral as $N$ approaches $\infty$ (see \eqref{eq:limitA}). From this point of view, one needs to pay attention to the existence of the limit. It should be noted that this perspective does not view the path integral formula as an integral over the space of paths. Due to this, the other attempt is to construct a measure on the space of paths to rewrite the formula as an integral over paths. This attempt has been achieved in quantum mechanics for the coherent state path integral formulation \cite{klauder1984quantum}. The process is quite similar to defining the integral $\int_{\mathbb R}e^{ix}\dd x$ as the limit of $\int_{\mathbb R}e^{- x^2/\lambda^2+ix}\dd x/(\lambda\pi)$ as $\lambda \to 0^+$. The difference is that, in the path integral formula, the term analogous to $e^{-x^2/\lambda^2}\dd x$ in the heuristic example finally gives rise to the Wiener measure in the space of paths. Then, it should be asked if this process can be applied to LQG to achieve a mathematically rigorous path integral formula.

\section{acknowledgement}
M.H. acknowledges Chen-Hung Hsiao for discussions at early stage of this work. M.H. receives support from the National Science Foundation through grants PHY-1912278 and PHY-2207763. M.H. also acknowledges funding provided by the Alexander von Humboldt Foundation for his visit at the Friedrich-Alexander-Universit\"at Erlangen-N\"urnberg. C.Z. acknowledges Jerzy Lewandowski for discussions. C.Z. is supported by the NSFC with Grants No. 11961131013 and No. 12275022.

\appendix
\section{semiclassical properties  of  $ \widetilde{\psi_{g}^{t}}$ and $|\widetilde{\Psi_\g^t},\gamma\rangle$}\label{app:semiclassicalpsit}
To begin with, we show that $|\widetilde{\Psi_\g^t},\gamma\rangle$ is normalized  up to some negligible term.  One has
\begin{equation}
\langle \widetilde{\psi_{g}^{t}}|\widetilde{\psi_{g}^{t}}\rangle=1-\frac{t^{3/2} }{2\pi^{\frac12} e^{ t/4}}\frac{\sinh(p)}{p}e^{-\frac{ p^2}{t}}=1+O(t^{\infty}).
\end{equation}
Thus, $1-\||\widetilde{\Psi_{\g}^t},\gamma\rangle\|^2$ can be estimated by
\begin{equation}\label{eq:normtildepsi}
\begin{aligned}
1-\langle\widetilde{\Psi_{\g}^t},\gamma|\widetilde{\Psi_{\g}^t},\gamma\rangle&=1-\prod_{e\in E(\gamma)}\left(1-\frac{t^{3/2} }{2\pi^{\frac12} e^{ t/4}}\frac{\sinh(p_e)}{p_e}e^{-\frac{ p_e^2}{t}}\right)\\
&\leq \prod_{e\in E(\gamma)}\left(1+\frac{t^{3/2} }{2\pi^{\frac12} e^{ t/4}}\frac{\sinh(p_e)}{p_e}e^{-\frac{ p_e^2}{t}}\right)-1.
\end{aligned}
\end{equation}
Leting $p_o$ be the minimum in $\{p_e\}_{e\in E(\gamma)}$, we can simplify \eqref{eq:normtildepsi} to be
\begin{equation}\label{eq:normtildepsi2}
1-\langle\widetilde{\Psi_{\g}^t},\gamma|\widetilde{\Psi_{\g}^t},\gamma\rangle\leq  \left(1+\frac{t^{3/2} }{2\pi^{\frac12} e^{ t/4}}\frac{\sinh(p_o)}{p_o}e^{-\frac{ p_o^2}{t}}\right)^{|E(\gamma)|}-1.
\end{equation}
To estimate the right hand side of this equation further, we may relate $|E(\gamma)|$ and the parameter $t$ as follows. At first, since we concern ourselves with  only the cubic graphs, $|E(\gamma)|$, i.e., the number of edges of a graph $\gamma$, is proportional to $1/\mu^3$ where $\mu$ is the lattice spacing of  $\gamma$. Then, we note $p_e\sim \mu^2$ in that  $p_e$ is  the integration of the classical flux over the surface dual to $e$. Thus, $\mu^4\gg t$  for $p_e^2\gg t$. As a consequence, $|E(\gamma)|\sim 1/\mu^3\ll 1/t^{3/4}$. Combining this result with \eqref{eq:normtildepsi2}, we finally obtain 
\begin{equation}\label{eq:normtildepsi3}
1-\langle\widetilde{\Psi_{\g}^t},\gamma|\widetilde{\Psi_{\g}^t},\gamma\rangle\leq \left(1+\frac{t^{3/2} }{2\pi^{\frac12} e^{ t/4}}\frac{\sinh(p_o)}{p_o}e^{-\frac{ p_o^2}{t}}\right)^{t^{3/4}}-1\leq  c\,\frac{\sinh(p_o)}{p_o} t^{3/4}e^{-\frac{p_o^2}{t}},
\end{equation}
for some constant $c$. This result implies
\begin{equation}\label{eq:compare2innera}
\langle\widetilde{\Psi_{\g}^t},\gamma|\widetilde{\Psi_{\g}^t},\gamma\rangle=1+O(t^{\infty}).
\end{equation}

Next, we show that the expectation values of monomials of holonomies, fluxes and volume operators in the states $\widetilde{\psi_{g}^{t}}$ and $|\widetilde{\Psi_{\g}^t},\gamma\rangle$ coincide with the classical values. That  is to say, the states $\widetilde{\psi_{g}^{t}}$ and $|\widetilde{\Psi_{\g}^t},\gamma\rangle$ have the desirable semiclassical limit. To show this, it is sufficient to do the calculations on a graph $\gamma$ comprising $n$ edges which intersects at a vertex $v$ as their common end point. The orientation of $\gamma$ will be chosen such that all edges are outgoing. Indeed, for $n=6$, this graph is just a single vertex of a cubic graph carrying the edges ending at it.  Define an operator $\hat \Fj_e$ on an edge $e$ as
\begin{equation}
\hat\Fj_e=\sqrt{\sum_{i=1}^3\left(\hat p_i^{v,e}\right)^2},
\end{equation}
and consider  the operators taking the form $\prod_{l=1}^m \hat p_{i_l}^{v_l,e_l}$.
Given a state $\psi\in \mathcal H_\gamma^G$, it can be expanded as a linear combination of the basis $\prod_{e\in E(\gamma)}D^{j_e}_{m_e n_e}(h_e)$,
\begin{equation}
\psi=\sum_{\vec j,\vec m,\vec n}\psi_{\vec j,\vec m,\vec n} \prod_{e\in E(\gamma)}D^{j_e}_{m_e n_e}(h_e). 
\end{equation} 
where $\vec j$, for instance, is the abbreviation of $\{j_e\}_{e\in E(\gamma)}$.  Then, for $\psi$ in the common  domain of $\hat p_3^{v',e}$ and $\hat\Fj_e$, one has
\begin{equation}
\| \hat p_3^{v',e}\psi\|^2=\sum_{\vec j,\vec m,\vec n}\frac{|\psi_{\vec j,\vec m,\vec n}|^2}{\prod_{\vec j}d_j}|tm_e|^2\leq \sum_{\vec j,\vec m,\vec n}\frac{|\psi_{\vec j,\vec m,\vec n}|^2}{\prod_{\vec j}d_{j_e}}t^2j_e(j_e+1)=\|\hat\Fj_e\psi\|^2,
\end{equation}
where the calculation is done by assuming $v'=s_e=v$ but the inequality still holds for $v'=t_e$. For the operators $ \hat p_i^{v',e}$ with $i=1,2$, we may take their eigenstates as the basis for the expansion so that  the same calculation is applied to result in the same conclusion.  Hence we conclude 
\begin{equation}
\| \hat p_i^{v',e}\psi\|\leq \|\hat\Fj_e\psi\|,\ \forall i=1,2,3, v'=v, t_e\text{ and } e\in E(\gamma).
\end{equation}
Moreover, due to
\begin{equation}\label{eq:allJes}
[\hat\Fj_e, \hat p_i^{v',e'}]=0,\forall i=1,2,3,v'=v,t_{e'} \text{ and } e,e'\in E(\gamma),
\end{equation}
one gets
\begin{equation}
\Bigg\|\prod_{l=1}^m  \hat p_{i_l}^{v_l,e_l}\psi\Bigg\|\leq \Bigg\|\hat\Fj_{e_1}\prod_{l=2}^m  \hat p_{i_l}^{v_l,e_l}\psi\Bigg\|=\Bigg\|\prod_{l=2}^m  \hat p_{i_l}^{v_l,e_l}\hat\Fj_{e_1}\psi\Bigg\|\leq \cdots \leq  \Bigg\|\prod_{l=1}^m \hat \Fj_{e_l}\psi\Bigg\|
\end{equation}
for all $\psi$ in the common domain.

The volume operator $\hat V_v$ at $v$ is given by 
\begin{equation}
\begin{aligned}
\hat V_v=&\kappa_0\sqrt{|\hat Q_v|},\\
\hat Q_v=&\frac{1}{6}\sum_{e,e',e''\text{ at } v}\epsilon_{ijk}\epsilon(e,e',e'') \hat p_i^{v,e} \hat p_j^{v,e'} \hat p_k^{v,e''},
\end{aligned}
\end{equation}
where $\kappa_0$ is a constant and $\epsilon(e,e',e'')=0,\pm 1$ depending on the orientation of $e\wedge e'\wedge e''$ \cite{ashtekar1997quantumII}.
Given a state $\psi\in \mathcal H_\gamma^G$ lying in the domain of $\hat V_v$, one has
\begin{equation}
\|\hat V_v\psi\|^2=\langle\psi,\hat V_v^2\psi\rangle\leq \|\psi\|\|\hat V_v^2\psi\|=\|\psi\|\sqrt{\langle\psi,\hat V_v^4\psi\rangle}.
\end{equation}
Substituting the definition of $\hat V_v$ to get  $\langle\psi,\hat V_v^4\psi\rangle=\kappa_0^4\langle \psi,\hat Q_v^2\psi\rangle=\kappa_0^4\|\hat Q_v\psi\|^2$, one obtains
\begin{equation}
\|\hat V_v\psi\|^2\leq \kappa_0^2\|\psi\|\|\hat Q_v\psi\|. 
\end{equation}
Then, employing the triangle inequality and \eqref{eq:allJes} result in
\begin{equation}\label{eq:Qinequ}
\begin{aligned}
\|\hat Q_v\psi\|
\leq &\sum_{e,e',e''\text{ at } v}|\epsilon(e,e',e'')|\, \|\hat \Fj_{e}\hat \Fj_{e'}\hat\Fj_{e''}\psi\|.
\end{aligned}
\end{equation}
For convenience, we introduce  an operator $\hat\Fq_{e,e',e''}$ as
\begin{equation}
\hat \Fq_{e,e',e''}=|\epsilon(e,e',e'')|(\hat \Fj_{e}+1)(\hat \Fj_{e'}+1)(\hat\Fj_{e''}+1).
\end{equation}
Because of $\hat \Fj_{e}>0$, one has $(|\epsilon(e,e',e'')|\hat \Fj_{e}\hat \Fj_{e'}\hat\Fj_{e''})^2\leq (\Fq_{e,e',e''})^2$ and $(\Fq_{e,e',e''})^2>1$, which implies
\begin{equation}\label{eq:QgeV2}
|\epsilon(e,e',e'')|\|\hat \Fj_{e}\hat \Fj_{e'}\hat\Fj_{e''}\psi\|\leq \|\hat \Fq_{e,e',e''}\psi\|
\end{equation}
and
\begin{equation}\label{eq:Qgerater1}
\|\psi\|\leq \|\hat \Fq_{e,e',e''}\psi\|.
\end{equation}
Combing \eqref{eq:Qinequ} and\eqref{eq:QgeV2}, we have
\begin{equation}
\|\hat  Q_v\psi\|\leq \sum_{e,e',e''}\|\hat \Fq_{e,e',e''}\psi\|
\end{equation}
This equation leads to
\begin{equation}\label{eq:Volumebound}
\|\hat V_v\psi\|^2\leq \kappa_0^2\sum_{e,e',e''\text{ at } v}\|\psi\|\|\hat \Fq_{e,e',e''}\psi\|\leq \kappa_0^2\sum_{e,e',e''}\|\hat \Fq_{e,e',e''}\psi\|^2\leq \kappa_0^2\left(\sum_{e,e',e''}\|\hat \Fq_{e,e',e''}\psi\|\right)^2
\end{equation}
where the last step uses $\|\hat \Fq_{e,e',e''}\psi\|>0$. 

Thanks to \eqref{eq:Volumebound}, we now can employ the relative bound of monomials of fluxes and volume operators.  Consider the operators 
$\hat K_1^{(m)}=\prod_{l=1}^m  \hat p_{i_l}^{v_l,e_l} \hat V_v$
and 
$\hat K_2^{(m)}=\hat V_v\prod_{l=1}^m  \hat p_{i_l}^{v_l,e_l}$. They satisfy
\begin{equation}\label{eq:Koperators}
\begin{aligned}
\|\hat K_1^{(m)}\psi\| \leq\Bigg\|\prod_{l=1}^m\hat\Fj_{e_l}\hat V_v\psi\Bigg\|=\Bigg\|\hat V_v\prod_{l=1}^m\hat\Fj_{e_l}\psi\Bigg\|
\leq\kappa_0\sum_{e,e',e''\text{ at } v}\Bigg\|\Fq_{e,e',e''}\prod_{l=1}^m\hat\Fj_{e_l}\psi\Bigg\|,\\
\|\hat K_2^{(m)}\psi\|\leq\kappa_0 \sum_{e,e',e''\text{ at } v}\Bigg\|\Fq_{e,e',e''}\prod_{l=1}^m  \hat p_{i_l}^{v_l,e_l}\psi \Bigg\|\leq \kappa_0\sum_{e,e',e''\text{ at } v}\Bigg\|\Fq_{e,e',e''}\prod_{l=1}^m\hat\Fj_{e_l}\psi\Bigg\|.
\end{aligned}
\end{equation}
Since a general  monomial of fluxes and volume operators takes the form
\begin{equation}\label{eq:genealKinequall}
\hat K=\prod_{i=1}^n\hat K_{j_i}^{(m_i)},
\end{equation}
we can obtain
\begin{equation}\label{eq:Krelativebound}
\begin{aligned}
\|\hat K\psi\|&\leq \kappa_0\sum_{e_1,e_1',e_1''}\Bigg\|\hat\Fq_{e_1,e_1',e_1''}\prod_{k=1}^{m_1}\hat \Fj_{e_k} \prod_{i=2}^n\hat K_{j_i}^{(m_i)}\psi \Bigg\|\\
&= \kappa_0\sum_{e_1,e_1',e_1''}\Bigg\|\prod_{i=2}^n\hat K_{j_i}^{(m_i)}\hat\Fq_{e_1,e_1',e_1''}\prod_{k=1}^{m_1}\hat \Fj_{e_k} \psi \Bigg\|\\
&\leq\cdots\\
&\leq \kappa_0^n\sum_{\substack{e_1,e_1',e_1''\\e_2,e_2',e_2''\\\cdots\\e_n,e_n',e_n''}}\Bigg\|\prod_{i=1}^n\left(\hat\Fq_{e_i,e_i',e_i''}\prod_{k_i=1}^{m_i}\hat \Fj_{e_{k_i}}\right) \psi \Bigg\|.
\end{aligned}
\end{equation}

Now let us investigate the relative bound of operators $\hat KD^{\frac12}_{ab}(h_e)$ and $D^{\frac12}_{ab}(h_e)\hat K$ by applying \eqref{eq:Krelativebound}. Because of $\|D^{\frac12}_{ab}(h_e)\|\leq 1$, 
we get
\begin{equation}\label{eq:DK1}
\|D^{\frac12}_{ab}(h_e)\hat K\psi\|\leq \|\hat K\psi\|\leq \kappa_0^n \sum_{\substack{e_1,e_1',e_1''\\e_2,e_2',e_2''\\\cdots}}\Bigg\|\prod_{i=1}^n\left(\hat\Fq_{e_i,e_i',e_i''}\prod_{k_i=1}^{m_i}\hat \Fj_{e_{k_i}}\right) \psi \Bigg\|.
\end{equation}
Moreover, for the operator $\hat KD^{\frac12}_{ab}(h_e)$, we have
\begin{equation}\label{eq:KD1}
\|\hat KD^{\frac12}_{ab}(h_e)\psi\|\leq \kappa_0^n \sum_{\substack{e_1,e_1',e_1''\\e_2,e_2',e_2''\\\cdots}}\Bigg\|\prod_{i=1}^n\left(\hat\Fq_{e_i,e_i',e_i''}\prod_{k_i=1}^{m_i}\hat \Fj_{e_{k_i}}\right) D^{\frac12}_{ab}(h_e)\psi \Bigg\|.
\end{equation}
For the term whose norm  is involved in the right hand side, we get
\begin{equation}\label{eq:QJDpsi}
\begin{aligned}
&\prod_{i=1}^n\left(\hat\Fq_{e_i,e_i',e_i''}\prod_{k_i=1}^{m_i}\hat \Fj_{e_{k_i}}\right) D^{\frac12}_{ab}(h_e)\psi\\
=&\sum_{\vec j,\vec m,\vec n}\psi_{\vec j,\vec m,\vec n}\sum_{J_e=j_e\pm \frac12}d_{J_e} \Ff(\{j_{\tilde e}\}_{\tilde e\neq e},J_e) (-1)^{a+m_e-b-n_e}\left(
\begin{array}{ccc}
\frac12&j_e&J_e\\
a&m_e&-(a+m_e)
\end{array}
\right)\left(
\begin{array}{ccc}
\frac12&j_e&J_e\\
b&n_e&-(b+n_e)
\end{array}
\right)\times\\
&D^{J_e}_{(a+m_e)(b+n_e)}(h_e)\otimes\bigotimes_{\tilde e\neq e}D^{j_{\tilde e}}_{m_{\tilde e},n_{\tilde e}}(h_e),
\end{aligned}
\end{equation}
where $\Ff(\{j_{\tilde e}\}_{\tilde e\neq e},J_e)$ denotes the eigenvalue of the operator $\prod_{i=1}^n\left(\hat\Fq_{e_i,e_i',e_i''}\prod_{k_i=1}^{m_i}\hat \Fj_{e_{k_i}}\right)$ acting on $D^{J_e}_{(a+m_e)(b+n_e)}(h_e)\otimes\bigotimes_{\tilde e\neq e}D^{j_{\tilde e}}_{m_{\tilde e},n_{\tilde e}}(h_e)$. 
Then a straightforward calculation gives 
\begin{equation}
\begin{aligned}
&\Bigg\|\prod_{i=1}^n\left(\hat\Fq_{e_i,e_i',e_i''}\prod_{k_i=1}^{m_i}\hat \Fj_{e_{k_i}}\right) D^{\frac12}_{ab}(h_e)\psi\Bigg\|^2
\leq & 4\sum_{\vec j,\vec m,\vec n}\frac{|\psi_{\vec j,\vec m,\vec n}|}{\prod_{\tilde e}d_{j_{\tilde e}}}\Big (\Ff(\{j_{\tilde e} \}_{\tilde e\neq e},j_e+\frac12)^2+\Ff(\{j_{\tilde e} \}_{\tilde e\neq e},j_e-\frac12)^2\Big )
\end{aligned}
\end{equation}
where it is necessary to employ the inequality 
\begin{equation}
\Bigg|d_j\left(
\begin{array}{ccc}
\frac12&j\pm \frac12&j\\
a&m&-(a+m)
\end{array}
\right)\left(
\begin{array}{ccc}
\frac12&j\pm \frac12&j\\
b&n&-(b+n)
\end{array}
\right)\Bigg|^2\leq 2
\end{equation}
Since $\Ff(\{j_{\tilde e} \}_{\tilde e\neq e},j_e\pm \frac12)$ is a polynomial of $\sqrt{j_{\tilde e}(j_{\tilde e}+1)}$ (for all $ \tilde e\neq e$) and $\sqrt{(j_e\pm \frac12)(j_e\pm \frac12+1)}$, there exists a positive constant $\beta$ such that
\begin{equation}
\Ff(\{j_{\tilde e} \}_{\tilde e\neq e},j_e+\frac12)^2+\Ff(\{j_{\tilde e} \}_{\tilde e\neq e},j_e-\frac12)^2\leq 2\, \Ff(\{j_{\tilde e} \}_{\tilde e\neq e},j_e)^2+\beta^2, \forall \vec j.
\end{equation}
Therefore, we have
\begin{equation}
\Bigg\|\prod_{i=1}^n\left(\hat\Fq_{e_i,e_i',e_i''}\prod_{k_i=1}^{m_i}\hat \Fj_{e_{k_i}}\right) D^{\frac12}_{ab}(h_e)\psi\Bigg\|^2
\leq 2 \Bigg\|\prod_{i=1}^n\left(\hat\Fq_{e_i,e_i',e_i''}\prod_{k_i=1}^{m_i}\hat \Fj_{e_{k_i}}\right)\psi\Bigg\|^2+\beta^2\|\psi\|^2
\end{equation}
This implies 
\begin{equation}
\Bigg\|\prod_{i=1}^n\left(\hat\Fq_{e_i,e_i',e_i''}\prod_{k_i=1}^{m_i}\hat \Fj_{e_{k_i}}\right) D^{\frac12}_{ab}(h_e)\psi\Bigg\|\leq \sqrt{2}\Bigg\|\prod_{i=1}^n\left(\hat\Fq_{e_i,e_i',e_i''}\prod_{k_i=1}^{m_i}\hat \Fj_{e_{k_i}}\right)\psi\Bigg\|+\beta\|\psi\|.
\end{equation}
Substituting this inequality into \eqref{eq:KD1}, one finally obtains
\begin{equation}\label{eq:DK2}
\|\hat KD^{\frac12}_{ab}(h_e)\psi\|\leq \sqrt{2}\kappa_0^n\sum_{\substack{e_1,e_1',e_1''\\e_2,e_2',e_2''\\\cdots}}\Bigg\|\prod_{i=1}^n\left(\hat\Fq_{e_i,e_i',e_i''}\prod_{k_i=1}^{m_i}\hat \Fj_{e_{k_i}}\right) \psi \Bigg\|+\beta'\|\psi\|. 
\end{equation}
for some constant $\beta'$. For a general monomial $\hat M$ of fluxes, holonomies and volume operator, one only  needs to do the derivations \eqref{eq:DK1} and \eqref{eq:DK2}  successively to get 
\begin{equation}\label{eq:Mpsi}
\|\hat M\psi\|\leq \alpha \kappa_0^N \sum_{\substack{e_1,e_1',e_1''\\e_2,e_2',e_2''\\\cdots\\e_N,e_N',e_N''}}\Bigg\|\prod_{i=1}^N\hat\Fq_{e_i,e_i',e_i''} \widehat{\mathscr{J}}\psi\Bigg\|+\beta''\|\psi\|,
\end{equation}
where $\alpha$ and $\beta''$ are some constants depending on the explicit expression of $\hat M$, $N$ is the number of volume operators in $\hat M$, and $\widehat{\mathscr{J}}$ is the operator
\begin{equation}
\widehat{\mathscr{J}}:=\hat M\Big|_{D_{ab}^{\frac12}(h_e)\to 1,\hat V_v\to 1,\hat p_i^{v',e}\to \hat\Fj_e}. 
\end{equation}

Now we can study the expectation value of $\hat M$ in the coherent state $|\widetilde{\Psi_{\g}^t},\gamma\rangle$. A straightforward calculation gives 
\begin{equation}\label{eq:MpsiPpsi}
\begin{aligned}
&\Big|\langle\Psi_\g^t,\gamma|\hat  M|\Psi_\g^t,\gamma\rangle-\langle \widetilde{\Psi_\g^t},\gamma|\hat  M| \widetilde{\Psi_\g^t},\gamma\rangle\Big|\\
\leq &2\Big\|(1-\p_\gamma)| \Psi_\g^t,\gamma\rangle\Big\|\,\Big\|\hat  M| \Psi_\g^t,\gamma\rangle\Big\|+\Big\|(1-\p_\gamma) | \Psi_\g^t,\gamma\rangle \Big\|\, \Big\| \hat  M(1-\p_\gamma)| \Psi_\g^t,\gamma\rangle\Big\|.
\end{aligned}
\end{equation}
Here $\Big\|\hat  M| \Psi_\g^t,\gamma\rangle\Big\|^2$ is just the expectation value of $\hat M^\dagger\hat M$ in the regular complexifier coherent state. Then employing the results in \cite{thiemann2001gauge} results in
\begin{equation}\label{eq:Mpsinorm}
\Big\|\hat  M| \Psi_\g^t,\gamma\rangle\Big\|^2=M(\g)^*M(\g)+O(t)
\end{equation}
where $M(\g)$ denotes the classical value of the observable $\hat M$ at $\g$. 
For  $\Big\| \hat  M(1-\p_\gamma)| \Psi_\g^t,\gamma\rangle\Big\|$ in \eqref{eq:MpsiPpsi}, \eqref{eq:Mpsi} leads to
\begin{equation}\label{eq:M1mppsi}
\begin{aligned}
\Big\| \hat  M(1-\p_\gamma)| \Psi_\g^t,\gamma\rangle\Big\|\leq \alpha \kappa_0^N \sum_{\substack{e_1,e_1',e_1''\\e_2,e_2',e_2''\\\cdots\\e_N,e_N',e_N''}}\Bigg\|\prod_{i=1}^N\hat\Fq_{e_i,e_i',e_i''} \widehat{\mathscr{J}}(1-\p_\gamma)| \Psi_\g^t,\gamma\rangle\Bigg\|+\beta''\|(1-\p_\gamma)| \Psi_\g^t,\gamma\rangle\|
\end{aligned}
\end{equation}
Due to $[\hat\Fj_e,P_\gamma]=0$ for all $e\in E(\gamma)$, one has
\begin{equation}\label{eq:normDJ1mp}
\begin{aligned}
\Bigg\|\prod_{i=1}^N\hat\Fq_{e_i,e_i',e_i''} \widehat{\mathscr{J}}(1-\p_\gamma)| \Psi_\g^t,\gamma\rangle\Bigg\|=\Bigg\|\prod_{i=1}^N\hat\Fq_{e_i,e_i',e_i''} \widehat{\mathscr{J}}| \Psi_\g^t,\gamma\rangle\Bigg\|-\Bigg\|\prod_{i=1}^N\hat\Fq_{e_i,e_i',e_i''} \widehat{\mathscr{J}}\p_\gamma| \Psi_\g^t,\gamma\rangle\Bigg\|.
\end{aligned}
\end{equation}
By definition, $\prod_{i=1}^N\hat\Fq_{e_i,e_i',e_i''} \widehat{\mathscr{J}}$ takes the form 
\begin{equation}\label{eq:Pjpsie0}
\prod_{i=1}^N\hat\Fq_{e_i,e_i',e_i''} \widehat{\mathscr{J}}=\prod_{e\in E(\gamma)}P_{k_e}(\hat\Fj_e)
\end{equation}
where $P_{k_e}(\hat\Fj_e)$ is a polynomial of $\hat\Fj_e$ with degree $k_e$. Let us define $\p_e$ as the projection 
\begin{equation}
\p_e:\bigoplus_{j_e\geq 0}\mathcal H_{j_e}\otimes \mathcal H_{j_e}^*\to \bigoplus_{j_e\neq 0}\mathcal H_{j_e} \otimes \mathcal H_{j_e}^*
\end{equation}
so that $\p_\gamma=\prod_{e\in E(\gamma)}\p_e$. We have
\begin{equation}\label{eq:Pjpsie1}
\begin{aligned}
\Bigg\|\prod_{i=1}^N\hat\Fq_{e_i,e_i',e_i''} \widehat{\mathscr{J}}\p_\gamma|=\prod_{e\in E( \gamma)}\|P_{k_e}(\hat\Fj_e)\p_e\psi_{g_e}^t\|
\end{aligned}
\end{equation}
Because of $[\p_e,\hat\Fj_e]=0$, one has
\begin{equation}\label{eq:Pjpsie2}
\begin{aligned}
\|P_{k_e}(\hat\Fj_e)\p_e\psi_{g_e}^t\|=&\|P_{k_e}(\hat\Fj_e)\psi_{g_e}^t\|-\|P_{k_e}(\hat\Fj_e)(1-\p_e)\psi_{g_e}^t\|\\
=&|P_{k_e}(p_e)|+O(t)-\frac{t^{3/4} e^{-\frac{ p_e^2}{2t}}}{\sqrt{2}\pi^{\frac14} e^{ t/8}}\sqrt{\frac{\sinh(p_e)}{p_e} }|P_{k_e}(0)|\\
=&|P_{k_e}(p_e)|-O(t^\infty),
\end{aligned}
\end{equation}
where we  used $\|P(\hat\Fj_e)\psi_{g_e}^t\|^2=\langle\psi_{g_e^t}P_{k_e}(\hat\Fj^e)P_{k_e}(\hat\Fj^e)^\dagger\psi_{g_e}^t\rangle=|P_{k_e}(p_e)|^2+O(t)$. Then combining \eqref{eq:normDJ1mp} \eqref{eq:Pjpsie0}, \eqref{eq:Pjpsie1}, and \eqref{eq:Pjpsie2}, we  get
\begin{equation}\label{eq:Pjpsie}
\Bigg\|\prod_{i=1}^N\hat\Fq_{e_i,e_i',e_i''} \widehat{\mathscr{J}}(1-\p_\gamma)| \Psi_\g^t,\gamma\rangle\Bigg\|=\prod_{e\in E(\gamma)}|P_{k_e}(p_e)|-\prod_{e\in E(\gamma)}(|P_{k_e}(p_e)|-O(t^\infty))=O(t^\infty). 
\end{equation}
Finally, substituting \eqref{eq:Mpsinorm}, \eqref{eq:Pjpsie} an \eqref{eq:M1mppsi} into \eqref{eq:MpsiPpsi}, we get
\begin{equation}
\Big|\langle\Psi_\g^t,\gamma|\hat  M|\Psi_\g^t,\gamma\rangle-\langle \widetilde{\Psi_\g^t},\gamma|\hat  M| \widetilde{\Psi_\g^t},\gamma\rangle\Big|= O(t^\infty)
\end{equation}
where we used
\begin{equation}\label{eq:1mpgamma}
\Big\|(1-\p_\gamma)|\Psi_\g^t,\gamma\rangle\Big\|=1-\langle\widetilde{\Psi}_\g^t,\gamma|\widetilde{\Psi}_\g^t,\gamma\rangle=O(t^\infty).
\end{equation}
Therefore, the expectation value of $\hat M$ in $|\widetilde{\Psi_\g^t},\gamma\rangle$ is the same as that in the regular compexifier coherent state up to some $O(t^\infty)$ term, i.e.,
\begin{equation}
\langle\Psi_\g^t,\gamma|\hat  M|\Psi_\g^t,\gamma\rangle=\langle \widetilde{\Psi_\g^t},\gamma|\hat  M| \widetilde{\Psi_\g^t},\gamma\rangle+O(t^\infty). 
\end{equation}

\section{Calculation of $\mathfrak{F}^{\vec N}_{\vec A}(\vec\tau,\vec v,\gamma)$}\label{app:calculatingF}
\subsection{The Hession matrix of $S_{\vec N}^F[\g_M^\gamma,\nu]$}

For the integral in \eqref{eq:GNABF}, we need the Hession matrix $\hf^{\vec N}$ given by
\begin{equation}\label{eq:hessionsf}
 S_{\vec N}^F[\g_M^\gamma,\nu]=\sum_{j,j'=0}^{N+3}\sum_{v,v'\in V(\gamma)}
 \nu^\dagger_{j'}(v')
 (\hf^{\vec N})_{(v',j'),(v,j)}
  \nu_{j}(v).
 \end{equation}
As shown in \eqref{eq:Sf}, $S_{\vec N}^F[\g_M^\gamma,\nu]$ contains the matrix element of the Hamiltonian operator $\p_\gamma\hat H_\gamma^F\p_\gamma$ between $|Z_j,\gamma\rangle$ and $\langle Z_{j+1},\gamma|$. Since the stationary phase of the gravity path integral has fixed the background geometry to the Minkowski geometry for all moments, 
$|Z_j,\gamma\rangle$ for each $j$ thus takes the form $|Z_j,\gamma\rangle=|\Psi_{\g_\gamma^M}^t,\gamma\rangle\otimes |\Phi_{\nu_j},\gamma\rangle$. As a consequence, sandwiched by $|Z_j,\gamma\rangle$ and $\langle Z_{j+1},\gamma|$, the holonomy and flux operators in $\p_\gamma\hat H_\gamma^F\p_\gamma$ becomes their expectation value in the states $\p_\gamma|\Psi_{\g_\gamma^M}^t,\gamma\rangle$. Since we are concerned with only the leading order of the integral \eqref{eq:GNABF}, we thus only need to calculate the leading order of the holonomy and flux operators'  expectation values. According to \eqref{eq:differenceexpect}, the leading order of these expectation values are the same as that in the states $|\Psi_{\g_\gamma^M}^t,\gamma\rangle$ and thus compatible with their classical value. We thus have
 \begin{equation}\label{eq:hfmatrixele}
 \begin{aligned}
 &\frac{\langle\Psi_{\g_M^\gamma}^t,\gamma|\otimes\langle\nu_{k+1}|\p_\gamma \hat  H_F\p_\gamma|\nu_k\rangle\otimes|\Psi_{\g_M^\gamma}^t,\gamma\rangle}{\langle\nu_{k+1}|\nu_k\rangle}\\
 =&\frac{i \hbar}{2}\frac{1}{a\sqrt{ \mathring{p}_{\gamma}\beta}} \sum_{v\in V(\gamma)}\sum_{e\in E_v(\gamma)}\left(\nu_{k+1}^\dagger(v) \sigma^{e}\nu_k(v+\delta_e)-\nu_{k+1}^\dagger(v+\delta_e)\sigma^{e} \nu_k (v) \right)
 \end{aligned}
 \end{equation}
 where we use the same notations as in \eqref{eq:HeffMaction}.  Substituting \eqref{eq:hfmatrixele} into \eqref{eq:Sf} and comparing the result with the definition \eqref{eq:hessionsf}, we get
 \begin{equation}\label{eq:HessionF}
 \begin{aligned}
 \frac{(\hf^{\vec N})_{(v',j'),(v,j)}}{t}=&\delta_{v',v}\delta_{j',j+1}-\delta_{j',j}\delta_{v',v}+\frac{\delta \tau_j}{2a\sqrt{\mathring{p}_{\gamma}\beta}}  \sum_{e\in E_v(\gamma)} (\delta_{v',v-\delta_e}-\delta_{v',v+\delta_e})\delta_{j',j+1}\sigma^e.
 \end{aligned}
 \end{equation} 
 The Hessian matrix $\hf^{\vec N}$ can be block diagonalized under the Bogoliubov transformation,
  \begin{equation}\label{eq:fouriertransformationnu}
\widehat{\nu}_j(\vec k)=\sqrt{\frac{\mu^3}{\ell_o^3}}\sum_{v\in V(\gamma)}\Theta(\vec k)\nu_j(v) e^{-i\frac{2\pi}{\ell_o}\vec k\cdot\vec x_v},
\end{equation}
which gives rise to the transform of $\hf^{\vec N}$
\begin{equation}\label{eq:Hession0}
 \begin{aligned}
\frac{ \widehat\hf^{\vec N}(\vec k,A)_{j',j}}{t}
= &-\delta_{j',j}+\delta_{j',j+1}\left(1-A\frac{i  \delta \tau_j}{a\sqrt{\mathring{p}_{\gamma}\beta}}\mathfrak{s}(\vec k)\right),\ \forall A=\pm,
 \end{aligned}
 \end{equation}
so that 
 \begin{equation}\label{eq:SfDFT}
 \begin{aligned}
 S_N^F
=& \sum_{\substack{\vec k\in K[\gamma]^3\\S=\pm}}\sum_{j,j'=0}^{N+3}
 \widehat{\nu}_{j',S}(\vec k)^*
  \widehat \hf^{\vec N}(\vec k,S)_{j',j}
   \widehat{\nu}_{j,S}(\vec k).
 \end{aligned}
 \end{equation}
 Substituting \eqref{eq:fouriertransformationnu} and \eqref{eq:SfDFT} into \eqref{eq:GNABF} leads to
\begin{equation}
\begin{aligned}
\mathfrak{F}^{\vec N}_{\vec A}(\vec\tau,\vec v,\gamma)=&\int \mathcal D[\hat \nu,\gamma] \exp(\frac{1}{t}\sum_{\substack{\vec k\in K[\gamma]^3\\S=\pm}}\sum_{j,j'=0}^{N+3}
 \widehat{\nu}_{j',S}(\vec k)^*
  \widehat \hf^{\vec N}(\vec k,S)_{j',j}
   \widehat{\nu}_{j,S}(\vec k))\times\\
&\mathcal F_{\vec A}\prod_{\vec k\in K[\gamma]^3}\widehat{\nu}_{N+3,-}(\vec k)\left(\prod_{\vec k\in K[\gamma]^3}\widehat{\nu}_{0,-}(\vec k)\right)^*,
\end{aligned}
\end{equation}
where $\mathcal D[\widehat\nu,\gamma]$ is
 \begin{equation}
 \begin{aligned}
 \mathcal D[\widehat \nu,\gamma]=\prod_{i=0}^{N+3}\prod_{\vec k\in K[\gamma]^3}\prod_{A=\pm } \dd\widehat{\nu}_{i,A}(\vec k)\dd\widehat{\nu}_{i,A}(\vec k)^*.
 \end{aligned}
 \end{equation}
 Let $\widehat{\mathcal F}_{\vec B}$ be the transform of $\mathcal F_{\vec A}$ under the transformation \eqref{eq:fouriertransformationnu}, i.e.,
 \begin{equation}
 \begin{aligned}
 \widehat{\mathcal F}_{\vec B}(\vec k_1,\vec k_2)=\hbar\, \widehat{\nu}_{\mathfrak n(1),B_1}(\vec k_1) \widehat{\nu}_{\mathfrak n(2)+1,B_{2}}^*(\vec k_{2}),\  \text{with }\vec k=(k_1,k_2).
 \end{aligned}
 \end{equation}
 With $\widehat{\mathcal F}_{\vec B}$, we defined $\widehat{\mathfrak{F}}^{\vec N}_{\vec B}(\vec\tau,\vec k_1,\vec k_2,\gamma)$ as
\begin{equation}\label{eq:Ghat00}
\begin{aligned}
\widehat{\mathfrak{F}}^{\vec N}_{\vec B}(\vec\tau,\vec k_1,\vec k_2,\gamma)=&\int \mathcal D[\hat \nu,\gamma] \exp(\frac{1}{t}\sum_{\substack{\vec k'\in K[\gamma]^3\\S=\pm}}\sum_{j,j'=0}^{N+3}
 \widehat{\nu}_{j',S}(\vec k')^*
  \widehat \hf^{\vec N}(\vec k',S)_{j',j}
   \widehat{\nu}_{j,S}(\vec k))\times\\
&\widehat{\mathcal F}_{\vec B}(\vec k_1,\vec k_2)\prod_{\vec k''\in K[\gamma]^3}\widehat{\nu}_{N+3,-}(\vec k'')\left(\prod_{\vec k'''\in K[\gamma]^3}\widehat{\nu}_{0,-}(\vec k''')\right)^*
\end{aligned}
\end{equation}
such that
\begin{equation}\label{eq:ftauv}
\begin{aligned}
\mathfrak{F}^{\vec N}_{\vec A}(\vec\tau,\vec v,\gamma)=&\frac{\mu^3}{\ell_o^3}\sum_{(\vec k_1,\vec k_2) \in K_2[\gamma]}\left( \Theta(\vec k_1)^\dagger_{A_1,B_1} \Theta(\vec k_{2})_{B_{2}A_{l'}} \right)\times\\ &\exp(i\frac{2\pi}{\ell_o}\left(\vec k_1\cdot\vec x_{v_1}-\vec k_{2}\cdot\vec x_{v_{2}}  \right))\widehat{\mathfrak{F}}^{\vec N}_{\vec B}(\vec\tau,\vec k_1,\vec k_2,\gamma).
\end{aligned}
\end{equation}
 where $K_2[\gamma]$ denotes $K[\gamma]^3\times K[\gamma]^3$. 
 
 To perform the integral  in \eqref{eq:Ghat00}, we employ the the result for Gaussian integral of complex fermions  \cite{caracciolo2013algebraic} as follows
\begin{equation}\label{eq:Gaussianfermion}
\int D[\psi,\psi^\dagger]\left(\prod_{\alpha=1}^r\psi_{i_\alpha}^*\psi_{j_\alpha}\right)\exp[\psi^\dagger A\psi]=\det(A)\det((A^{-T})_{IJ}),
\end{equation}
with $\mathcal D[\psi,\psi^\dagger]:=\prod_{j}\dd\psi_j\dd\psi_j^*$, $I:=(i_1,i_2,\cdots,i_r)$ and $J:=(j_1,j_2,\cdots,j_r)$. 
According to this formula, we study the determinant and the inverse of the Hessian matrix $\widehat\hf^{\vec N}$. 
 Writing the matrix in the matrix form, we get
 \begin{equation}
 \begin{aligned}
 \frac{\widehat\hf^{\vec N}(\vec k,S)}{t}=
 \begin{pmatrix}
 -1 &0&0&0&\cdots&0\\
\am_1(\vec k,S)&-1&0&0&\cdots&0\\
0&\am_2(\vec k,S)&-1&0&\cdots&0\\
\vdots&\vdots&\vdots&\ddots&\vdots&\vdots\\
0&0&\cdots&\am_{N+2}(\vec k,S)&-1&0\\
0&0&\cdots&0&\am_{N+3}(\vec k,S)&-1
 \end{pmatrix}
 ,
 \end{aligned}
 \end{equation}
 with 
 \begin{equation}
 \begin{aligned}\label{eq:defineaj}
  \am_j(\vec k,S)=1-S\frac{i  \delta \tau_j}{a\sqrt{\mathring{p}_{\gamma}\beta}}\mathfrak{s}(\vec k).
  \end{aligned}
 \end{equation} 
The fact that this matrix is lower triangular results in
 \begin{equation}\label{eq:detHf}
 \det(\frac{\widehat\hf^{\vec N}(\vec k,S)}{t})=(-1)^{N+4}=\pm 1.
 \end{equation}
 Moreover, one can check that the inverse of $\widehat\hf^{\vec N}(\vec k,S)$ is given by
 \begin{equation}\label{eq:inverseHf}
 \begin{aligned}
 t(\widehat\hf^{\vec N}(\vec k,S)^{-1})_{mn}=\left\{
 \begin{array}{cc}
0,&\  m<n,\\
1,&\ m=n,\\
 -\prod_{l=n+1}^m\am_l(\vec k,S),&\  m> n,
 \end{array}
 \right.
 \end{aligned}
 \end{equation}

 \subsection{result of $\widehat{\mathfrak{F}}^{\vec N}_{\vec B}(\vec\tau,\vec k_1,\vec k_2,\gamma)$}
 Let us now turn to the integral in \eqref{eq:Ghat00}. Since the Hession matrix has been block diagonalized with respect to modes, the integral  in \eqref{eq:Ghat00} thus can be calculated mode by mode by applying \eqref{eq:Gaussianfermion}. At first, for $\vec k_1\neq \vec k_2$ the integral of $\prod_{i=0}^{N+3}\prod_{A=\pm }\dd\widehat{\nu}_{i,A}(\vec k_1)\dd\widehat{\nu}_{i,A}(\vec k_1)^*$ vanishes $\widehat{\mathfrak{F}}^{\vec N}_{\vec B}(\vec\tau,\vec k_1,\vec k_2,\gamma)$ clearly. We thus only need to consider the case with $\vec k_1=\vec k_2$. Then, for the mode $\vec k$ satisfying $\vec k\neq \vec k_1$, the integral of $\prod_{i=0}^{N+3}\prod_{A=\pm}\dd\widehat{\nu}_{i,A}(\vec k)\dd\widehat{\nu}_{i,A}(\vec k)^*$, denoted by $I_{\vec k}$, is 
 $$I_{\vec k}=-\prod_{S=\pm}\det(\frac{\widehat\hf^{\vec N}(\vec k,S)}{t} )t(\widehat\hf^{\vec N}(\vec k,-)^{-1})_{N+3,0}=\prod_{l=1}^{N+3}\am_l(\vec k,-).$$ 
 According to \eqref{eq:defineaj}, for $\delta\tau\ll 1$ and, equivalently, $N\gg 1$, we have 
 \begin{equation}\label{eq:aapexp}
 I_{\vec k}=\exp( i\frac{\mathfrak{s}(\vec k)}{a\sqrt{\mathring{p}_\gamma\beta}}\sum_{j=1}^{N+3}\delta\tau_j)+O(\delta\tau)=1+O(\delta\tau)
 \end{equation}
 where in the last step we applied $\sum_{j=1}^{N+3}\delta\tau_j=0$ resulting from \eqref{eq:deltatauj}. For the mode $\vec k_1$, using $I_{\vec k_1,\vec B}$ to denote the result, we have
 \begin{equation}
 \begin{aligned}
\hbar^{-1} I_{\vec k_1,\vec B}=&\det
 \begin{pmatrix}
\delta_{B_1,B_2}t(\widehat\hf^{\vec N}(\vec k,B_1)^{-1})_{\mathfrak n(1),\mathfrak n(2)+1}& \delta_{B_1,-}t(\widehat\hf^{\vec N}(\vec k,-)^{-1})_{\mathfrak n(1),0}\\
\delta_{B_2,-}t(\widehat\hf^{\vec N}(\vec k,-)^{-1})_{N+3,\mathfrak n(2)+1}& t(\widehat\hf^{\vec N}(\vec k,-)^{-1})_{N+3,0}\\
 \end{pmatrix}\\
=&\det
 \begin{pmatrix}
\delta_{B_1,B_2}\theta(\tau_1-\tau_2)e^{-B_1i\frac{\mathfrak{s}(\vec k_1)}{a\sqrt{\mathring{p}_\gamma\beta}} (\tau_1-\tau_2)}& \delta_{B_1,-}e^{i\frac{\mathfrak{s}(\vec k_1)}{a\sqrt{\mathring{p}_\gamma\beta}} \tau_1}\\
\delta_{B_2,-}e^{-i\frac{\mathfrak{s}(\vec k_1)}{a\sqrt{\mathring{p}_\gamma\beta}} \tau_2}&1\\
 \end{pmatrix}
 +O(\delta\tau)
  \end{aligned}
 \end{equation}
 where we used the same arguments as in \eqref{eq:aapexp} and the step function $\theta(\tau_1-\tau_2)$ results from that $t(\widehat\hf^{\vec N}(\vec k,B_1)^{-1})_{\mathfrak n(1),\mathfrak n(2)+1}$ is proportional to  $\theta(\mathfrak n(1)-\mathfrak n(2))$. Performing the determinant and  doing some straightforwards calculations, we finally get
 \begin{equation}\label{eq:Ik1}
 \begin{aligned}
 \hbar^{-1}I_{\vec k_1,\vec B}=&\delta_{B_1,+}\delta_{B_2,+}\theta(\tau_1-\tau_2)e^{-i\frac{\mathfrak{s}(\vec k_1)}{a\sqrt{\mathring{p}_\gamma\beta}} (\tau_1-\tau_2)}-\delta_{B_1,-}\delta_{B_2,-}\theta(\tau_2-\tau_1)e^{-i\frac{\mathfrak{s}(\vec k_1)}{a\sqrt{\mathring{p}_\gamma\beta}}(\tau_2- \tau_1)}
 \end{aligned}
 \end{equation}
 Combining the results \eqref{eq:aapexp} and \eqref{eq:Ik1}, one obtains 
 \begin{equation}
 \begin{aligned}
 \widehat{\mathfrak{F}}^{\vec N}_{\vec B}(\vec\tau,\vec k_1,\vec k_2,\gamma)=\delta_{k_1,k_2}\hbar \left(\delta_{B_1,+}\delta_{B_2,+}\theta(\tau_1-\tau_2)e^{-i\frac{\mathfrak{s}(\vec k_1)}{a\sqrt{\mathring{p}_\gamma\beta}} (\tau_1-\tau_2)}-\delta_{B_1,-}\delta_{B_2,-}\theta(\tau_2-\tau_1)e^{-i\frac{\mathfrak{s}(\vec k_1)}{a\sqrt{\mathring{p}_\gamma\beta}}(\tau_2- \tau_1)}
 \right)+O(\delta\tau).
 \end{aligned}
 \end{equation}
 Substituting this result into \eqref{eq:ftauv} and noting that 
\begin{equation}
\begin{aligned}
\pm\Theta(\vec k)_{A_1,\pm}^\dagger\Theta(\vec k)_{\pm,A_2}=\frac{\pm \mathfrak{s}(\vec k)(\sigma^0)_{A_1,A_2}-\sum_{m=1}^3\sin(\frac{2\pi\mu}{\ell_o}k^m)(\sigma^m)_{A_1,A_2}}{2\mathfrak{s}(\vec k)},
\end{aligned}
\end{equation}
we get the result of $\mathfrak{F}^{\vec N}_{\vec A}(\vec\tau,\vec v,\gamma)$,
\begin{equation}\label{eq:FN}
\begin{aligned}
\mathfrak{F}^{\vec N}(\vec\tau,\vec v,\gamma)=&\frac{\mu^3}{\ell_o^3}\hbar\sum_{\vec k \in K[\gamma]^3}\exp(i\frac{2\pi}{\ell_o}\vec k\cdot \left(\vec x_{v_1}-\vec x_{v_{2}}  \right))\times\\
&\Bigg(\theta(\tau_1-\tau_2)\frac{\mathfrak{s}(\vec k)\sigma^0-\sum_{m=1}^3\sin(\frac{2\pi\mu}{\ell_o}k^m)\sigma^m}{2\mathfrak{s}(\vec k)} e^{-i\omega(\vec k)(\tau_1-\tau_2)}\\
&-\theta(\tau_2-\tau_1)\frac{\mathfrak{s}(\vec k)\sigma^0+\sum_{m=1}^3\sin(\frac{2\pi\mu}{\ell_o}k^m)\sigma^m}{2\mathfrak{s}(\vec k)}e^{-i\omega(\vec k)(\tau_2-\tau_1)}\Bigg)+O(\delta\tau).
\end{aligned}
\end{equation}
where $\mathfrak{F}^{\vec N}(\vec\tau,\vec v,\gamma):=\lim_{N_i\to\infty}\mathfrak{F}^{\vec N}(\vec\tau,\vec v,\gamma)$ with $\mathfrak{F}^{\vec N}(\vec\tau,\vec v,\gamma)$ denoting the matrix taking $\mathfrak{F}^{\vec N}_{\vec A}(\vec\tau,\vec v,\gamma)$ for $A_1,A_2=\pm$ as its entries and we defined $\omega(\vec k)$ as
\begin{equation}\label{eq:omegak1}
\omega(\vec k)=\frac{\mathfrak{s}(\vec k)}{a\sqrt{\mathring{p}_\gamma\beta}}.
\end{equation}



\providecommand{\href}[2]{#2}\begingroup\raggedright\endgroup

\end{document}